\documentclass[fleqn,usenatbib]{mnras}
\usepackage{soul}
\usepackage{newtxtext,newtxmath}

\usepackage[T1]{fontenc}
\usepackage{ae,aecompl}
\usepackage{subfig}
\usepackage[dvipdfmx]{graphicx}	
\usepackage{mwe}
 
\usepackage{amsmath}	
\usepackage{amssymb}	
\let\oldAA\AA
\usepackage[normalem]{ulem}

\usepackage{natbib}
\usepackage{ulem}
\usepackage{color}
\usepackage{subfig}
\usepackage{float}
\usepackage[usenames,dvipsnames]{xcolor}
\usepackage{multirow}
\usepackage{footnote}
\usepackage[bottom]{footmisc}
\usepackage{ulem}
\usepackage{xltabular}
\usepackage{longtable}
\usepackage{lipsum}
\usepackage{pdflscape}
\usepackage[export]{adjustbox}
\usepackage{color,soul}

\newcommand{\pion}[2]{{#1}\,{\sc {#2}}}

\renewcommand{\AA}{\text{\normalfont\oldAA}}

\title[Optical and near infrared spectroscopy of nova V2891 Cyg]{Optical and
near-infrared spectroscopy of Nova V2891 Cygni: evidence for shock-induced dust formation}

\author[Kumar et al.]{	
	Vipin Kumar$^{1,2}$\thanks{vipin@prl.res.in},
	Mudit K. Srivastava$^{1}$\thanks{mudit@prl.res.in}, Dipankar P.K. Banerjee$^{1}$, 
	C. E. Woodward$^{3}$\thanks{Visiting Astronomer at the Infrared Telescope Facility, which is operated by the University of Hawaii under contract 80HQTR19D0030 with the National Aeronautics and Space Administration}, 
	\newauthor
	Ulisse Munari$^{4}$, Aneurin Evans$^{5}$, Vishal Joshi$^{1}$,
	Sergio Dallaporta$^{6}$, 
	and Kim L. Page$^{7}$
	\\
	$^{1}$Astronomy \& Astrophysics Division, Physical Research Laboratory, Ahmedabad 380009, India\\
	$^{2}$Indian Institute of Technology, Gandhinagar, 382335, India\\
	$^{3}$Minnesota Institute for Astrophysics, University of Minnesota, 116 Church Street SE, Minneapolis, MN 55455, USA\\
	$^{4}$INAF Astronomical Observatory of Padova, 36012 Asiago (VI), Italy \\
	$^{5}$Astrophysics Group, Keele University, Keele, Staffordshire, ST5 5BG, UK\\
	$^{6}$ANS Collaboration, c/o Astronomical Observatory, 36012 Asiago (VI), Italy \\
	$^{7}$School of Physics \& Astronomy, University of Leicester, Leicester LE1 7RH, UK\\
}

\begin{document}
	
\date{Accepted YYYY Month DD.  Received YYYY Month DD; in original form YYYY Month DD}

\maketitle
	
\label{firstpage}
	
\begin{abstract}
	
We present multi-epoch optical and near-infrared observations of the highly reddened, \pion{Fe}{ii}  class slow nova V2891 Cygni. The observations span 15 months since its discovery.  The initial rapid brightening from quiescence, and the presence of a $\sim$35 day long pre-maximum halt, is well documented. The evidence that the current outburst of V2891 Cyg has undergone several distinct episodes of mass ejection is seen through time-varying P Cygni profiles of the O\,{\sc i} 7773\,$\AA$ line. A highlight is the occurrence of a dust formation event centred around $\sim$+273d, which coincides with a phase of coronal line emission. The dust mass is found to be $\sim0.83-1.25 \times 10^{-10} M_{\odot}$. There is strong evidence to suggest that the coronal lines are created by shock heating rather than by photoionization. The simultaneous occurrence of the dust and coronal lines (with varying velocity shifts) supports the possibility that dust formation is shock-induced.  Such a route for dust formation has not previously been seen in a nova, although the mechanism has been proposed for dust formation in some core-collapse supernovae. Analysis of the coronal lines indicates a gas mass and temperature of 8.35--8.42$\times10^{-7}$ M$_\odot$ and $\sim(4.8-9.1)\times10^{5}$~K respectively, and an overabundance of aluminium and silicon. A Case B analysis of the hydrogen lines yields a mass of the ionized gas of ($8.60\pm1.73)\times10^{-5}$ M$_{\odot}$. The reddening and distance to the nova are estimated to be $E(B-V)$ = 2.21$\pm$0.15 and $d$ = 5.50 kpc respectively.
\end{abstract}


\section{Introduction}
\label{sec-intro}
\par
%
Nova V2891 Cygni (also known as AT 2019qwf, PGIR 19brv and ZTF19abyukuy) was  discovered at a $J$ band magnitude of 11.3$\pm$0.02 on 2019 September 17.25 UT by 
\cite{De2019} during regular survey operations of the Palomar Gattini-IR telescope. The source was not detected to a five sigma limit of $J = 13.8$~mag on 2019 Sep 14.27 UT, implying a rise of at least $\Delta J\geq2.5$~mag in $\leq$3.0 days. It was classified as a Galactic classical nova on the basis of its optical spectrum \citep{De2019}, which showed a reddened continuum and broad emission lines of H$\alpha$, H$\beta$, and \pion{O}{i}. It was subsequently pointed out \citep{Lee2019} that the nova was first detected on 2019 Sep 15 by the Zwicky Transient Factory (ZTF), as ZTF19abyukuy.  However, our search of the ZTF archival data\footnote{lasair.roe.ac.uk/object/ZTF19abyukuy/} shows that the first detection was not on 15 September, but rather on JD 2458740.66 (2019 September 14.16 UT), at $r = 19.34\pm0.18$, when the nova had just begun to rise from quiescence.  We henceforth use this date as the reference point for measuring time in all our analyses ($t_0=\mbox{JD~}2458740.66$). 
\par 
Spectroscopy by \cite{Lee2019} on 2019 Sep 21(+7d) showed emission lines of H$\alpha$ having a full width at half maximum (FWHM) of 820 kms$^{-1}$, H$\beta$, {O}\,{\sc i} (at 7774, 8446, and 9266\,$\AA$), several Paschen lines between 8500 and 10050\,$\AA$; the $R$ magnitude was measured to be 15.1 mag. Optical spectra, recorded by \cite{Srivastava2019} on 2019 November 01.71 UT (+48d), showed a significant narrowing of the H$\alpha$ emission (FWHM$\sim$300 kms$^{-1}$) compared to the preceding reports. In addition, the {O}\,{\sc i} 7773\,$\AA$ line showed a P Cygni profile, even $\sim$48 days after the outburst. The prolonged mass-loss suggested that this would be a slow nova. A {Fe}\,{\sc ii} classification of the nova was later made by \cite{Munari2019} with optical spectrum recorded on 2019 November 5.94 UT(+52d). The {Fe}\,{\sc ii} classification was confirmed with near-infrared (NIR) spectroscopy \citep{Joshi2019} on 2019 November 17.85 UT(+64d). Based on the presence of low-excitation Na and Mg lines in the NIR spectrum, it was predicted that the nova is likely to form dust \citep{das2008}, which did transpire though late in the nova's evolution.
\par 
Subsequent evolution of the light curve showed that V2891 Cyg was indeed a slow nova whose brightness fluctuated around maximum for a considerable time (section~\ref{subsec_Lightcurve}), followed by a slow decline. During this extended phase around maximum, while the photometric variations were minimal, substantial spectroscopic changes apparent in regular (mostly optical) spectroscopic monitoring. These changes encompassed variations in the FWHMs and equivalent widths of the emission lines, changes and later disappearance of P Cygni profiles, the appearance of time-varying structures in the line profiles, and the emergence of forbidden lines of [O\,{\sc i}] etc. \citep{Munari2019b}. The nova was also followed at other wavelengths, and spectral changes were reported, e.g. optically thick free-free emission from the expanding nova ejecta \citep{Sokolovsky2020}, fading of carbon lines (a hallmark of the Fe\,{\sc ii} class of novae), and the emergence of NIR coronal lines \citep{De2020, Woodward2020} etc.
\par 
In this work, we present optical and NIR observations of V2891 Cyg, carried out 
between 2019 November 1 (+48d) to 2020 December 12 (+455d). These observations comprise optical photometry and spectroscopy from the 1.2m Mt. Abu Telescope (India), the Asiago 1.22m and 1.82m telescopes (Italy), the ID0310 telescope from the ANS (Asiago Novae and Symbiotic stars) collaboration \citep{Munari2012}, and NIR spectroscopy from Mt.Abu 1.2m telescope, the 3.6m ARIES-DOT telescope (India), and the 3.2m NASA-IRTF facility. 

\section{Observations}
\label{sec-observations}

\subsection{Optical Observations from Mt. Abu}
\label{subsec-ObsMtAbu}

Optical photometry and spectroscopy of V2891 Cyg were carried out with the Mount-Abu Faint Object Spectrograph and Camera-Pathfinder (MFOSC-P) instrument on the 1.2 m Mount Abu telescope \citep{Srivastava2018,Rajpurohit2020,Srivastava2021}. The instrument provides seeing-limited imaging in the Bessell $BV\!RI$  filters over a $5.2\times5.2$~arc-minute$^2$ field of view, with  a sampling of 3.3 pixels per arc-second. MFOSC-P is equipped with three plane reflection gratings that yield resolutions of $R = \Delta\lambda/\lambda = 2000, 1000$  and $500$ (referred to as
the R2000, R1000, and R500 modes hereafter). The standard spectral coverage of these modes are $\sim$6000-7000$\AA$, $\sim$4700-6650$\AA$, and $\sim$4500-8500$\AA$ respectively. In the R500 mode, the spectral region beyond 7400$\AA$ is covered by rotating the grating while simultaneously using a Bessel-$I$ blocking filter to avoid second-order contamination. The response corrected spectra for both settings of the grating were co-joined together around 7200$\AA$, after flux matching in the common overlap region, to obtain a complete spectrum free from second-order contamination. The spectroscopic observations presented here were made with a 75$\mu$m slit width (equivalent to $1.0\arcsec$ on the sky). Wavelength calibration was done using Xenon and Neon calibration lamps. For the purpose of instrument response correction, spectrophotometric standard stars from the ESO catalogue \footnote{https://www.eso.org/sci/observing/tools/standards/spectra/stanlis.html} were observed, with settings identical to those for the nova observations.
\par 
The raw data were reduced using in-house-developed data analysis routines in Python\footnote{https://www.python.org/} using open-source image processing libraries (ASTROPY\footnote{https://www.astropy.org/}, etc.). Spectroscopy data reduction steps involve bias subtraction, cosmic ray removal, tracing and extracting the spectra, sky background subtraction, etc. Pixel-to-pixel response variation was found to be less than 1$\%$  using the halogen lamp, and hence the correction is not applied to the reduced spectra.  The reduced spectra were 
flux calibrated using broadband photometric magnitudes obtained by us on the same night, or when these were not available, by using contemporaneous ANS magnitudes. The log of the MFOSC-P spectroscopic and photometric observations are given in Table~\ref{table-ObsMFOSCP}.

\subsection{Optical Spectroscopy from Asiago}
\label{subsec-ObsAsiago}

Low-resolution spectra of nova V2891 Cyg were recorded in 2019 November and 
December with the 1.22-m telescope with a B\&C spectrograph\footnote{This is operated in Asiago by the Department of Physics and Astronomy of the University of Padova}.  A 300 line/mm grating, blazed to 5000~\AA, provided a wavelength coverage from 3300 to 8000~\AA\ at a dispersion of 2.31~\AA/pixel, and a resolution of FWHM=1.9~pixel. The spectrograph is equipped with an ANDOR iDus DU440A CCD camera having a back-illuminated E2V 42-10 sensor ($2048\times512$ array of 13.5-$\mu$m pixels). Its high efficiency at ultraviolet (UV) wavelengths, together with the high UV transparency of the whole optical train, 
allows it to observe down to the limit of the atmospheric transmission for the observing site ($\sim$3100~\AA). The very red colour of V2891 Cyg, its faintness, and the relatively short exposure times, limited the usefulness of the recorded spectra to wavelengths $\geq$4000~\AA.
\par 
High resolution ($R\sim20000$) profiles of the H-$\alpha$ emission line in V2891 Cyg were recorded in 2019 November and December 2019 with the REOSC Echelle spectrograph mounted on the 1.82~m telescope operated in Asiago by the National Institute of Astrophysics of Italy (INAF). All spectroscopic reductions were carried out in IRAF \citep{Tody1986}, following the data reduction recipes discussed in \cite{Zwitter2000}. The spectrophotometric standards have been taken from the internal Asiago spectrophotometric database \citep{Moro2000,Sordo2006}. A log of the spectroscopic observations at the Asiago telescopes is provided in Table~\ref{table-ObsSpecAsiago}.

\subsection{BVRIz'Y photometry with ANS Collaboration telescopes}
\label{subsec-PhotANS}

The photometric evolution of V2891 Cyg from day $t=+67$d to $t=+230$d was monitored with the ANS Collaboration telescope ID 0310, which is a 30~cm f/10 instrument located in Cembra (on the Italian Alps). The telescope is equipped with a SBIG-10 CCD camera and photometric filters from Custom Scientific ($UBV\!RI$) and the Baader Planetarium ($g'r'i'z'Y$).  We have measured the transmission profiles of all the filters in the laboratory \citep{Munari2012b}; 
in particular, to exclude that any red-leak up to 1.1~$\mu$m had no detrimental effect, given the very red spectral energy distribution of V2891 Cyg.  
\par 
As is the case for all telescopes operated by the ANS Collaboration
\citep{Munari2012}, the collected photometry is transformed from the
instantaneous local photometric system to a standard one via color-equations
extending over a wide range of stellar colours. The reference standard systems are those of \cite{Landolt1992} for the $UBV\!RI$ bands (as realised locally by APASS All Sky Survey \citep{Henden2014} in its DR8 data release \citep{Munari2014b}; and PanSTARSS PS1 \citep{Chambers2016b} for the $g'r'i'z'Y$ bands. The collected photometry is presented in Table~\ref{table-ObsPhotANS}, where the quoted errors are the quadratic sum of the Poissonian component on the variable, and the error in the transformation from the local to the standard system. For continuity reasons with the other photometric bands, the $z'Y$ magnitudes in Table~\ref{table-ObsPhotANS} are expressed on the Vega-scale rather than the native AB-mag of the PanSTARSS PS1 by adopting the transformations given 
by \cite{Tonry2012}.


\begin{table*}
	\centering
	\caption{Log of the optical photometry and spectroscopy observations from Mt. Abu.}	
	\begin{tabular}{ccccccccc}
		\hline
		Date of      & Days$^a$      &            &        &      & Spectroscopy Exposure time & Spectro-photometric \\
		Observation  & since         &  $V$       &    $R$ &  $I$ & (seconds)                  & standard star$^b$\\
		(UT)         & outburst      & (mag)      & (mag)  & (mag)& (R500,R1000,R2000)         & \\
		\hline
		\hline
		
		2019-11-01.67	& 48.51  &  -		   & -  	    &	-		 &  ( 840, --- , 1200)  &  HR 718\\
		2019-11-11.66	& 58.50  &  -		   & -  	    &	-		 &  ( 800, --- , 1200)  &  HR 718\\
		2019-11-12.65	& 59.49  &  -		   & -  	    &	-		 &  ( 800, 1200 , 1800) &  HR 718\\ 
		2019-11-19.73	& 66.57  &  15.32$\pm$0.03 & 13.50$\pm$0.03 &	11.83$\pm$0.05   &  ( 600, --- , 600)& HR 718\\
		2019-12-01.66	& 78.50  &  -		   & 12.46$\pm$0.04 &	10.83$\pm$0.04   &  ( 1200, --- , ---)  &  HR 3454\\
		2019-12-07.63	& 84.47  &  -		   & -  	    &	-		 &  ( 1400, 1800 ,1800) &  HR 3454\\		    
		2019-12-08.57	& 85.41  &  16.22$\pm$0.03 & 13.82$\pm$0.04 &	12.10$\pm$0.03   &  --- 		&  ---  \\ 
		2019-12-09.60	& 86.44  &  -		   & -  	    &	12.19$\pm$0.03   &  --- 		&  ---  \\ 
		2019-12-10.61 	& 87.45  &  16.13$\pm$0.04 & -  	    &	12.26$\pm$0.03   &  --- 		&  ---  \\ 
		2019-12-13.63	& 90.47  &  -		   & -  	    &	-		 &  ( 600, 1500 , ---)  &  HR 718 \\
		2019-12-14.60 	& 91.44  &  15.61$\pm$0.03 & 13.74$\pm$0.02 &	12.12$\pm$0.02   &  ( ---, ---, 900)	&  HR 718\\
		2019-12-15.65	& 92.49  &  -		   & -  	    &	-		 &  ( 1800, 3600 , 900) &  HR 718\\ 
		2019-12-19.61	& 96.45  &  15.35$\pm$0.02 & -  	    &	11.77$\pm$0.03   &  ( 600, ---, 900)	&  HR 718 \\
		2020-01-04.60	& 112.44 &  15.88$\pm$0.02 & -  	    &	12.34$\pm$0.03   &  ( 600, ---,  900)	&  HR 1544 \\
		2020-01-05.59 	& 113.43 &  -		   & -  	    &	12.50$\pm$0.02   &  --- 		&  ---  \\
		2020-01-13.57   & 121.41 &  -		   & -  	    &	-		 &  ( 300, ---,  600)	&  HR 1544 \\		    
		2020-01-14.57 	& 122.41 &  14.71$\pm$0.02 & 12.90$\pm$0.02 &	11.28$\pm$0.02   &  --- 		&  ---  \\
		2020-05-06.99	& 235.83 &  16.50$\pm$0.06 & -  	    &	13.09$\pm$0.05   &  ( 300, ---,  ---)	&  HR 4468 \\
		2020-05-10.98	& 239.82 &  17.15$\pm$0.12 & 14.90$\pm$0.16 &	13.26$\pm$0.06   &  ( 450, ---, 1800)	&  HR 4468\\
		2020-05-20.98	& 249.82 &  17.70$\pm$0.14 & -  	    &	13.90$\pm$0.04   &  ( 600, ---,  ---)	&  HR 5191 \\
		2020-05-22.98   & 251.82 &  -		   & -  	    &	-		 &  ( 600, ---, 900)	&  HR 5191 \\
		2020-05-24.95   & 253.79 &  -		   & -  	    &	-		 &  ( 1200, ---, 1800)  &  HR 5191 \\ 
		2020-10-23.69	& 405.53 &  18.36$\pm$0.03 & 16.78$\pm$0.02 &	16.10$\pm$0.01   &  ( 1200, ---, ---) & HR 718  \\ 
		2020-11-01.68	& 414.52 &  18.59$\pm$0.14 & 16.99$\pm$0.04 &	16.50$\pm$0.07   &  ( 1200, ---, ---)	&  HR 1544  \\
		2020-11-02.73	& 415.57 &  18.48$\pm$0.06 & -  	    &	-		 &  ( 3600, ---, ---)	&  HR 1544\\
		2020-11-03.72	& 416.56 &  18.52$\pm$0.06 & 16.96$\pm$0.02 &	16.24$\pm$0.03   &  ( 900, ---, ---)	&  HR 1544\\
		2020-11-16.62	& 429.46 &  18.60$\pm$0.02 & 17.17$\pm$0.02 &	16.48$\pm$0.04   &  ( 3300, ---, ---)	&  HR 718\\
		2020-11-17.58	& 430.42 &  18.58$\pm$0.02 & 17.16$\pm$0.02 &	16.44$\pm$0.02   &  ( 3300, ---, ---)	&  HR 718\\
		2020-11-19.65	& 432.49 &  18.72$\pm$0.05 & -  	    &	-		 &  ( 1800, ---, ---)	&  HR 718 \\
		2020-11-27.58	& 440.42 &  18.71$\pm$0.07 & 17.32$\pm$0.03 &	16.57$\pm$0.03   &  ( 1200, ---, ---)	&  HR 718\\
		2020-11-29.61	& 442.45 &  18.50$\pm$0.06 & 17.36$\pm$0.02 &	16.57$\pm$0.03   &  ( 2400, ---, ---)	&  HR 718\\
		2020-12-01.62	& 444.46 &  18.81$\pm$0.07 & 17.45$\pm$0.02 &	16.67$\pm$0.04   &  --- 		&  ---  \\
		2020-12-08.56	& 451.40 &  18.83$\pm$0.03 & 17.47$\pm$0.02 &	16.75$\pm$0.02   &  ( 2700, ---, ---)	&  HR 718\\ 
		2020-12-12.58	& 455.42 &  18.78$\pm$0.05 & 17.59$\pm$0.07 &	16.85$\pm$0.03   &  --- 		&  ---  \\
								
		\hline
		\hline
	\end{tabular}
	\label{table-ObsMFOSCP}
	
	\begin{list}{}{}
		\item a: Time of discovery, 2019 September 14.160 UT (JD 2458740.660), is taken as time of outburst. All spectra were recorded and  photometry was done using MFOSC-P instrument.	
		\item b: Spectrophotometric standard stars were observed on the same, or on contemporaneous, nights.
	\end{list}
\end{table*}


\begin{table}
	\centering
	\caption{Log of the spectroscopic observations from the ANS
		telescopes.}
	
	\begin{tabular}{lllllll}
		\hline
		Date of         & Days since        & Instrument & Wavelength & IT$^a$  \\
		Observation & outburst          &  Name      & range      & (s)
		\\
		(UT)        & (days)            &            & ($\AA$)    &
		\\
		\hline
		\hline
		
		2019-11-05.94  &  52.78  & B\&C    &  4000-7840 & 480  \\
		2019-11-13.80  &  60.64  & Echelle &  6445-6700 & 1800 \\
		2019-11-13.85  &  60.69  & B\&C    &  4000-7840 & 600  \\
		2019-11-27.77  &  74.61  & B\&C    &  4000-7840 & 120  \\
		2019-12-06.79  &  83.63  & B\&C    &  4000-7840 & 360  \\
		2019-12-07.86  &  84.70  & B\&C	   &  4000-7840 & 1080 \\
		2019-12-08.74  &  85.58  & Echelle &  6445-6700 & 1800 \\
		2019-12-09.72  &  86.56  & Echelle &  6445-6700 & 1800 \\
		2019-12-14.81  &  91.65  & B\&C	   &  4000-7840 & 1980 \\
		2019-12-14.82  &  91.66  & Echelle &  6445-6700 & 1800 \\				
		\hline
		\hline
	\end{tabular}
	\label{table-ObsSpecAsiago}
	\begin{list}{}{}
		\item a: IT = Integration Time
	\end{list}
\end{table}


\begin{table*}
	\centering
	\caption{$BV\!RIz'Y$ photometry of V2891 Cyg from ANS Collaboration telescope ID 0310.}
	\begin{tabular}{@{}c@{~~}c@{~~}c@{~~}c@{~~}c@{~~}c@{~~}c@{~~}c@{~~}c@{~~}c@{~~}c@{}}
		\hline
		Date of      &          & Days$^a$     &       &            &        &     &  &  \\
		Observation  &   HJD    & since        &   $B$ &  $V$       &    $R$ &  $I$ & $z^\prime$ & $Y$  \\
		(UT)         &(-2450000)& outburst     &       &            &        &     &  &  \\
		\hline
		\hline
   2019-11-20.725 & 8808.225 &  67.57 &              -     & 15.330  $\pm$0.023 &               -     & 11.798  $\pm$0.010 &             -     &             -       \\
   2019-11-25.795 & 8813.295 &  72.64 &              -     & 14.903  $\pm$0.010 &  13.179  $\pm$0.008 & 11.540  $\pm$0.009 & 10.746 $\pm$0.012 & 10.281 $\pm$0.012   \\
   2019-11-29.862 & 8817.362 &  76.70 &              -     & 14.138  $\pm$0.010 &  12.487  $\pm$0.009 & 10.893  $\pm$0.008 & 10.151 $\pm$0.012 &  9.584 $\pm$0.011   \\
   2019-11-30.802 & 8818.302 &  77.64 & 16.546  $\pm$0.027 & 14.051  $\pm$0.007 &  12.419  $\pm$0.008 & 10.816  $\pm$0.008 & 10.011 $\pm$0.012 &  9.514 $\pm$0.010   \\
   2019-12-02.740 & 8820.240 &  79.58 & 17.295  $\pm$0.030 & 14.802  $\pm$0.008 &  12.955  $\pm$0.007 & 11.236  $\pm$0.008 & 10.466 $\pm$0.014 &  9.958 $\pm$0.014   \\
   2019-12-03.763 & 8821.263 &  80.60 &              -     & 15.486  $\pm$0.010 &  13.496  $\pm$0.007 & 11.667  $\pm$0.009 & 10.886 $\pm$0.012 & 10.524 $\pm$0.012   \\
   2019-12-04.786 & 8822.286 &  81.63 &              -     & 15.531  $\pm$0.010 &  13.451  $\pm$0.007 & 11.620  $\pm$0.009 & 10.841 $\pm$0.013 & 10.507 $\pm$0.013   \\
   2019-12-05.738 & 8823.238 &  82.58 &              -     & 15.358  $\pm$0.010 &  13.344  $\pm$0.007 & 11.598  $\pm$0.009 &             -     &             -       \\
   2019-12-06.758 & 8824.258 &  83.60 &              -     & 15.700  $\pm$0.016 &  13.535  $\pm$0.007 & 11.764  $\pm$0.011 & 10.993 $\pm$0.013 & 10.698 $\pm$0.014   \\
   2019-12-07.739 & 8825.239 &  84.58 &              -     &              -     &               -     &              -     & 11.174 $\pm$0.014 & 10.913 $\pm$0.014   \\
   2019-12-09.773 & 8827.273 &  86.61 &              -     &              -     &               -     &              -     & 11.420 $\pm$0.014 & 11.201 $\pm$0.013   \\
   2019-12-10.751 & 8828.251 &  87.59 &              -     & 16.176  $\pm$0.019 &  13.987  $\pm$0.008 & 12.275  $\pm$0.009 & 11.485 $\pm$0.013 & 11.231 $\pm$0.014   \\
   2019-12-12.812 & 8830.312 &  89.65 &              -     & 16.360  $\pm$0.027 &  14.131  $\pm$0.010 & 12.428  $\pm$0.009 & 11.602 $\pm$0.012 & 11.356 $\pm$0.012   \\
   2019-12-14.726 & 8832.226 &  91.57 &              -     & 15.644  $\pm$0.011 &  13.773  $\pm$0.008 & 12.123  $\pm$0.009 & 11.317 $\pm$0.015 & 10.903 $\pm$0.014   \\
   2019-12-14.750 & 8832.250 &  91.59 &              -     & 15.629  $\pm$0.015 &  13.778  $\pm$0.007 & 12.163  $\pm$0.009 &             -     &             -       \\
   2019-12-27.744 & 8845.244 & 104.58 &              -     & 16.107  $\pm$0.024 &  14.207  $\pm$0.010 & 12.610  $\pm$0.009 &             -     &             -       \\
   2019-12-27.783 & 8845.283 & 104.62 &              -     & 16.109  $\pm$0.016 &  14.234  $\pm$0.009 & 12.580  $\pm$0.010 &             -     &             -       \\
   2019-12-29.773 & 8847.273 & 106.61 &              -     & 16.123  $\pm$0.015 &  14.278  $\pm$0.010 & 12.617  $\pm$0.010 & 11.784 $\pm$0.013 & 11.374 $\pm$0.014   \\
   2020-01-01.760 & 8850.260 & 109.60 & 17.636  $\pm$0.029 & 15.218  $\pm$0.010 &  13.486  $\pm$0.008 & 11.872  $\pm$0.010 & 11.086 $\pm$0.012 & 10.651 $\pm$0.011   \\
   2020-01-02.713 & 8851.213 & 110.55 &              -     & 15.474  $\pm$0.010 &  13.710  $\pm$0.009 & 12.061  $\pm$0.010 & 11.247 $\pm$0.011 & 10.764 $\pm$0.011   \\
   2020-01-05.713 & 8854.213 & 113.55 &              -     & 16.001  $\pm$0.012 &  14.154  $\pm$0.010 & 12.484  $\pm$0.010 & 11.708 $\pm$0.011 & 11.299 $\pm$0.013   \\
   2020-01-06.734 & 8855.234 & 114.57 &              -     & 16.080  $\pm$0.013 &  14.213  $\pm$0.009 & 12.569  $\pm$0.010 & 11.759 $\pm$0.012 & 11.338 $\pm$0.012   \\
   2020-01-07.719 & 8856.219 & 115.56 &              -     & 15.494  $\pm$0.011 &  13.761  $\pm$0.009 & 12.215  $\pm$0.009 & 11.329 $\pm$0.011 & 10.899 $\pm$0.011   \\
   2020-01-09.719 & 8858.219 & 117.56 &              -     & 14.678  $\pm$0.009 &  13.021  $\pm$0.010 & 11.401  $\pm$0.008 & 10.532 $\pm$0.012 & 10.018 $\pm$0.012   \\
   2020-01-10.728 & 8859.228 & 118.57 &              -     & 15.742  $\pm$0.015 &  13.797  $\pm$0.009 & 12.031  $\pm$0.009 & 11.242 $\pm$0.011 & 10.870 $\pm$0.012   \\
   2020-01-12.728 & 8861.228 & 120.57 &              -     & 14.386  $\pm$0.007 &  12.722  $\pm$0.008 & 11.142  $\pm$0.009 & 10.329 $\pm$0.011 &  9.834 $\pm$0.011   \\
   2020-01-13.817 & 8862.317 & 121.66 &              -     & 15.225  $\pm$0.012 &  13.299  $\pm$0.009 & 11.583  $\pm$0.011 &             -     &             -       \\
   2020-01-14.749 & 8863.249 & 122.59 &              -     & 14.638  $\pm$0.009 &  12.869  $\pm$0.009 & 11.283  $\pm$0.009 & 10.467 $\pm$0.012 & 10.011 $\pm$0.010   \\
   2020-01-14.828 & 8863.328 & 122.67 &              -     & 14.639  $\pm$0.009 &               -     & 11.282  $\pm$0.010 &             -     &             -       \\
   2020-01-15.739 & 8864.239 & 123.58 &              -     & 14.506  $\pm$0.008 &  12.758  $\pm$0.008 & 11.144  $\pm$0.010 & 10.323 $\pm$0.012 &  9.888 $\pm$0.011   \\
   2020-01-16.762 & 8865.262 & 124.60 &              -     & 14.244  $\pm$0.008 &  12.523  $\pm$0.008 & 10.912  $\pm$0.011 & 10.079 $\pm$0.011 &  9.618 $\pm$0.008   \\
   2020-01-20.735 & 8869.235 & 128.57 &              -     & 15.123  $\pm$0.016 &  13.120  $\pm$0.009 & 11.529  $\pm$0.013 & 10.662 $\pm$0.011 & 10.303 $\pm$0.012   \\
   2020-01-23.750 & 8872.250 & 131.59 &              -     &              -     &               -     &              -     & 10.852 $\pm$0.012 & 10.512 $\pm$0.011   \\
   2020-02-06.732 & 8886.232 & 145.57 &              -     & 16.275  $\pm$0.027 &               -     & 12.527  $\pm$0.012 &             -     &             -       \\
   2020-02-08.199 & 8887.699 & 147.04 &              -     & 16.458  $\pm$0.026 &  14.363  $\pm$0.010 & 12.623  $\pm$0.010 &             -     &             -       \\
   2020-02-13.187 & 8892.687 & 152.03 &              -     &              -     &  14.504  $\pm$0.015 & 12.740  $\pm$0.010 &             -     &             -       \\
   2020-02-13.187 & 8892.687 & 152.03 &              -     &              -     &  14.556  $\pm$0.015 & 12.770  $\pm$0.010 &             -     &             -       \\
   2020-02-16.196 & 8895.696 & 155.04 &              -     & 15.982  $\pm$0.015 &  14.144  $\pm$0.011 & 12.456  $\pm$0.008 &             -     &             -       \\
   2020-02-24.189 & 8903.689 & 163.03 &              -     & 16.966  $\pm$0.036 &               -     & 13.132  $\pm$0.013 &             -     &             -       \\
   2020-02-27.194 & 8906.694 & 166.03 &              -     & 16.308  $\pm$0.019 &  14.422  $\pm$0.010 & 12.731  $\pm$0.008 &             -     &             -       \\
   2020-02-28.194 & 8907.694 & 167.03 &              -     & 16.347  $\pm$0.021 &  14.414  $\pm$0.015 & 12.778  $\pm$0.009 &             -     &             -       \\
   2020-03-08.161 & 8916.661 & 176.00 &              -     & 15.795  $\pm$0.021 &  13.891  $\pm$0.011 & 12.203  $\pm$0.010 &             -     &             -       \\
   2020-03-16.139 & 8924.639 & 183.98 &              -     & 15.862  $\pm$0.011 &  14.058  $\pm$0.007 & 12.425  $\pm$0.009 &             -     &             -       \\
   2020-03-24.133 & 8932.633 & 191.97 &              -     & 15.744  $\pm$0.014 &  13.842  $\pm$0.007 & 12.175  $\pm$0.009 & 11.338 $\pm$0.010 & 10.931 $\pm$0.011   \\
   2020-04-02.113 & 8941.613 & 200.95 &              -     & 15.382  $\pm$0.009 &  13.591  $\pm$0.007 & 11.924  $\pm$0.010 & 11.089 $\pm$0.009 & 10.632 $\pm$0.008   \\
   2020-04-05.070 & 8944.570 & 203.91 &              -     & 15.954  $\pm$0.018 &  14.021  $\pm$0.008 & 12.373  $\pm$0.009 & 11.493 $\pm$0.009 & 11.117 $\pm$0.010   \\
   2020-04-10.074 & 8949.574 & 208.91 &              -     & 15.909  $\pm$0.020 &  13.991  $\pm$0.009 & 12.336  $\pm$0.009 & 11.476 $\pm$0.010 & 11.051 $\pm$0.010   \\
   2020-04-16.027 & 8955.527 & 214.87 &              -     & 16.430  $\pm$0.014 &  14.440  $\pm$0.009 & 12.754  $\pm$0.010 & 11.890 $\pm$0.011 & 11.566 $\pm$0.010   \\
   2020-04-22.098 & 8961.598 & 220.94 &              -     & 16.938  $\pm$0.013 &  14.892  $\pm$0.007 & 13.207  $\pm$0.009 &             -     &             -       \\
   2020-04-24.022 & 8963.522 & 222.86 &              -     & 17.212  $\pm$0.022 &  15.122  $\pm$0.010 & 13.476  $\pm$0.011 & 12.616 $\pm$0.012 & 12.405 $\pm$0.016   \\
   2020-04-30.039 & 8969.539 & 228.88 &              -     & 16.076  $\pm$0.011 &  14.254  $\pm$0.007 & 12.632  $\pm$0.008 & 11.727 $\pm$0.012 & 11.340 $\pm$0.011   \\
   2020-05-02.107 & 8971.608 & 230.95 &              -     & 16.035  $\pm$0.016 &  14.201  $\pm$0.009 & 12.513  $\pm$0.009 &             -     &             -       \\
		\hline
		\hline
	\end{tabular}
	\label{table-ObsPhotANS}

	\begin{list}{}{}
		\item a: Time of discovery, 2019 September 14.160 UT (JD 2458740.660), is taken as time of outburst. 		
	\end{list}
\end{table*}
\color{black}


\subsection{NIR observations}
\label{subsec-ObsIRTF}

A single epoch of NIR photometry and spectroscopy from Mt.  Abu was obtained
with the Near-Infrared Camera and Spectrometer (NICS) on 2019 November 17 (+64d). NICS is equipped with a $1024\times1024$ HgCdTe Hawaii array, and operates in the 0.85--2.45$\mu$m wavelength range with a resolution of $R\sim1000$. The spectra were recorded in three settings of the grating that cover the $IJ, JH$, and $HK$ regions. A standard star HD 209932 (spectral type A0V) was also observed at a similar airmass and in similar conditions to remove the telluric lines. The spectra were recorded at two dithered positions of the object along the slit; these were subtracted, one from the other, to remove sky and dark contributions. The data reduction was done using the procedures described in \cite{Joshi2014, Srivastava2016}. The wavelength calibration was done using OH skylines, and telluric lines that register with the stellar spectra. Hydrogen Paschen and Brackett absorption lines in the $JH$ and $K$  bands were identified and removed from the standard star's spectra, and subsequently the nova spectra were ratioed with these spectra to remove telluric features. These telluric-corrected spectra were then multiplied by a blackbody at the standard star's effective temperature to obtain the resultant spectra. Photometric observations in $JH$ and $K_{s}$ 
bands were also obtained, along with the photometric standard stars; the nova's magnitudes were obtained using aperture photometry.
\par 
Four spectra of nova V2891 Cyg were obtained using SpeX \citep{Rayner2003} on the 3.2m IRTF telescope between 2020 May 20 (+249d) and September 19 (+371d). The SpeX data were reduced and calibrated using Spextool \citep{Cushing2004}, and the IDL tool xtellcor \citep{Vacca2003} was used for the corrections for telluric absorptions. The spectra were flux calibrated using AO-standard HD199217 observed at a comparable airmass. Observations used either $0\farcs5 \times 15\farcs0$ or a $0\farcs8 \times 15\farcs0$ slit, and both short-crossed dispersed (SXD) and long-cross dispersed (LXD) spectrograph modes.
\par 
A single epoch of NIR spectrum was obtained on 2020 Oct 26 (+408d) from the TANSPEC instrument on 3.6m ARIES-DOT telescope \citep{Ojha2018}. The data have been reduced using in-house-developed data analysis routines in Python. Details of the NIR observations are given 
in Table~\ref{table-ObsNIR}.

\begin{table*}
	\centering
	\caption{Log of the NIR photometric/spectroscopic observations.}
	\begin{tabular}{lcccccc}
		\hline
		Date of         & Days since        & Instrument & Wavelength &Spectral& Integration   time   \\
		Observation     & outburst          &  Name      & range ($\mu$m)&Resolution    & (s)              \\
		(UT)        & (days)            &            &  /Magnitudes  & &             \\
		\hline
		\hline
		2019-11-17.85 &  64.69   & NICS - Spec & 0.85-2.4 &1000 &450  \\
		2019-11-17.91 &  64.75   & NICS$^a$ - Phot & $J= 9.98\pm0.02$& - &(250,90,56)   \\					
		&    &  & $H = 8.87\pm0.03$, $K_s = 7.66\pm0.02$ & &   \\								
		2020-05-20.25 &  249.43  & SpeX - Spec & 0.7-2.55 &1200 &1434.53\\
		2020-06-07.62 &  267.46  & SpeX - Spec & 0.7-4.2 &1200 &SXD$^b$ = 637.570, LXD = 667.224\\
		2020-06-30.53 &  290.37  & SpeX - Spec & 0.7-2.55 &1200 &478.180\\
		2020-09-19.26 &  371.10  & SpeX - Spec & 0.7-2.55 &750 &1438.24\\
		2020-10-26.59 &  408.43  & TANSPEC-Spec& 0.86-2.5  & 2750  &  2947.348  \\
		
		\hline
		\hline
	\end{tabular}
	\label{table-ObsNIR}
	\begin{list}{}{}
		\item (a) : NICS- Near infrared Camera Spectrograph, Mt Abu. Phot: indicates photometry; Spec: Spectroscopy (b) SXD and LXD: Short (0.7-2.5 microns) and long wavelength (1.7-4.2 microns) spectral regions covered by SpeX (at IRTF) in the cross-dispersed mode.
	\end{list}
\end{table*}


\section{Parameters from the light curve and a possible progenitor}
\label{sec_Lightcurve}

\par
\subsection{Light curve}
\label{subsec_Lightcurve}

\begin{figure}
	\centering
	\includegraphics[angle=0,width=0.49\textwidth]{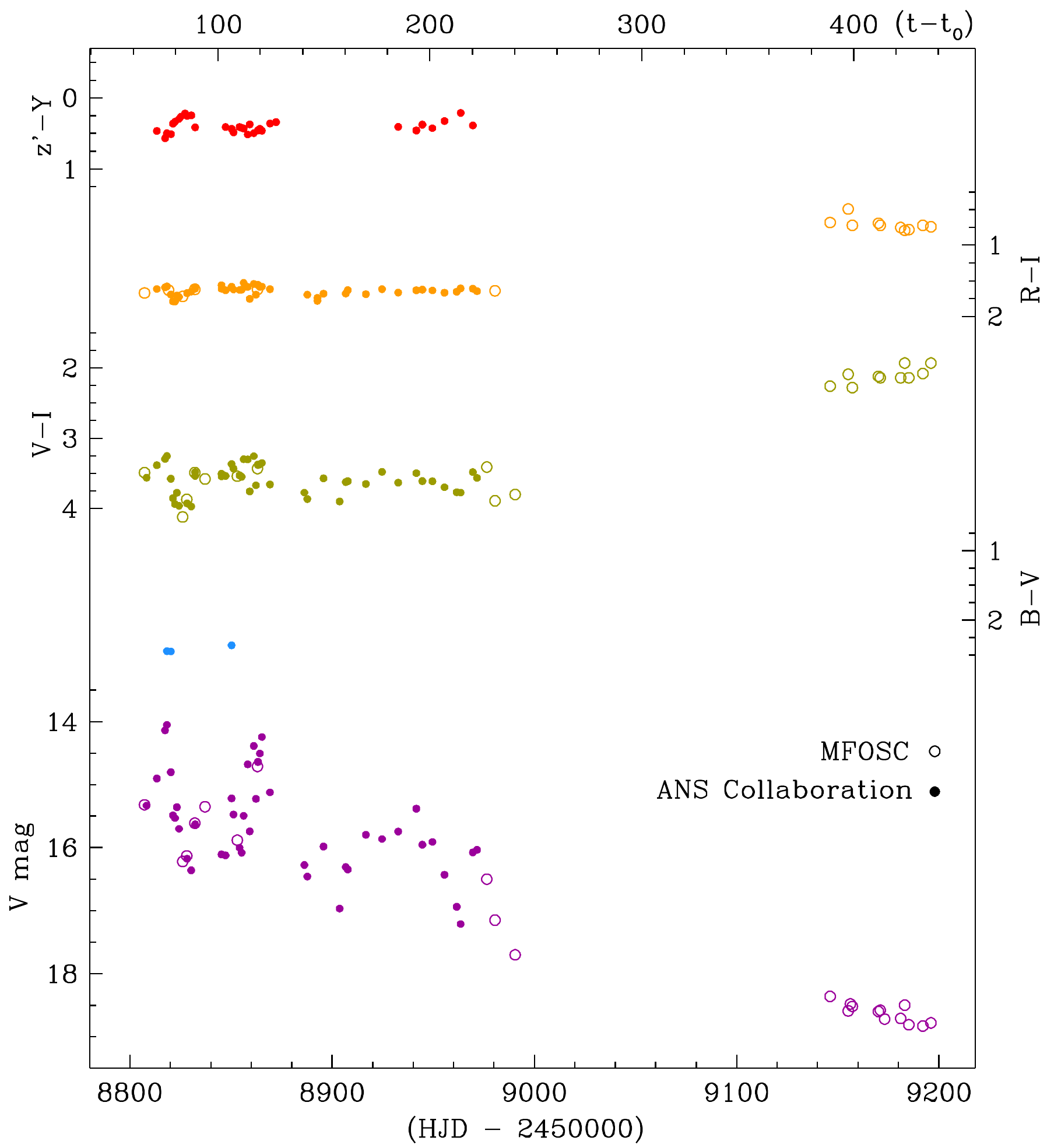}
	
	\caption{Photometric evolution of Nova V2891 Cyg in the $BV\!RIz'Y$ bands. 
	The time of discovery, 2019 September 14.160 UT (JD 2458740.660), 
	is taken as $t_0$ on the abscissae at the top panel. Open and filled 
	symbols refer to MFOSC-P and ANS Collaboration data, respectively}.  
	\label{fig-LC1}
\end{figure}

\begin{figure}
	\centering
	\includegraphics[angle=0,width=0.49\textwidth]{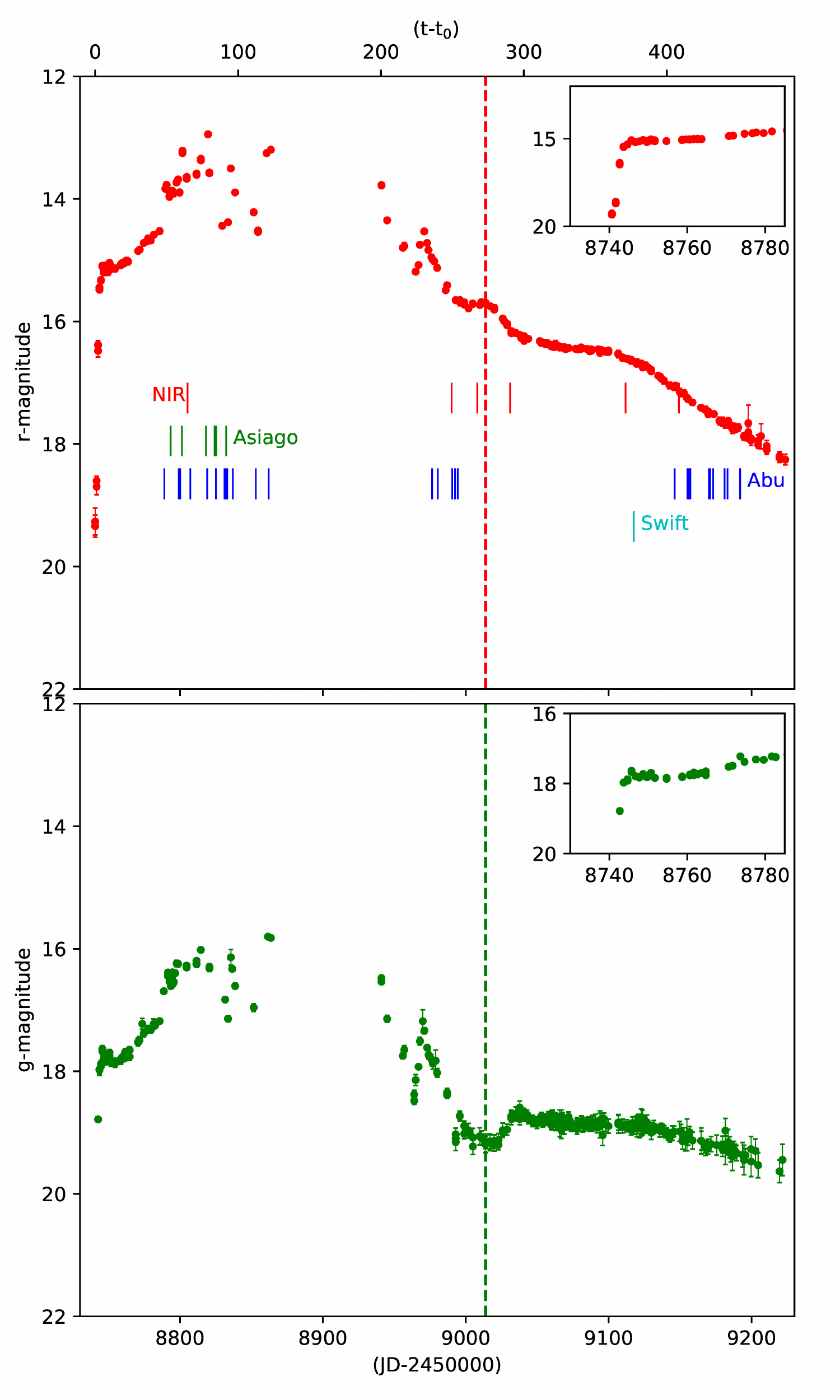}
	\caption{The ZTF light curve of Nova V2891 Cyg in the $g$ and $r$ bands. The insets show the well-documented rise from quiescence towards maximum, and also the prolonged pre-maximum halt, lasting for $\sim$ 35 days. The epochs of our spectroscopy observations, as well as a Swift target of opportunity observation, are marked in top panel by vertical lines. The dotted vertical lines show a phase of dust formation in the nova, characterized by a drop in the $g$ band flux accompanied by a simultaneous increment in the $r$ band flux.} 
	
	\label{fig-LC2}
\end{figure}

The light curve (LC) and colour plots of the nova V2891 Cyg using data from Mt. Abu and the ANS telescopes are presented in Fig.~\ref{fig-LC1}. The $g$ and $r$ band light curves from the ZTF are presented separately, for clarity, in Fig.~\ref{fig-LC2}. The distinctive features of the LCs include the recording of the nova's fast climb from quiescence to maximum, the presence of a fairly long pre-maximum halt, and three pronounced large-amplitude ``jitters'' around the maximum. There is also a phase of dust formation centred around +273d that is marked by dotted lines in Fig.~\ref{fig-LC2}. The LC is of a slightly distinctive nature, which does not easily belong to any known LC classes \citep{Strope2010}.

\par 
Within the grouping scheme of nova LC proposed by \cite{Strope2010}, V2891 Cyg shows partial resemblance to both the F (flat-topped) and J (Jitters) classes.  The LC of nova V 2891 Cyg shows a slow rise of duration $\sim120$ days, including several episodes of re-brightening with $\delta$magnitude$\sim$1.2, before it begins its decline. There is also a reasonably ``flat'' phase, of duration $\sim$180 days, between 30-210 days after outburst, during which the nova stays at a mean value of $V\sim16$~mag, accompanied by modulations with amplitude ranging up to $\pm1.2$~mags. Defining a $t_2$ (the time interval in days for the LC to decline 2 magnitudes from the peak) for an LC of this nature is difficult, so the $V$ band data were smoothed with 5 point running averages to annul the smaller amplitude jitters (\cite{Munari2017} and references therein). From the resultant smoothed curves, we find that, if the first peak is considered as the principal maximum, then t$_2$ is $\sim150$~days. If the second peak is taken as 
the maximum, then t$_2\sim100$~days. A similar estimate of t$_2$ is also obtained from ZTF $g$ band light curve. For our analysis, an exact value of $t_2$ is not essential.  It is sufficient to note that $t_2$ is large, and a range of 100 to 150 days may be considered appropriate  -- we will use this approximation in the 
distance estimation that follows.  Similarly, $t_3$ (the time interval in days for the LC to decline 3 mags from the peak) is constrained to be in the range 180--230~days. Again, the exact value of $t_3$ is not critical for any of the subsequent analyses.


\subsection{Extinction and distance estimates}
\label{subsec_ExtinctionDistance}	

The nova lies in a direction of high extinction, with the total extinction along the entire line-of-sight being  estimated as $A_v$ = 11.224 from \cite{Schlafly2011ApJ}, and 13.051 from \cite{Schlegel1998}. The intrinsic colour of novae at peak, following \cite{VanDen1987}, is $+0.23(\pm 0.06)$.  Both maxima attained by V2891 Cyg have been covered in our observations, as shown in Fig.~\ref{fig-LC1}, averaging $(B-V)=2.47$ ($\pm0.021$ as error of the mean), leading to $E(B-V)=2.24$. We also estimated the reddening following the empirical law determined by \cite{Munari2014}, by using the equivalent width (EW) of the Diffuse Interstellar Band (DIB) at 6614\AA\ in our spectra ($E(B-V) = 4.40 \times EW(\AA)$). In the echelle spectra, the measured EWs on the different days are rather dispersed, given the low signal-to-noise. The average equivalent width is 0.492\AA, leading to $E(B-V)=2.16$. The MFOSC-P spectrum at $R = 2000$, obtained on 2019 Nov 1 (+48d), gives a closely similar value.  Averaging over the two independent estimates gives $E(B-V)=2.20\pm0.03$, or $A_{v} = 6.82$ (for a value of the total-to-selective extinction $R_v$=3.1) which is the value we adopt for the rest of our analysis. It was not possible to use the O\,{\sc i}  0.8446 $\mu$m and 1.2187$\mu$m line ratios to estimate the reddening because the spectra recording these lines (observed between 2020 May and September) are affected to varying degrees by nova dust, which clearly causes additional extinction.

\par
\begin{figure}
	\centering
	\includegraphics[width=0.49\textwidth]{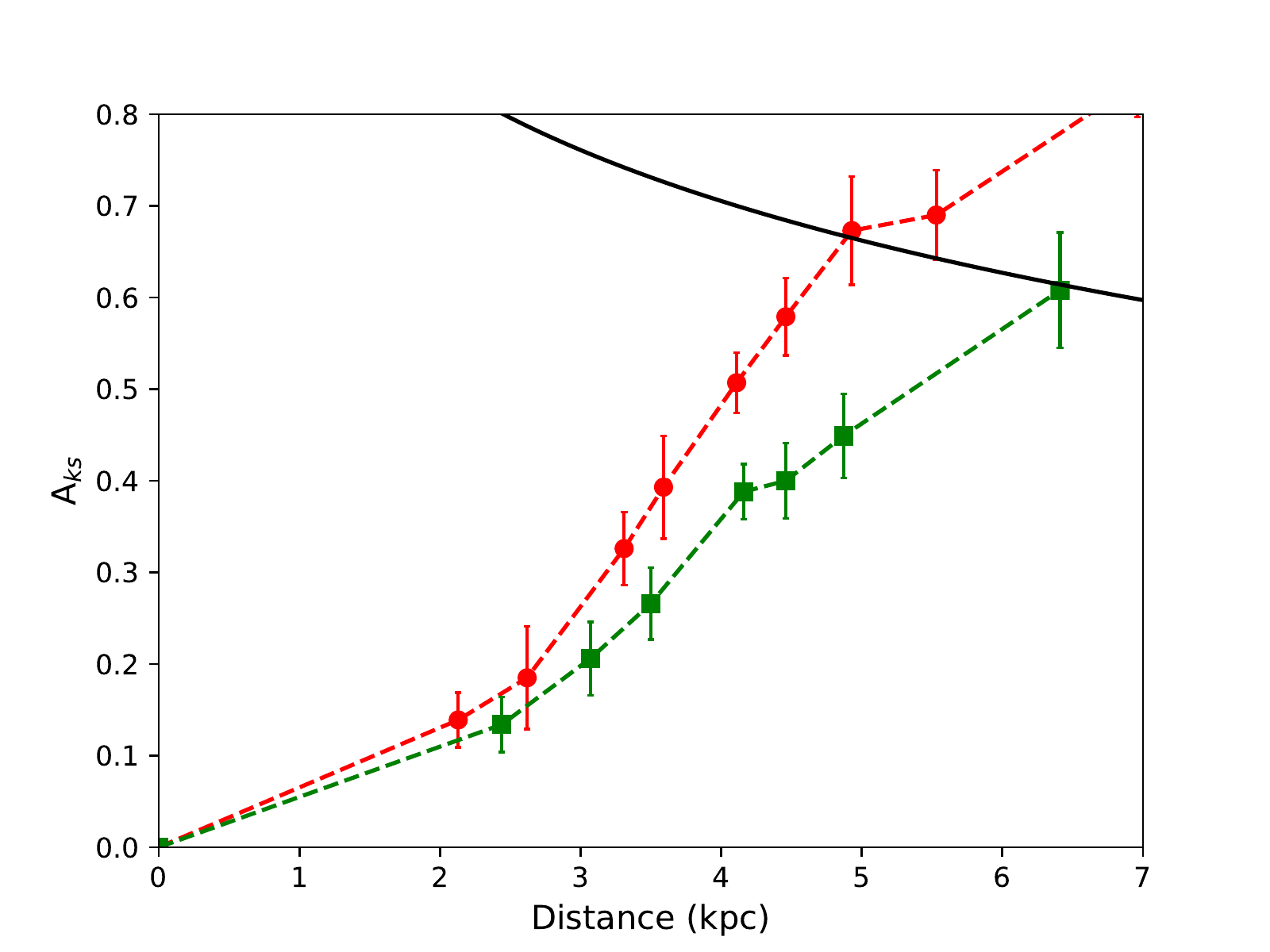}
	\caption{The dashed lines connecting the data points (red circles and 
	green squares) show the extinction towards nova V2891 Cyg based on the
	results of Marshall et al. (2006) along two nearest lines-of-sight 
	(Galactic coordinates  $l = 89.5, b= +0.25$ for red and $l = 89.75, b= +0.25$ for green). V2891 Cyg's Galactic coordinates are $l = 89.59686$, $b = +0.21382$. The continuous black curve is a plot of extinction $A_{ks}$ versus distance $d$ (kpc) from the equation $m_V- M_V = 5\log{d}-5 + A_V$, where $m_V$ is known from observations. and $M_V$ is estimated from 
	the MMRD relation \citep{DellaValle1995}. A relation $A_{ks}/A_V = 0.089$
	is used for the necessary conversion \citep{Glass1999, Marshall2006}. 
	The intersection of the two curves permits simultaneous estimation of 
	the extinction and distance to the nova. See text for details.}
	\label{fig-ak-dist}
\end{figure}

 
Following \cite{Das2015} and \cite{Banerjee2018}, we also use an independent approach with the additional benefit that it allows  the simultaneous estimates of the the extinction and distance, by using the distance-dependence of the extinction towards the nova, and the absolute magnitude derived from the Maximum Magnitude Rate of Decline (MMRD) relationship \citep[and references therein]{Cohen1985}. The extinction towards the nova, over a 5~arc-minute field \citep{Marshall2006}, is plotted in Fig.~\ref{fig-ak-dist} for the two closest lines of sight available. We also plot the extinction versus distance as derived from the distance modulus equation using $m_V(\mbox{max})=14.05$ from observations, and $M_V = (-6.86$ to $-6.91)$ $\pm$ 0.20 from the MMRD relation of \cite{DellaValle1995}; we use $t_2$ = 100d to 150d as derived earlier. The intersection of this curve (solid black in Fig.~\ref{fig-ak-dist}) with the two extinction curves (dashed lines)  allows us simultaneously to determine the distance and extinction for the nova. The two different intersections (for red and green lines) yield $A_{ks} =0.67\pm0.02$ and $A_{ks} =0.61\pm0.02$ respectively. The corresponding distances are $d = 4.90 \pm 0.15$~kpc, and $d = 6.45 \pm 0.14$ kpc respectively. The average values from this method are $A_{ks} =0.64\pm0.02$ and $d = 5.68 \pm 0.78$ kpc. A recent MMRD relation of \cite{Selvelli2019}, based on revised distances from Gaia \citep{Gaia2016}, uses $t_3$ values. For $t_3$ in the range 180 to 230 days, we find  $M_v = -6.30\pm0.56$ to $-6.07\pm0.58$. Following the same method as above, this corresponds to an average $A_{ks} =  0.59\pm0.05$ and distance $d = 5.32 \pm 0.37$ kpc. 
\par 
We choose the average of the values derived from these two MMRD relations for $A_{ks}$ ($= 0.61 \pm 0.04$) and $d$ ($=5.50\pm0.86$) kpc. This implies $A_v$ = A$_{ks}$/0.089 = 6.85$\pm$0.45 (or $E(B-V)=2.21\pm0.15$) which is a good match with the value $A_v$ = 6.82 derived above. We shall use these values for further analysis.

\subsection{Identification of a possible progenitor}

\begin{figure}
	\centering
	\includegraphics[width=0.49\textwidth]{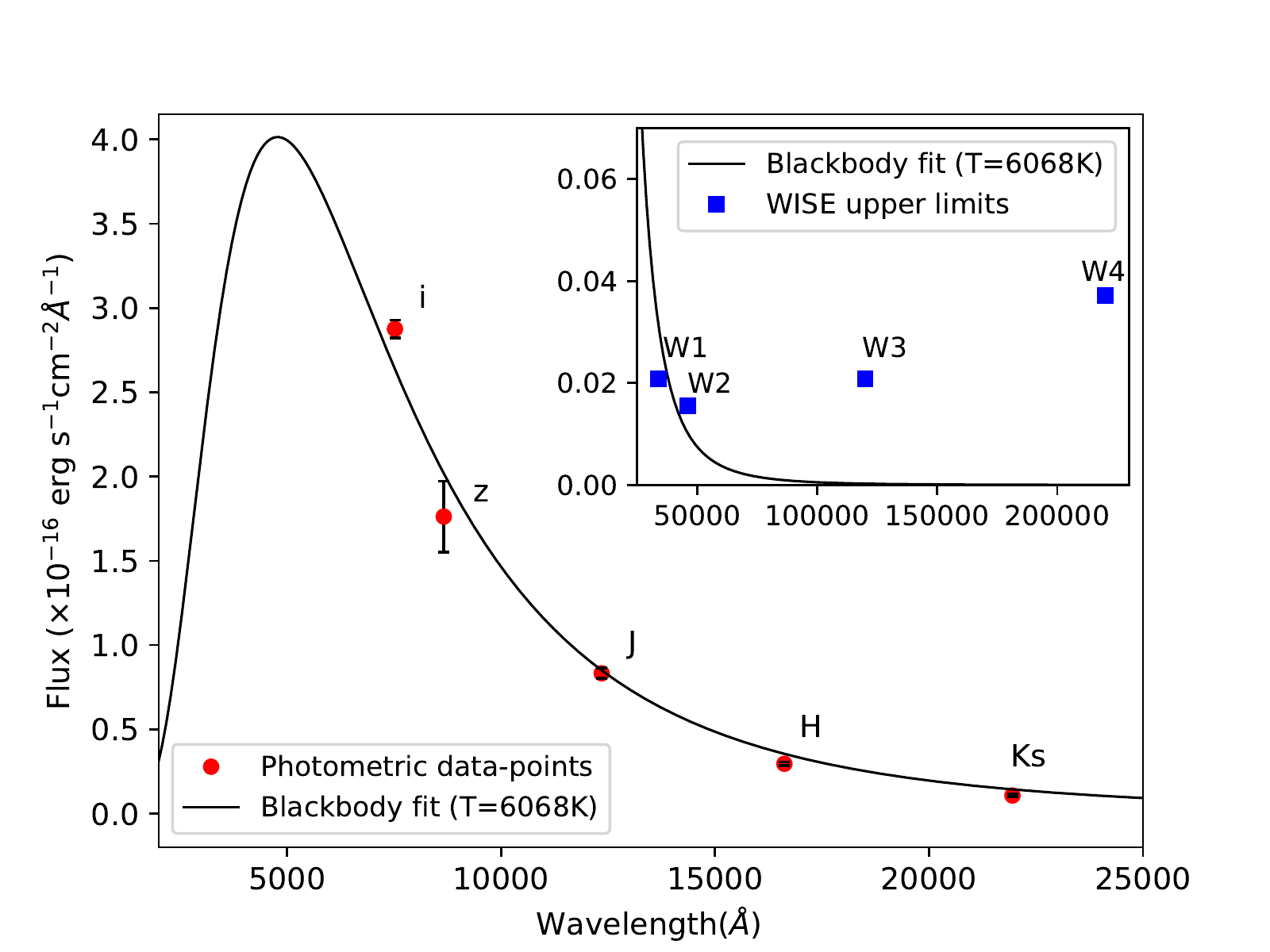}
	\caption{The blackbody fit with the temperature of 6068$\pm$97K. Filled circles (red) shows the dereddened flux corresponding to the Pan-STARRS-$i$,z and UKIDSS-DR6-J,H,K magnitudes. Inset shows the  WISE 5-$\sigma$ detection limits \citep{Wright2010}.}
	\label{fig-Blackbody_fit}
\end{figure}

\begin{figure}
	\centering
	\includegraphics[width=0.49\textwidth]{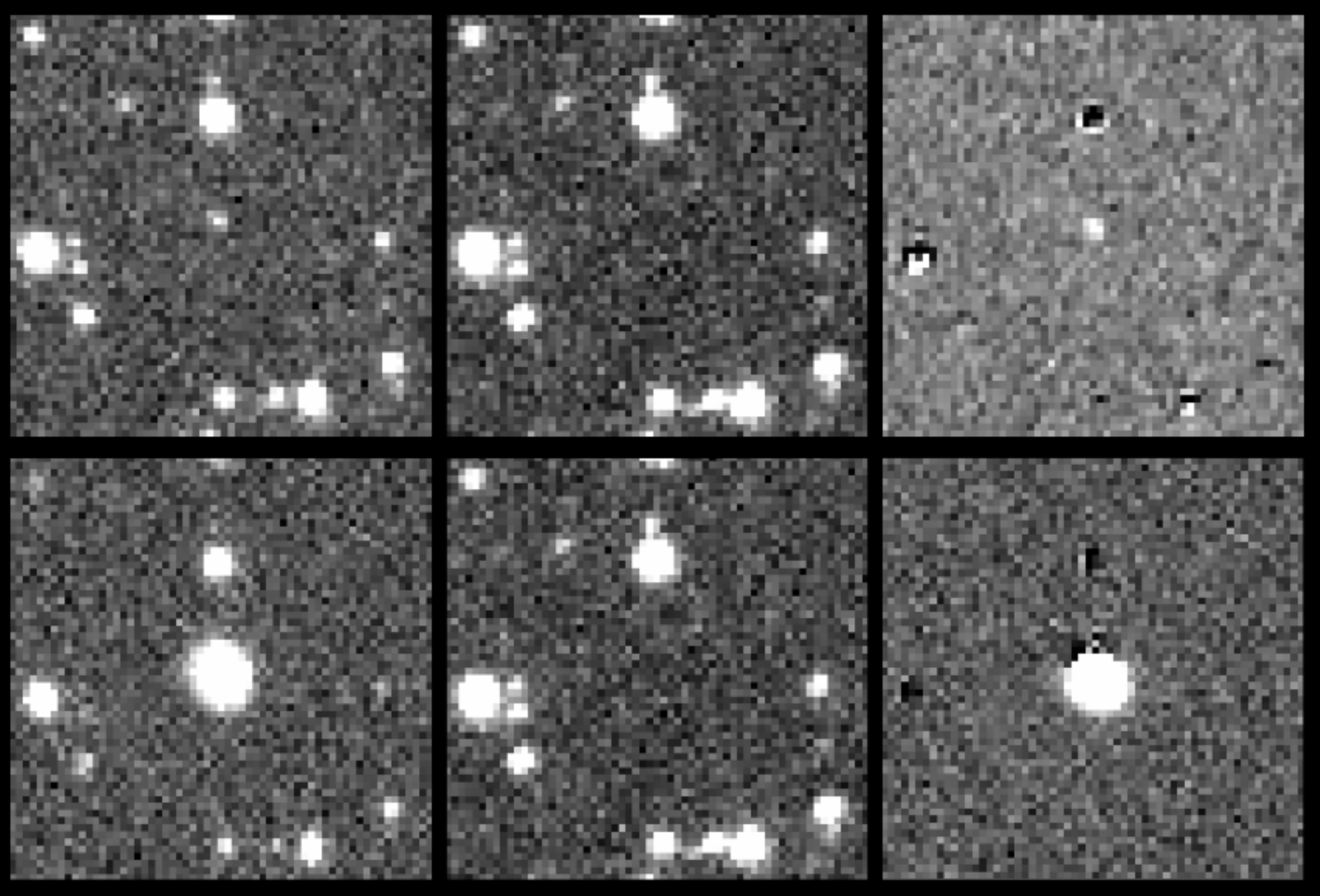}
	\caption[]{The top 3 panels show the ZTF first detection of the nova on ZTF images on  MJD 58740.160 (2019 Sept 14.16 UT) in $r$ at 19.34 $\pm$ 0.18 when it was ascending from quiescence to maximum (lasair.roe.ac.uk/object/ZTF19abyukuy/). The left panel is the raw image, middle is the reference template and the right panel is the difference image from the ZTF archive. Image size is 60 arc-seconds a side with pixel scale of $\sim1$ arc-seconds per pixel. The nova is seen in both the raw and difference image (above 5$\sigma$). The bottom panels show the nova close to maximum light on MJD 58819.1230 (2019 Dec 2) at $r$ = 12.945 $\pm$ 0.02.}

	\label{fig-ZTFDetection}
\end{figure}

A possible progenitor could be a faint star, detected both in the optical
and in the NIR in the Pan-STARRS \citep{Chambers2016} and UKIDSS \citep{UKIDSS2012} surveys, respectively, which shows a good positional match with the nova. The Gattini discovery coordinates of the nova are RA J2000: 21:09:25.52, Dec J2000: +48:10:51.9 \citep{De2019}. These coordinates were refined to the end figures of 25.524, 52.248 after Gaia detected the source on 2019 Nov 4 (+51d) and designated it as Gaia19ext.  The UKIDSS-DR6 Galactic Plane Survey \citep{UKIDSS2012} shows a source, just 0\farcs1 away from the Gaia coordinates, with end figures of 25.5278, 52.309 and NIR magnitudes of $J = 18.41\pm 0.04$, $H = 17.66 \pm 0.04$, $K = 17.25 \pm 0.09$.  Similarly, the Pan-STARRS release 1 (PS1) Survey -- DR1 \citep{Chambers2016} identifies the same 
source with end figures of 25.5343, 52.227, just 0\farcs2 from the nova's 
position, with $z = 20.80 \pm 0.13$~mag and $i = 21.65\pm 0.02$. Nothing is seen at the nova's position in the WISE W1, W2, W3 and W4 bands \citep{Wright2010}. After de-reddening the Pan-STARRS and UKIDSS magnitudes (using $E(B-V) = 2.20$),
a blackbody fit (see Fig.~\ref{fig-Blackbody_fit}) yields a temperature of $T = 6068\pm 98$~K for the potential progenitor. The luminosity and radius of the donor star are determined to be $2.76\pm0.31\,L{_\odot}$ and $1.50\pm0.07$~$R{_\odot}$, respectively. Although these values may correspond to the spectral type of F9V for the potential progenitor, we also note that an overall temperature of 6000--7000~K is also suitable for an accretion disc. Pre-novae in quiescence can also be disc-dominated systems \citep{Selvelli2013}. Therefore, we are cautious about drawing any conclusion about the nature of the donor. Extrapolating this black-body curve gives an $r$ band magnitude of $\sim22$, which, when compared to the peak value of $r = 12.95$ recorded by the ZTF, indicates an outburst amplitude $\cal{A}$ of approximately 10 magnitudes. This is consistent with the amplitude expected for a slow nova \cite[see, e.g., the $\cal{A}$ versus $t_2$ correlation plot in][]{Warner1995}. The close positional coincidence, the estimated $\cal{A}$ value and the spectral type of the UKIDSS/Pan-STARRS source are all consistent with that expected of a nova progenitor. An image of the earliest ZTF detection of the nova when it was still rising from quiescence to maximum is interesting to see; this is shown in Fig.~\ref{fig-ZTFDetection} along with its image at maximum.

\section{Results from optical spectroscopy}
\label{sec_OptSpec}
Nova V2891 Cyg presents a good case study for those nova systems that show multiple outbursts, and the interactions between ejected material. The claim that the current outburst of nova V2891 Cyg had undergone multiple episodes of mass ejection can be seen in the evolution of the H-$\alpha$ and O\,{\sc i} 7773$\AA$ lines. NIR spectroscopy offers further clues for the interactions of ejecta from these multiple outbursts in the form of the occurrence and evolution of coronal lines, a brief epoch of dust formation, etc. We present the results from the optical and NIR spectroscopy separately for ease of readability, although there are areas of overlap. Optical spectroscopy is presented here, and NIR spectroscopy is presented later in section~\ref{sec_NIRSpecMaySep2020}.

\subsection{Evolution of optical spectra}
\label{subsec_OptSpecEvolution}

\begin{figure*}
	\centering
	\includegraphics[width=0.99\textwidth]{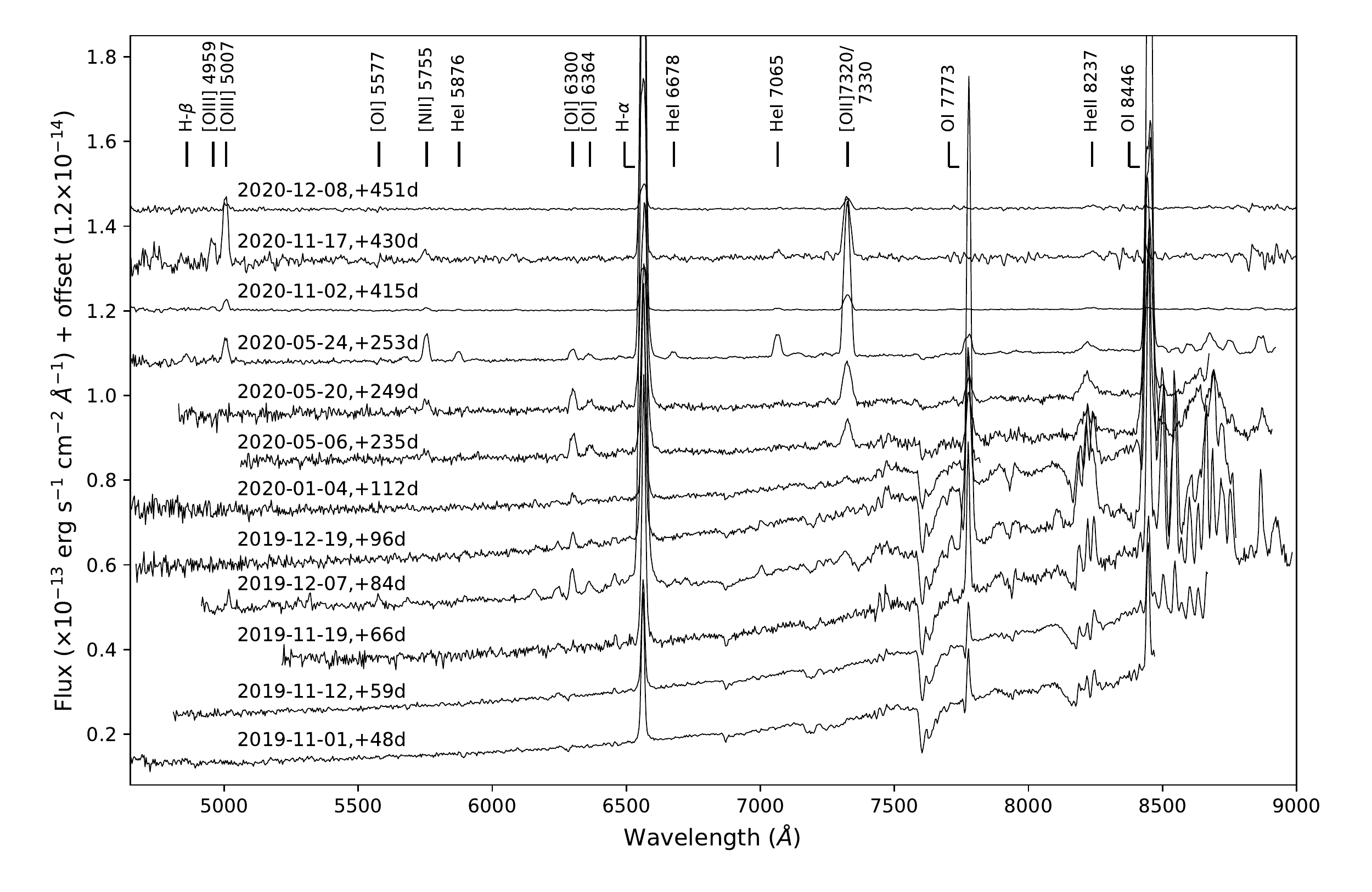}
	\caption[]{Evolution of the optical spectrum of nova V2891 Cyg as seen with
	the MFOSC-P instrument. The spectra are not reddening corrected. See 
	section~\ref{subsec_OptSpecEvolution} for discussion.}
	\label{fig-OpticalSpec_MFOSCP_allEpochs}
\end{figure*}

Fig.~\ref{fig-OpticalSpec_MFOSCP_allEpochs} shows the evolution of the low-resolution optical spectra of V2891 Cyg from MFOSC-P. Our first optical spectrum was recorded on 2019 Nov 1 (+48d) with MFOSC-P in R500 and R2000 modes, when the nova was still rising. The spectrum showed a steeply rising continuum towards longer wavelengths, signifying a large amount of reddening, which was also clear from the photometry. Prominent emission lines of H-$\alpha$, O\,{\sc i} 7773$\AA$, and 8446$\AA$ were present. The O\,{\sc i} 7773$\AA$ line exhibited a P cygni profile \citep{Srivastava2019}. 
\par
\begin{figure}
	\centering
	\includegraphics[width=0.49\textwidth,angle=00]{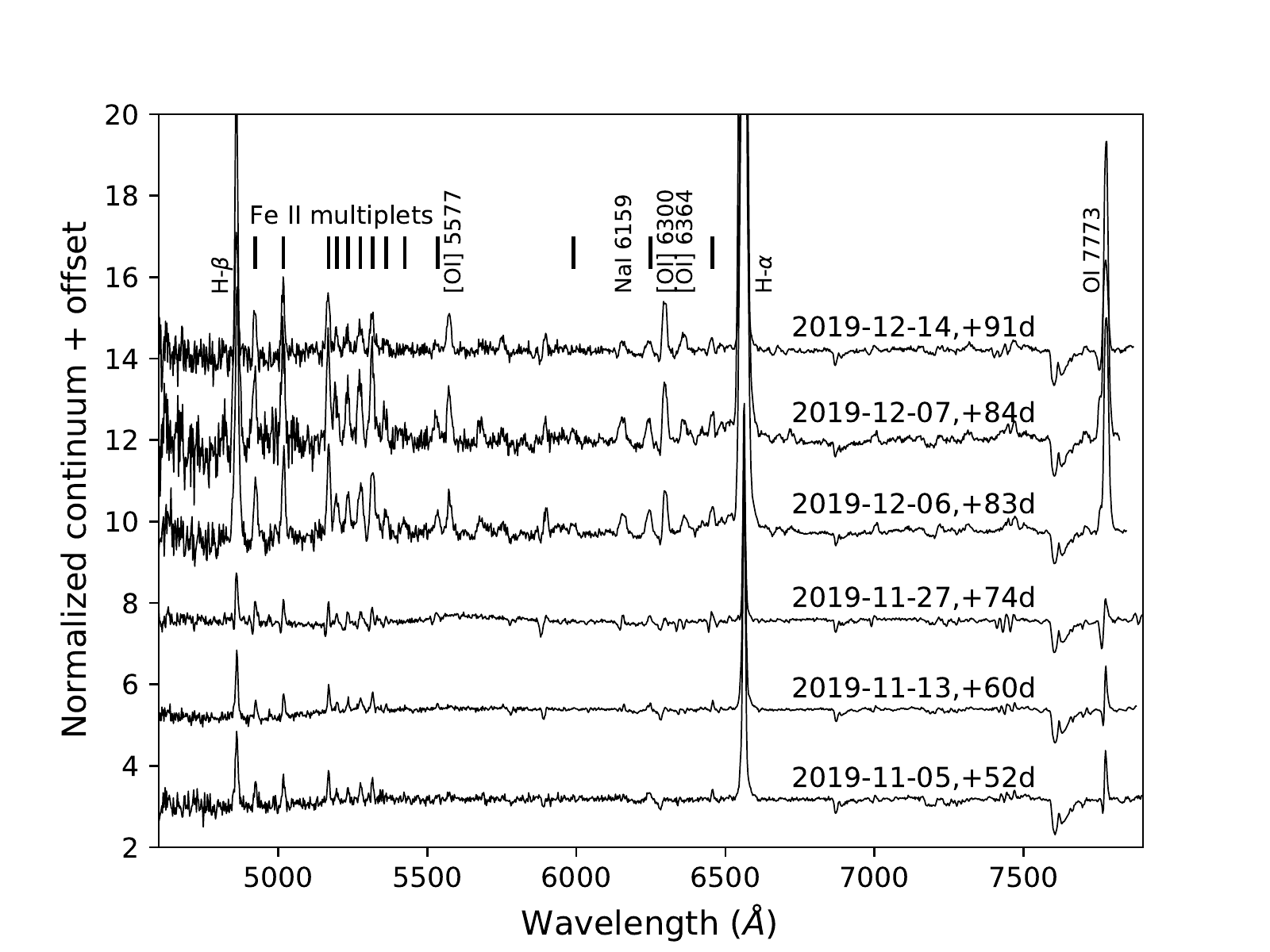}
	\caption[]{The continuum normalised spectra of nova V2891 Cyg from the B$\&$C 
	spectrograph. The Fe\,{\sc ii} multiplets on the blue end are prominent here.}
	\label{fig-OpticalSpec-BandC-NovDec}
\end{figure}

 
The continuum normalised spectra recorded with the B$\&$C spectrograph are shown in Fig.~\ref{fig-OpticalSpec-BandC-NovDec}. The H-$\beta$ and other emission features can be seen on the blue side of the spectra, which aided in the nova classification. Fe\,{\sc ii} emission at 4924$\AA$ (42\footnote{Fe II multiplet number}), 5018$\AA$ (42), 5169$\AA$ (42), 5198$\AA$ (49), 5235$\AA$ (49), 5276$\AA$ (49), 5317$\AA$ (49,48) and 6456$\AA$ (74) are clearly present. Later optical spectra, recorded on 2019 Nov 11(+58d), 12(+59d), 19(+66d), and Dec 1(+78d) also show the same features. A noticeable change was the disappearance of the P Cygni profile in O\,{\sc i} 7773$\AA$ between 2019 Nov 12 (+59d) and Nov 19 (+66d). This will be discussed later in section~\ref{subsec-OxygenLines}.


\begin{figure}
	\centering
	\includegraphics[width=0.49\textwidth]{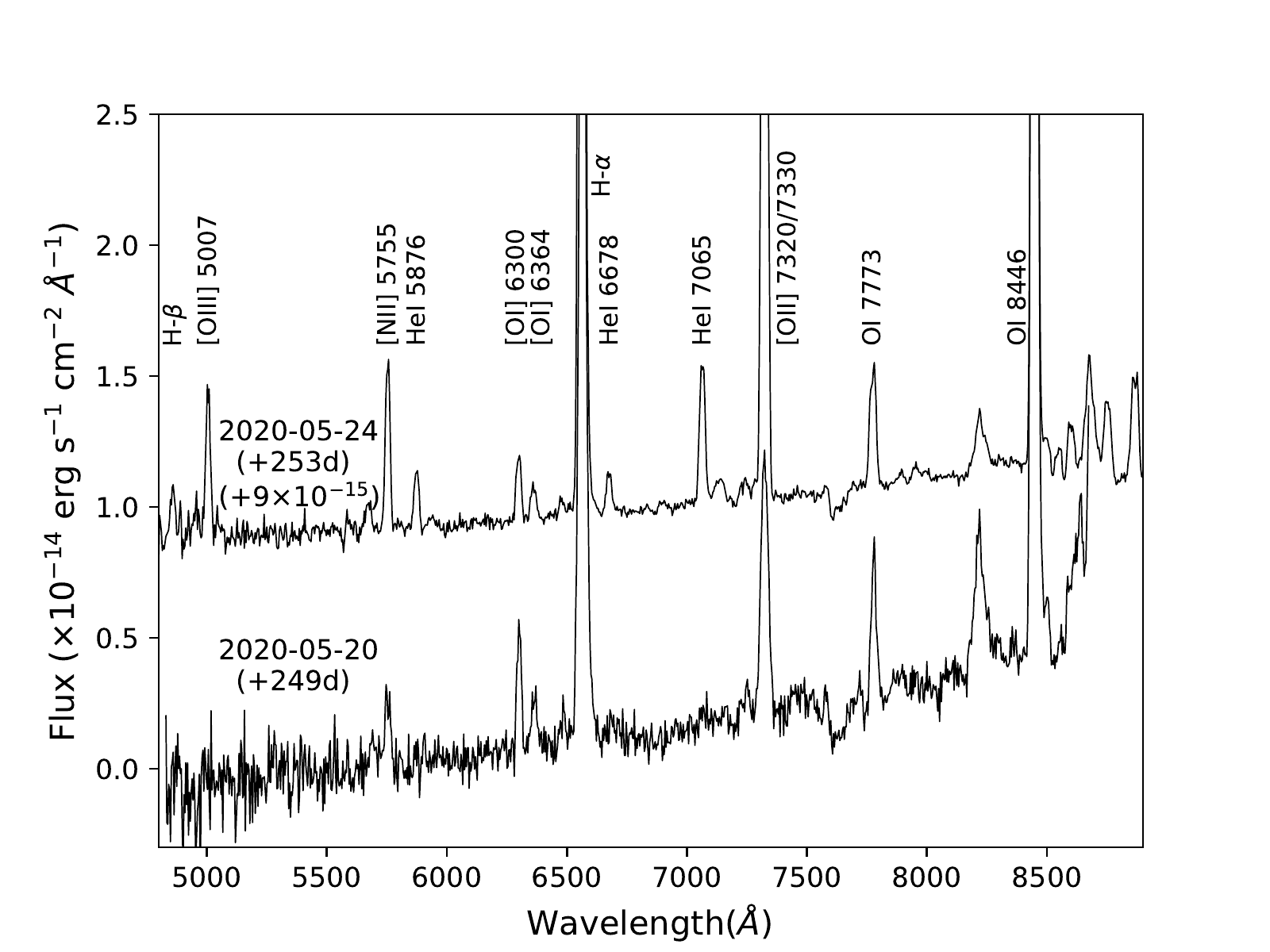}
	\caption[]{Change in spectra between 2020 May 20 (+249d) and May 24(+253d). 	The emergence of [O\,{\sc iii}] 5007$\AA$, [N\,{\sc ii}] 5755$\AA$, He\,{\sc i} 5876, 6678, 7065$\AA$ is evident. The spectra are not reddening corrected.}
	\label{fig-nebular-SpecChange}
\end{figure}


The nova spectra in 2019 December to 2020 January (+84d to +122d) show the emergence of [O\,{\sc i}] 5577, 6300, 6364$\AA$ emission. As the object entered solar conjunction in 2020 January, and due to subsequent COVID-19 lock-down restrictions, our next observations of the nova were not obtained until 2020 May. During the period between 2020 January and 2020 May, as the light curve shows, the nova largely remained at $V\sim16.5$~mag, varying irregularly within $\sim$1.5~mag. However, a decline had begun towards quiescence.
\par	
The optical spectra of 2020 May (+235d to +253d) showed the onset of the nebular phase, as typically characterised by the presence of [N\,{\sc ii}] 5755$\AA$, [O\,{\sc i}] 6300, 6364$\AA$, and [O\,{\sc ii}] 7320$\AA$ emission. The appearance of these new lines is highlighted in Fig.~\ref{fig-nebular-SpecChange} which shows the spectral changes between 2020 May 20 (+251d) and 24 (+253d). Several new lines have appeared, e.g. [O\,{\sc iii}] 5007$\AA$, neutral helium lines at 5876$\AA$, 6678$\AA$ and 7065$\AA$ etc. The H-$\alpha$/H-$\beta$ ratio, from the de-reddened spectrum on May 24 (+253d), is $\sim 7.2$, which is significantly greater than the value 2.7 as deduced from recombination theory \citep{Osterbrock1989} for optically thin conditions. As the critical densities for [O\,{\sc iii}] and [N\,{\sc ii}] are in the range of $\sim10^{5}-10^{6} cm^{-3}$ \citep{Osterbrock1989}, this shows that the hydrogen emission lines probably arise in the high-density clumps embedded in lower density surroundings.
\par
Few low-resolution spectra were recorded in 2020 Oct-Dec (+405d to +451d), when nova was around $\sim$18.5, 17, and 16.5 mags in the $V, R$ and $I$ bands, respectively. Given the moderate aperture of PRL 1.2m telescope, and the efficiency on the blue side of MFOSC-P spectral range, the spectra are noisy on the bluer side. Nevertheless, the emission features of [O\,{\sc iii}] 4959, 5007$\AA$, [N\,{\sc ii}] 5755$\AA$, O\,{\sc i} 8446$\AA$ are evident.  [O\,{\sc ii}] $\lambda\lambda$ 7320,7330 $\AA$ are strong and distinguishable, along with H-$\alpha$. He\,{\sc i} 6678$\AA$ had by then weakened, and He\,{\sc i} 7065$\AA$ is noticeable. He\,{\sc ii} 8237$\AA$ is still present, as in the spectra of 2020 May.


\subsection{Evolution of the  H$-\alpha$ : The absorption systems}
\label{subsec_Halpha_Evolution}

\begin{figure}
	\centering
	\includegraphics[angle=0,width=0.49\textwidth]{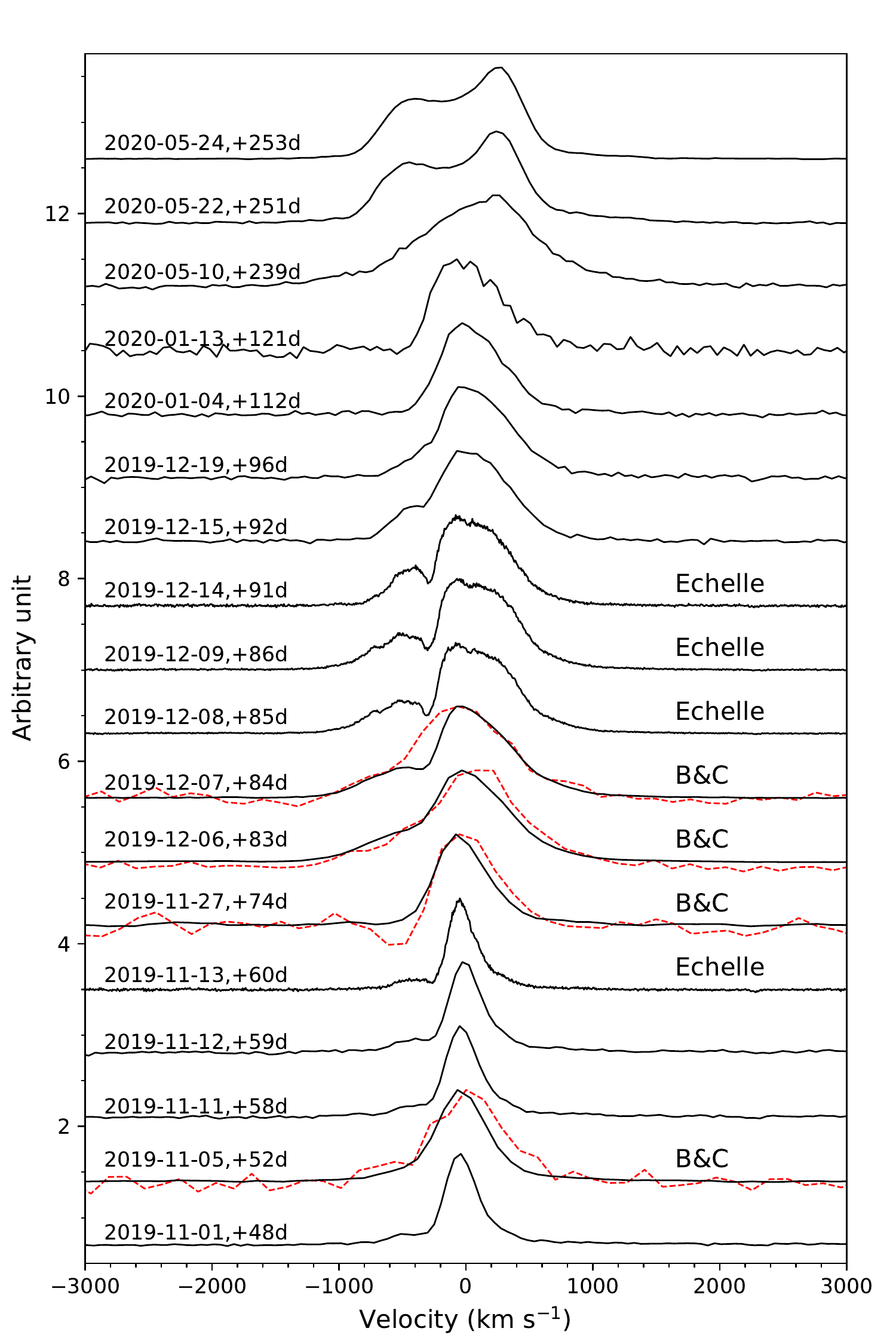}
	\caption[]{H-$\alpha$ emission profile variations in V2891 Cyg. 
	Profiles from R$\sim$2000 resolution MFOSC-P and B\&C spectrographs,
	and high resolution echelle spectrograph data, are presented. 
	H-$\beta$ profiles, obtained with the B\&C spectrograph, are shown as
	dashed red lines.}
	\label{fig-spec_H-alpha}
\end{figure}

\begin{figure}
	\centering
	\includegraphics[angle=0,width=0.49\textwidth]{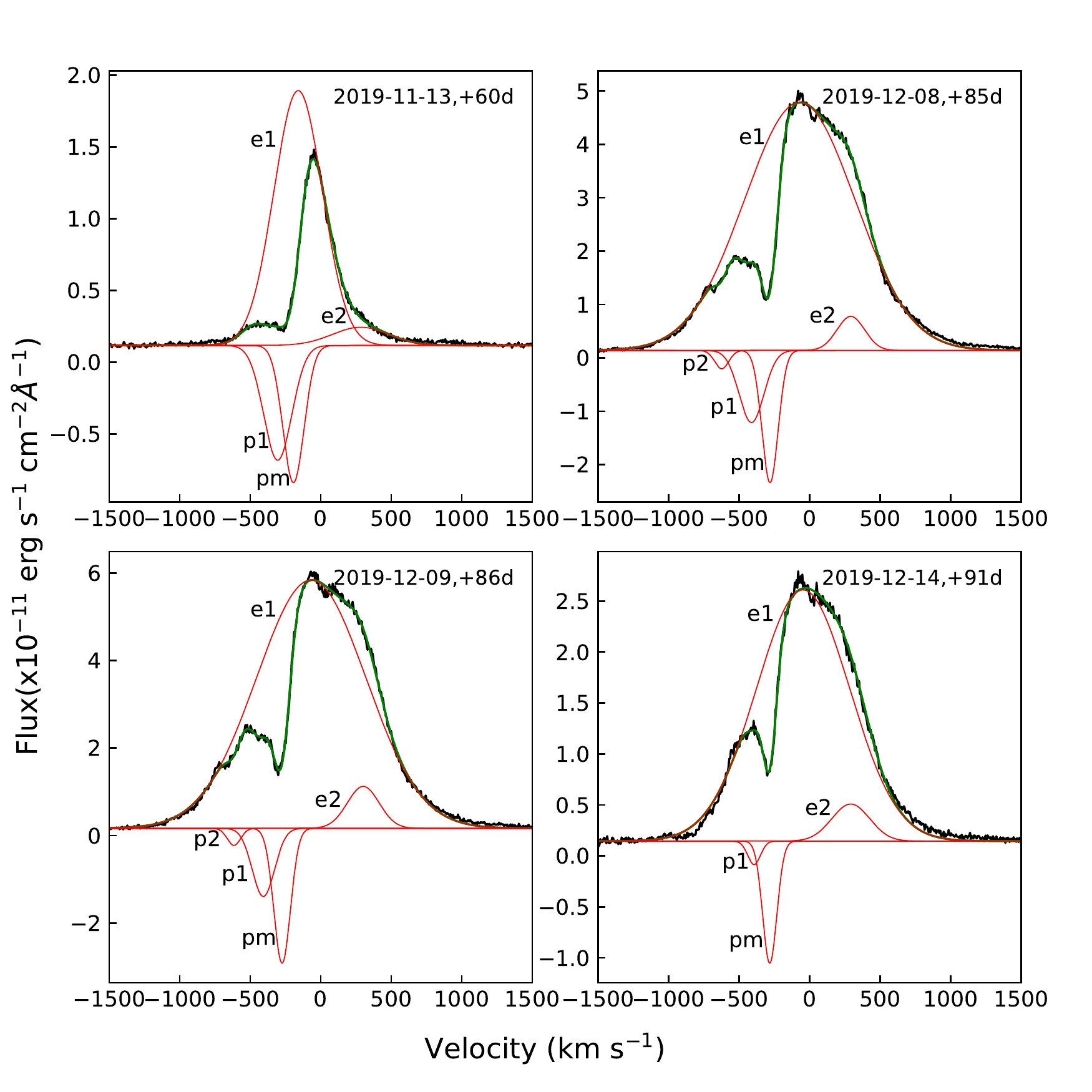}	
	\caption[]{The absorption systems superimposed on the (dereddened) H-$\alpha$ profiles from the echelle spectra. The profiles for 2019 Nov 13, Dec 8, Dec 9 and Dec 14 are shown. The constituents profiles are in red, while the combined profiles are in green, superimposed on the observed profiles (black). See section~\ref{subsec_Halpha_Evolution} for discussion.}
	\label{fig-Halpha-Absorption}
\end{figure}

The absorption systems in novae have long been known. \cite{McLaughlin1964} noted four classes: (a) pre-maximum, seen during the rise to maximum until just after maximum light,  (b) the principal component, which replaces the pre-maximum component a few days after the maximum; (c) the diffuse-enhanced component, which appears later than the pre-maximum and (d) the Orion component, which appears when the diffuse enhanced component is strongest. We notice the presence of such systems in nova V2891 Cyg, superimposed on the H-$\alpha$ profile (Fig.~\ref{fig-spec_H-alpha}).  We decomposed the H-$\alpha$ profiles of the high-resolution echelle spectra recorded on 2019 Nov 13(+60d), Dec 8(+85d), 9(+86d) and 14(+91d) for these superimposed absorption components, as shown in Fig.~\ref{fig-Halpha-Absorption} in the velocity space of H-$\alpha$. While H-$\alpha$ can be blended with [N\,{\sc ii}] 6548, 6584 $\AA$, the first instance of [N\,{\sc ii}] emission was not noted until 2020 May 22-24 (+251d,+253d), with the appearance of the [N\,{\sc ii} 5755$\AA$ line. Thus, the H-$\alpha$ emission on these earlier epochs is not expected to be blended with [N\,{\sc ii}]. The absorption and emission profiles were fitted with symmetric Gaussian functions. The derived heliocentric velocity (HV), velocity width at half maximum (VHM), equivalent width (EW) and integrated flux are given in Table~\ref{table-Halpha-components}.  

\begin{table}
	\centering
	\caption{Details of the decomposed emission and absorption components of H-$\alpha$ from the echelle spectra. The decomposed profiles are shown in Fig.~\ref{fig-Halpha-Absorption} for emission (e1 and e2), pre-maximum (pm) and principal absorption (p1 and p2) components. 
	Heliocentric velocity (HV), the velocity width at half maximum (VHM), equivalent width (EW), and integrated flux (reddening corrected) for emission (e1), pre-maximum (pm), principal (p), and one red-shifted emission (e2) systems are given. The listed values are not corrected for instrumental resolution. Errors are derived from the numerical fit. Fluxes are in units of $10^{-11}$ erg s$^{-1}$ cm$^{-2}$.}
	\begin{tabular}{lllll}
		\hline
		Component       &  HV     &   VHM           &   EW      &   Flux  \\
		&  (km s$^{-1}$)   & (km s$^{-1}$)   &  (\AA)    &   \\
		\hline
		\hline
		2019-11-13       &        &                 &         &                       \\
		(+60d) &       &                 &         &                       \\
		e1     &  -160$\pm$7  &  406$\pm$7  &   -145.1$\pm$5.6    & 
		16.77$\pm$0.65  \\
		pm &  -194$\pm$7  &  177$\pm$11 &     34.0$\pm$12.2   &  
		3.93$\pm$1.41  \\
		p1   &  -305$\pm$33 &  234$\pm$33 &     37.7$\pm$12.0   &  
		4.35$\pm$1.39  \\
		e2     &   282$\pm$43 &  471$\pm$52 &    -12.1$\pm$1.8    &  
		1.39$\pm$0.21  \\
		&        &                 &         &                       \\
		2019-12-08      &        &                 &         &                     \\
		(+85d)  &        &                 &         &                       \\
		e1     &  -69$\pm$2   &  924$\pm$2  &   -855.3$\pm$2.4 & 118.77$\pm$0.33\\
		pm &  -280$\pm$3  &  136$\pm$5  &   66.9$\pm$7.2   & 9.29$\pm$1.00         \\
		p1   &  -411$\pm$17 &  208$\pm$32 &   55.8$\pm$7.8   & 7.75$\pm$1.09         \\
		p2   &  -622$\pm$7  &  116$\pm$15 &   8.0$\pm$1.2    & 1.11$\pm$0.17         \\
		e2     &  293$\pm$3   &  231$\pm$8  &   -29.4$\pm$1.1  &
		4.08$\pm$0.15  \\
		&        &                 &         &                       \\
		2019-12-09      &        &                 &         &                       \\
		(+86d)&        &                 &         &                       \\
		e1 & -68$\pm$2   &  913$\pm$2  &   -731.8$\pm$2.6 & 120.63$\pm$0.43\\
		pm & -274$\pm$3  &  139$\pm$4  &   60.3$\pm$4.8   & 9.95$\pm$0.79        \\
		p1 & -407$\pm$12 &  185$\pm$20 &   40.8$\pm$4.9   & 6.73$\pm$0.81        \\
		p2 & -616$\pm$6  &  111$\pm$13 &   6.1$\pm$0.9    & 1.01$\pm$0.14        \\
		e2 & 301$\pm$2   &  266$\pm$7  &   -36.0$\pm$1.3  & 
		5.93$\pm$0.21        \\
		&        &                 &         &                       \\
		2019-12-14   &        &                 &         &         \\
		(+91d)&        &                 &         &                       \\
		e1 &  -46$\pm$3   &  779$\pm$4  &   -308.3$\pm$2.2 &  44.72$\pm$0.31  \\
		pm &  -282$\pm$3  &  119$\pm$5  &   22.8$\pm$1.1   & 3.30$\pm$0.15    \\
		p1 &  -393$\pm$14 &  106$\pm$21 &   3.9$\pm$1.0    & 0.57$\pm$0.15    
		\\
		e2 &  293$\pm$5   &  316$\pm$18 &   -18.5$\pm$1.6 &	2.68$\pm$0.23       \\
		\hline
	\end{tabular}
	\label{table-Halpha-components}
\end{table}


\par 
The H-$\alpha$ emission profile of 2019 Nov 13 (+60d) can be reconstructed with two additional absorption components at -194 and -305 kms$^{-1}$.  We identify these as pre-maximum and principal components. The main emission component (e1) had a velocity width of 406 kms$^{-1}$. Another faint emission component (e2) is also seen on the red wing, at a velocity of +282 km$s^{-1}$. Subsequent profiles can also be approximated by the same emissions (e1 and e2) and,
together with pre-maximum and principal absorption components. On 2019 Dec 8 (+85d), the pre-maximum component was seen at --280 kms$^{-1}$, while the principal absorption system now had two components (p1 and p2), at --411 and --622 kms$^{-1}$. The profile of Dec 9 (+86d) is similar 
to that of Dec 8. In the decomposed profile on Dec 14 (+91d), the pre-maximum absorption is at --282kms$^{-1}$. The p2 component of the principal absorption is not seen, while the p1 component persists at --393kms$^{-1}$.

\par 
The origin of absorption components in novae have been explored in recent studies \citep[see, e.g.,][and references therein]{Williams2008, Williams2010, Arai2016} in the context of {\it Transient Heavy Element Absorption} (THEA) systems, where the presence of ``principal'' and ``diffuse enhanced'' systems was seen as an indication of the presence of multiple envelopes/shells in the nova ejecta. In nova V2891 Cyg, these absorption components were evident until 2019 mid-December, i.e. around 13 weeks after the outburst. The MFOSC-P (R$\sim2000$) spectrum of Dec 19 (+96d) also shows the presence of these components. However, these absorption components are not present in the MFOSC-P spectrum ($R\sim2000$) recorded on 2020 Jan 4(+112d), i.e. around 16 weeks after the outburst detection. If we consider the velocity difference of $\sim$700 kms$^{-1}$ between the ejecta and such absorption system, the distance of these systems from the central source is estimated as $\sim$43~au. This is consistent with the projection of \cite{Williams2008}, who postulated that the disappearance is due to the collision between the two gaseous components.
\par 
The emission component e1 shows the evolution of its heliocentric velocity, as well as a spread in its velocity profile (Table~\ref{table-Halpha-components}), while the red-shifted e2 emission remains almost constant. The e2 component is weak in the spectrum of Nov 13 (+60d), but strengthens by Dec 8 (+85d). This could be due to a change in the optical depth of the gas between +60 and +85 days, so that receding material (positive velocity) -- previously not visible
-- has now gained in strength. This e2 component again becomes weaker by Dec 14 (+91d). It is interesting to note that the flux of the e1 and e2 components dropped by factors of $\sim$2.7 
and $\sim$2.2, respectively, in a 5-day interval (between +86d and +91d). Such speedy recombination rates point to very high electron densities. However, at the time of these H-$\alpha$ observations, the light curve of the nova was undergoing the largest and and most rapid up-and-down changes, and the flux of the e1 component evolved in parallel with the $V$ band flux of the nova. Therefore, evidently, the nebular material responsible for the emission in the $V$
band was at high density, and capable of recombining in a matter of very few days and rapidly following the changing photo-ionising input from the central star. This behaviour is mirrored by that of the e1 H-$\alpha$ component, being emitted by the same gas as that responsible for the continuum emission at optical wavelengths. 
\par 
Optical depth and line of sight effects, the complex geometry of the ejecta (e.g. bipolar flow), interaction with surrounding pre-existing material around the nova etc., can give rise to the asymmetry in line profiles, as also seen in other novae e.g. Nova Sco 2015 \citep{Srivastava2015}, Nova Del 339 \citep{DeGennaroAquino2015} etc. As the nova attained its first maximum on 2019 Nov 30(+77d), the relatively narrower width, and larger blue-shifted 
velocity of the e1 component (on +60d), may possibly be due to the emission coming from a very thin outer layer of the optically thick ejecta. Whereas its evolution between +84d to +91d corresponds to the optically thinning of the ejecta as a larger volume is now available for the emission profile. This could also explain why the fluxes (or equivalent widths) of both emission components e1 and e2 were not in synchronisation between +60d and +84d, whereas between +84d and +91d, e2 followed the behaviour of e1.


\subsection{Behaviour of O\,{\sc i} lines: variability of P Cygni 
profile of O\,{\sc i}  7773$\AA$ line}
\label{subsec-OxygenLines}

V2891 Cyg showed prominent oxygen lines at various stages of its evolution, from initial fireball to coronal phase, a period of 15 months. We determined the flux ratios of various optical and NIR lines (e.g., O\,{\sc i} 0.8446$\mu$m, 1.1287$\mu$m, and 1.3164$\mu$m), which suggested the action of Ly-$\beta$ fluorescence. The [O\,{\sc i}] lines  (5577, 6300, 6364 $\AA$) are used to determine the electron temperature $T_e$ of the neutral gas, and the  mass M$_{OI}$ of the neutral O\,{\sc i}, following the prescription of \cite{Williams1994}. These lines were detected in the spectra of 2019 Dec 7(+83d) and Dec 15(+92d). The mass of neutral oxygen is estimated  to be $\sim8.6[\pm1.3]\times10^{-6}$M$_\odot$, which is consistent with the typical mass of oxygen in a nova eruption \citep[$\sim1.1\times10^{-4} - 5.3\times10^{-8}$M$_{\odot}$;][]{Williams1994}. The corresponding electron temperature was determined to be $\sim$5600 K.
 \par
\begin{figure*}
	\centering
	\includegraphics[angle=0,width=0.99\textwidth]{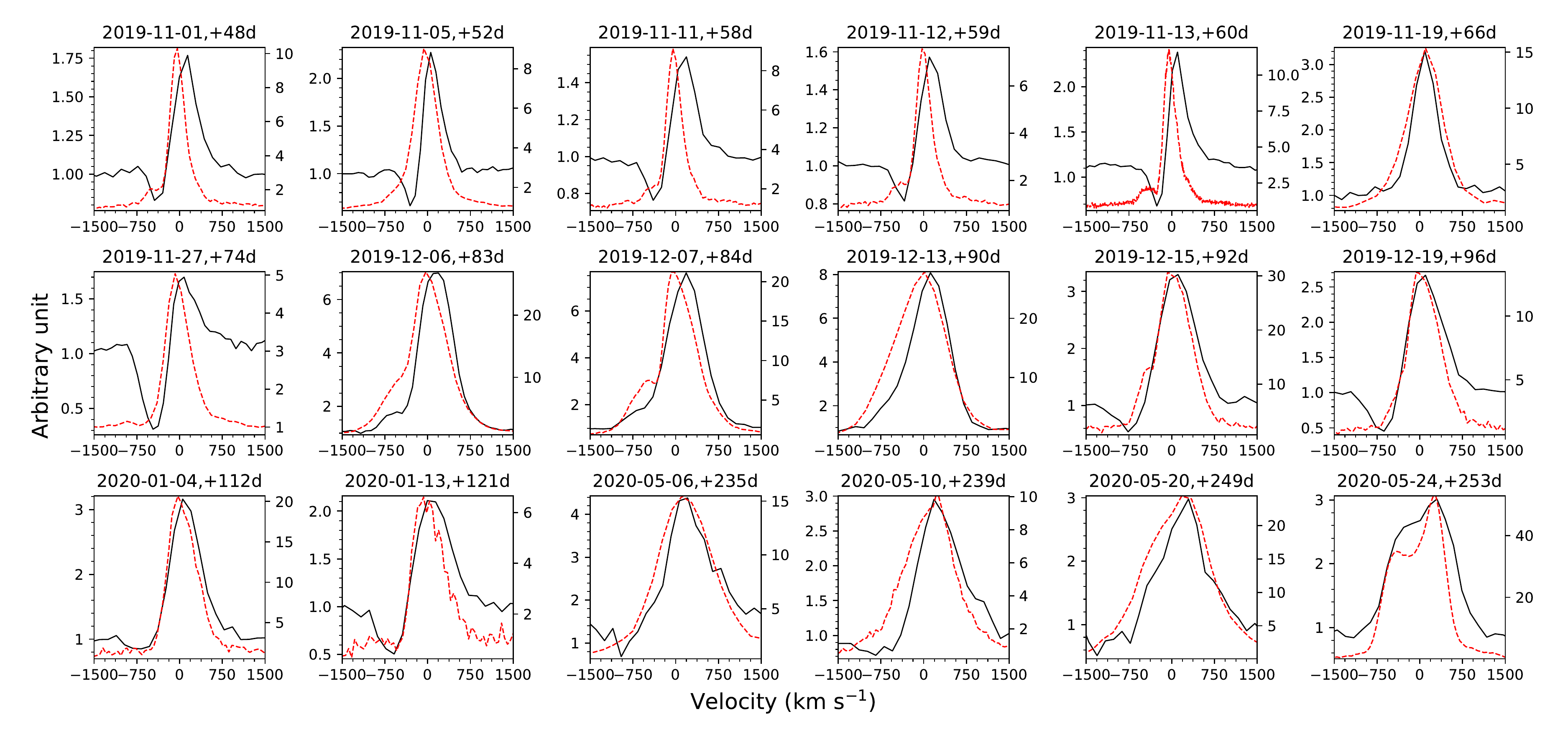}
	\caption[]{The line profile variation seen in the O\,{\sc i} 7773$\AA$ (solid black) and H$\alpha$ (dashed red) lines during 2019 Nov - 2020 May. The relative flux for O\,{\sc i} 7773$\AA$ and H$\alpha$ are given on the left and right axes for each of the sub-plots respectively. The appearance and disappearance of P Cygni features is evident. Profiles for Nov 5, Nov 13, Nov 27 and Dec 6 are from the B\&C spectrograph; the remainder all are from MFOSC-P. 
	See section~\ref{subsec-OxygenLines} for discussion}
	\label{fig-OI7773-VeloPlot}
\end{figure*}

 
An interesting development was the variability in the line profile of the O\,{\sc i} 7773$\AA$ during its early evolution of during 2019 Nov--2020 Jan. The velocity profiles of O\,{\sc i} 
7773$\AA$ are shown in Fig.~\ref{fig-OI7773-VeloPlot}, from 2019 Nov to 2020 May. A P Cygni profile in the O\,{\sc i} 7773$\AA$ line was seen in our first optical spectrum, taken on 2019 Nov 1(+48d), and on the spectrum taken on 2019 Nov 12(+59d). It had disappeared by Nov 19(+66d).  The absorption feature was at $\sim$500 kms$^{-1}$ on Nov 1(+48d), while on Nov 12(+59d), it was at $\sim$300 kms$^{-1}$. The subsequent change in the O\,{\sc i} 7773$\AA$ line profile was found to be correlated with its re-brightening. It can be seen from the LC (Fig.~\ref{fig-LC1}), and Tables~\ref{table-ObsMFOSCP}, and ~\ref{table-ObsPhotANS}, that the first instance of brightening had begun around 2019 Nov 25(+72d). The nova attained its first maximum on Nov 30(+77d) ($m{_V}\sim$14.05), faded to a minimum on Dec 7(+84d), and brightened on Dec 17(+94d). Correspondingly, the O\,{\sc i} 7773$\AA$ profile of Nov 27(+74d) shows deep P Cygni absorption (absorption component at $\sim$450 kms$^{-1}$), which disappeared by Dec 13(+90d). The  P Cygni feature in O\,{\sc i} 7773$\AA$ had again re-appeared in the spectra taken on Dec 15(+92d) and Dec 19(+96d) (absorption feature at $\sim$700-750 kms$^{-1}$). Another re-brightening occurred on 2020 Jan 12(+120d), whose imprint can be seen in the re-appearance of a deep P Cygni profile of the O\,{\sc i} 7773$\AA$ line on 2020 Jan 13(+121d), with absorption feature at $\sim$600 kms$^{-1}$. It is of further interest to note that the H-$\alpha$ profile of Jan 13(+121d) also shows the P Cygni feature. In both cases, the blue-shifted absorption occurred at a velocity of $\sim$500-600km$s^{-1}$, indicating that the blue absorptions
in H-$\alpha$/O\,{\sc i} 7773$\AA$ are likely to be due to the outburst. 
\par 
The disappearance and later re-appearance of P Cygni profiles also fit the  pattern seen in other novae \citep{Tanaka2011}. \cite{Tanaka2011} have studied six slow/moderate fast novae, which showed re-brightening in the early phase after maximum. However, these novae showed a similar 
re-appearance of P Cygni profiles in multiple emission lines, e.g. H-$\alpha$, H-$\beta$, He\,{\sc i}, Na\,{\sc i}, Fe\,{\sc ii}, O\,{\sc i} etc. The typical blue-shifted velocity separation of absorption components was found to be in the range $\sim750-2500$~km$s^{-1}$. Several consistent reasons for re-brightening, and the re-appearance of P cygni features, in novae have been suggested, such as hydrogen-burning instabilities \citep{Pejcha2009}, re-expansion of the photosphere \citep{Tanaka2011}, mass ejection at the re-brightening \citep{Csak2005} etc. The re-appearance of P Cygni absorption at different epochs, around the re-brightening episodes, and at higher velocities confirm the scenario that periodic mass loss is happening during the re-brightening phase. Similar behaviour was also seen during the epoch of re-brightening, for example, in nova V4745 Sgr \citep{Tanaka2011}, in which the P Cygni profiles appeared after the second brightness peak.

\section{Results from NIR Spectoscopy}
\label{Sec_NIRSpec}

\begin{figure*}
	\centering
	\includegraphics[width=0.99\textwidth]{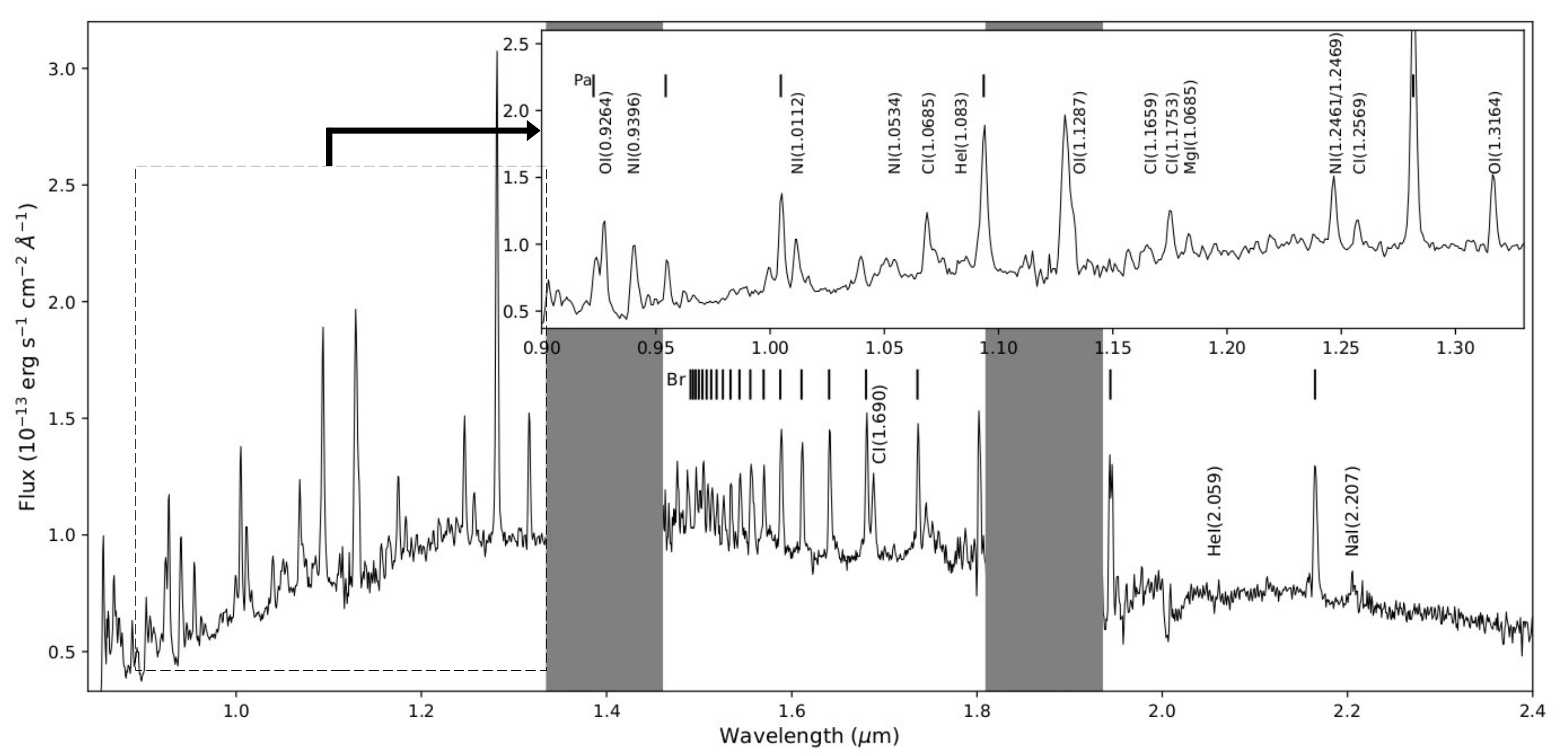}	
	\caption[]{The NIR spectrum of Nova V2891 Cyg covering 0.84--2.4 $\mu$m 
	obtained from Mount Abu on 2019 Nov 17 (+64d). The region between 
	0.90--1.35$\mu$m is shown in the inset, with features identified. 
	The $H$ band ($\sim$1.4-1.8$\mu$m) region shows numerous emission 
	lines of the hydrogen Brackett series, which are marked by vertical
	ticks. Regions of poor atmospheric transmission are shaded.
	The spectrum is not corrected for reddening.}
	\label{fig-SpecNIR-2019Nov18}
\end{figure*}

\begin{figure}
	\centering
	\includegraphics[width=0.49\textwidth]{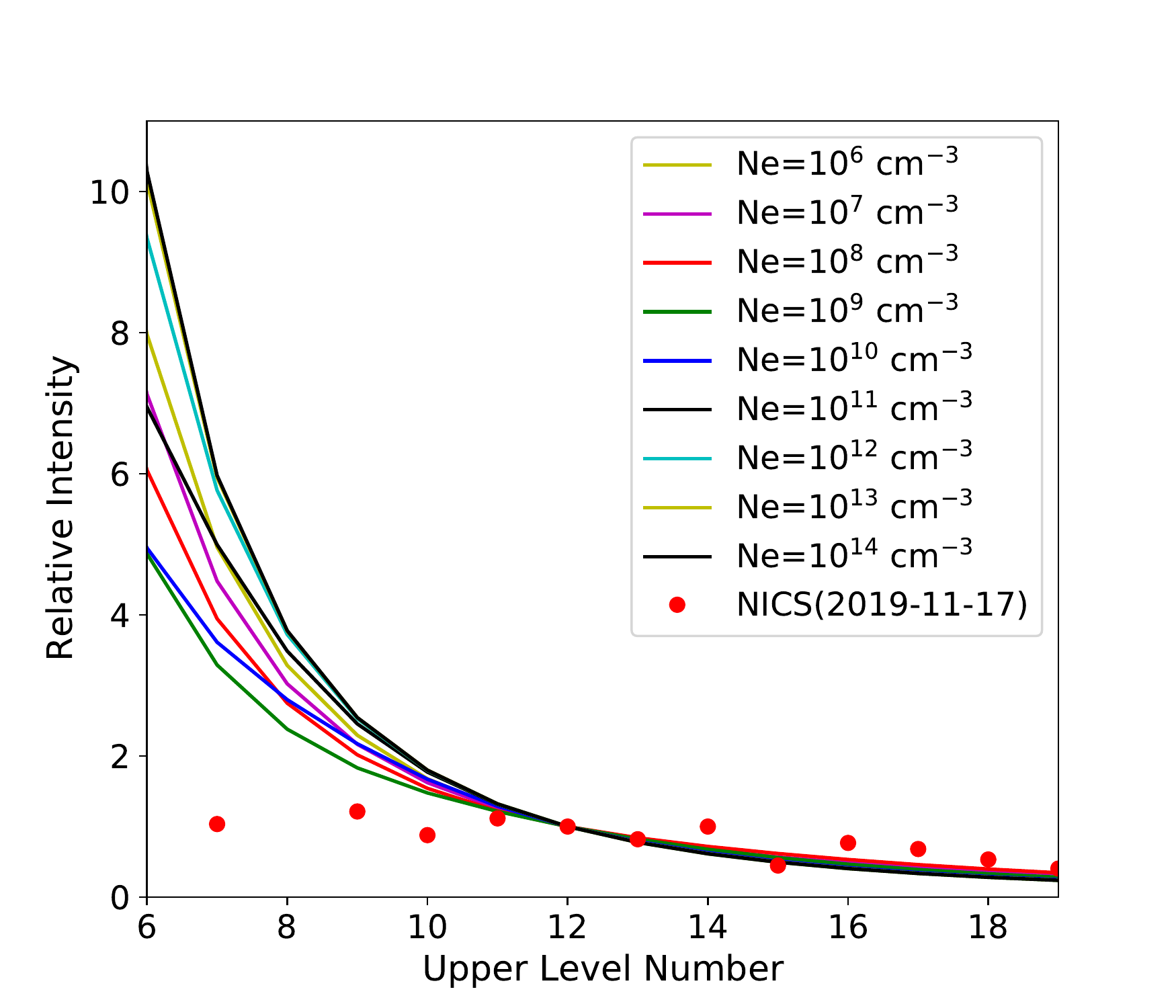}
	\caption[]{Recombination Case B analysis for the Brackett lines on 
	$\sim$+64d. The abscissa  gives the upper level of the line transition. 
	The line intensities  are normalized with respect to Br12. The model 
	Case B predictions are also shown for temperature = 10000 K 
	and electron number densities of 10$^6$-10$^{14}$ cm$^{-3}$.}
	\label{fig-CaseB}
\end{figure}	

\subsection{NIR Spectroscopy: the pre-maximum phase}
\label{subsec_NIRSpec}

Our earliest NIR spectrum of V2891 Cyg, obtained on 2019 Nov 17 (+64d) from Mt. Abu, is shown in Fig.~\ref{fig-SpecNIR-2019Nov18} with prominent lines marked in the inset. The spectrum displays emission lines of H, C, N and O, typically seen in the spectra of the Fe\,{\sc ii} class of novae, several of which are shown in \cite{Banerjee2012}. The He\,{\sc i} 1.083, 2.059$\mu$m lines are present, but weak. The lines are narrow (Pa$\beta$ has a FWHM of $\sim$640 km$s^{-1}$), with no P Cygni features. Although no first overtone CO emission is seen in the $K$ band, lines of Na and Mg are present, which suggest \citep{das2008} the presence of cool, low-excitation regions or clumpy material, conducive for molecules and dust to form later in the evolution; dust did indeed condense subsequently. The NIR photometric observations yielded $J = 9.98\pm0.02$, $H = 8.87\pm0.03$ and $Ks = 7.66\pm0.02$.
\par 
A recombination Case B analysis of the de-reddened hydrogen Brackett (Br) lines shows that they are optically thick. Fig.~\ref{fig-CaseB} shows the Br line strengths normalised to  Br-12 (the typical error is $\sim10$\% in the line flux measurements). The Case B emissivities at $T_e$ = 10000 K, and various electron densities from \cite{Storey1995}, are also shown. A significant deviation is seen in the Br-7 (Br$\gamma$) line strength, which is expected, under optically thin 
conditions, to be of higher intensity compared to the higher Br series \citep{Storey1995}. Such optical depth effects are expected in the early epoch spectra of novae \citep[e.g., Nova Cep 2014, and Nova Sco 2015;][]{Srivastava2015}. During these epochs,  V2891 Cyg was undergoing the episodes of re-brightenings (Fig.~\ref{fig-LC1}), so such optical depth effects are plausible.
\par 
Since the Br$\gamma$ line is optically thick, only constraints on the electron density and emission measure are possible. The optical depth at the line-center, $\tau_{n,n'}$, is given by 
$\tau = n_e n_i \Omega(n,n') D$, where $n_e$, $n_i$, $D$ and $\Omega(n,n')$ are the electron and ion number densities, path length, and opacity corresponding to the transition from upper level $n$ to lower level $n'$ respectively \citep{Hummer1987,Storey1995}.  Using opacities from \cite{Storey1995}, the condition that the optical depth in the Br-$\gamma$ line, $\tau$, be  $> 1 $, and using the same procedure as in \cite{Srivastava2015}, the emission measure $n_e^2D$ can be constrained to lie between between 1.34$\times$10$^{33}$ and 3.83$\times$10$^{34}$cm$^{-5}$. 
Further, the electron density can be constrained by taking $D = v\times$t as the kinematical distance travelled by the ejecta, where $v$ is the ejecta velocity, and $t$ is the time after outburst. Considering half of the Full Width at Zero Intensity (FWZI) of the H-$\alpha$ line 
($\sim1000$ km$s^{-1}$) as the expansion velocity, and $t$ as $\sim60$ days after the outburst, this puts a lower limit on electron density in the ejecta in the range of 1.14$\times$10$^{9}$-6.08$\times$10$^{9}$ cm$^{-3}$. 
\par 
The four other NIR epochs of observation between May -- October 2020 could be mildly affected by a dust formation event that occurred during that period. The formation of dust affects the line strengths through additional reddening, and hence the dust event is discussed first, followed by a Case B analysis of these May -- October 2020 NIR observations.


\subsection{Dust formation}
\label{subsec_dust}

\begin{figure}
	\centering
	\includegraphics[bb = 116 179 458 787, width=3.2in, clip]{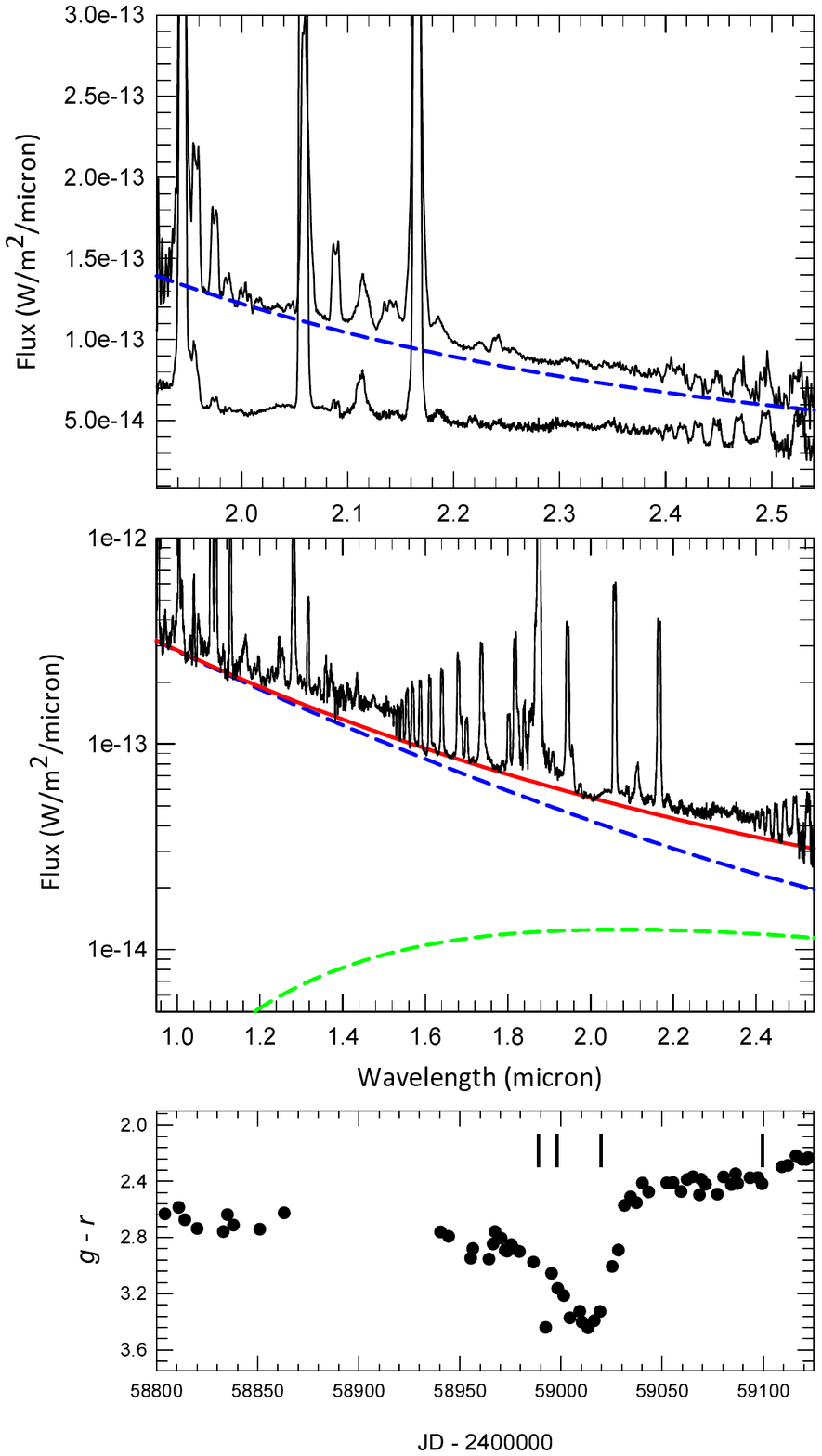}
	\caption[]{Top panel: how the slope of the spectral continuum flattened out 
	between May 20(+249d) (represented by the spectrum whose continuum is fitted
	by the blue dashed line) and the spectrum of June 7(+267d) just below, 
	thereby indicating the development of an IR excess. The four vertical 
	ticks on the lower panel are the days marking the spectra of May 20(+249d), 
	June 7(+267d), June 30(+290d) and September 19(+371d), respectively. 
	Middle panel: 1400K blackbody curve (green), whose contribution needs 
	to be added to the blue-dashed continuum, to reproduce the observed 
	continuum of June 7 (shown in red). Details are given in Section~\ref{subsec_dust}. The spectra are dereddened.}
	\label{fig-DustFormation}
\end{figure}

The evolution of the $g - r$ colour from the ZTF lightcurves (bottom panel of Fig.~\ref{fig-DustFormation}; also Fig.~\ref{fig-LC2} of the ZTF LC's presented initially) clearly shows a reddening of the nova over a period of $\pm$ 50 days, centred around JD 2459013 
(2020 June 13 or $\sim$ +273d after outburst). We believe this behaviour indicates a short episode of dust formation and destruction. This assertion is supported by a dip in the $g - r$ colour at around this time, indicating reddening by dust. The epochs of our NIR spectroscopy, relative to the dust event, are shown by vertical ticks in the lower panel of Fig.~\ref{fig-DustFormation}. The top panel of Fig.~\ref{fig-DustFormation} shows the change in the slope of the $K$ band continuum for 2020 May 20(+249d), when the dust had just started to form, relative to the spectrum of 2020 June 7(+267d). The latter shows a significant flattening with respect to the former, indicating the development of a clearly discernible IR excess due to 
dust. To estimate the dust mass, we fit a black-body curve (green curve; middle panel) to the excess IR emission of June 7(+267d) relative to the blue-dashed continuum of May 20(+249d).  
A black-body temperature of $1400\pm200$~K is estimated.
\par 
We estimate the dust mass following \cite{Evans2017} and \cite{Banerjee2018}, assuming that the grains are spherical, and that the dust is composed of carbonaceous material. It has been shown 
\citep{Evans2017} that the dust masses for amorphous carbon (AC) and graphitic (GR) grains, assuming optically thin emission, are given by:
\begin{equation}
\frac{M_{\rm dust(AC)}}{M_{\odot}} \simeq 5.63 \times 10^{17} 
\left(\frac{d}{5.5~{\rm kpc}}\right)^2 
\frac{(\lambda f_{\lambda})_{\rm max}}{T^{4.754}_{dust}} 
\end{equation}
\begin{equation}
\frac{M_{\rm dust(GR)}}{M_{\odot}} \simeq 5.01 \times 10^{19} 
\left(\frac{d}{5.5~\mbox{kpc}}\right)^2 \frac{(\lambda f_{\lambda})_{\rm max}}{T^{5.315}_{dust}} 
\end{equation}
where a value of $d$ = 5.5 kpc has been used for the distance, $\rho$ = 2.25 gm cm$^{-3}$ has been taken for the density of the carbon grains and ($\lambda{f}_{\lambda}$)$_{\rm max}$ is obtained from the blackbody fit, and measured in units of W/m$^{2}$. The dust mass, which is independent of grain size \citep{Evans2017}, is estimated to be $\sim 0.83 \times 10^{-10}$ M$_{\odot}$, and $1.25 \times 10^{-10}$ M$_{\odot}$, for AC and GR grains, respectively;
these are very modest amounts.
\par 
Classical novae, especially of the Fe\,{\sc ii} class, commonly show evidence of rapid dust formation within months of the outburst. In this particular case, we propose a specific origin for the dust condensation. Following \cite{Derdzinski2017}, we propose that dust formation occurred within the cool, dense shell behind shocks that had formed in the ejecta.  We offer proof in the coming subsection that strong shocks were likely present at this stage in time, and that the nebular and coronal lines that were seen between 2020 May to September arose due to shock-heating of the gas rather than due to the presence of a hot central photoionising source. 

\subsection{NIR spectroscopy at late stages between $\sim$ 250 to 370 days}
\label{sec_NIRSpecMaySep2020}

Near and mid-IR observations in 2020 May \citep{De2020, Woodward2020} showed that the carbon lines (a hallmark of the Fe\,{\sc ii} class of novae), that had initially been present, had faded. The spectra were rich in the emission lines of H, He, O, N, Fe etc., along with prominent lines of the Paschen and Brackett series. The Paschen and Brackett lines now showed a double-peaked structure having a separation of $\sim 480-540$ km$s^{-1}$, with the red peak stronger than the blue. Subsequent monitoring further showed the emergence of the coronal lines [Si\,{\sc vi}] 1.962$\mu$m, [Ca\,{\sc viii}] 2.322$\mu$m, and [Si\,{\sc vii}] 2.483$\mu$m in June-September \citep{Woodward2020b}; we discuss these later in this section.

\subsubsection{Estimate of the ejecta mass }
\label{subsec_ejectaMass}

\begin{table}
	\centering
	\caption{Mass of the ionised hydrogen in the ejecta as determined
	from H lines. H line fluxes are estimated from the spectrum of 
	2020 June 07 (+267d). See section~\ref{subsec_ejectaMass} for details.}
	\begin{tabular}{lcccccc}
		\hline
		Wavelength         & Upper level        & Line Flux  & Mass ($\times$10$^{-5}$$M_{\odot}$)     \\
		(micron)     & of transition          &  ($\times$10$^{-13}$      & (for $n_e=10^{7.5}$   \\
		&    &  erg/s/cm$^{2}$)  &  $cm^{-3}$)  \\
		
		\hline
		\hline				
		
		Brackett	&		&		&		\\
		4.0500	&	5 (Br$\alpha$)	&	73.60	&	8.35 \\
		2.1650	&	7 (Br$\gamma$)	&	27.40	&	8.09 \\
		Pfund	&		&		&		\\
		3.7385	&	8	&	10.10	&	8.18 \\
		3.2950	&	9	&	7.42	&	8.35 \\
		3.0376	&	10	&	5.80	&	8.68 \\
		2.8714	&	11	&	4.42	&	8.52 \\
		2.4940	&	17	&	1.68	&	9.02 \\
		2.4686	&	18	&	1.45	&	8.94 \\
		2.4470	&	19	&	1.21	&	8.52 \\		
		2.4300	&	20	&	1.02	&	8.18 \\
		2.4150	&	21	&	0.72	&	6.60 \\
		2.4023	&	22	&	0.55	&	5.59\\		
		Humphreys	&		&		&    \\
		3.9054	&	15	&	1.34	&	9.27 \\
		3.8170	&	16	&	1.00	&	8.02 \\
		3.6910	&	18	&	0.87	&	9.19 \\
		3.6440	&	19	&	0.65	&	7.85 \\
		3.6050	&	20	&	0.56	&	7.77 \\
		3.5720	&	21	&	0.56	&	8.67 \\
		3.5200	&	23	&	0.48	&	9.44 \\
		
		\hline
		\hline
	\end{tabular}
	\label{table-ejectaMass}
\end{table}

\begin{figure*}
	\centering
	\includegraphics[width=0.99\textwidth]{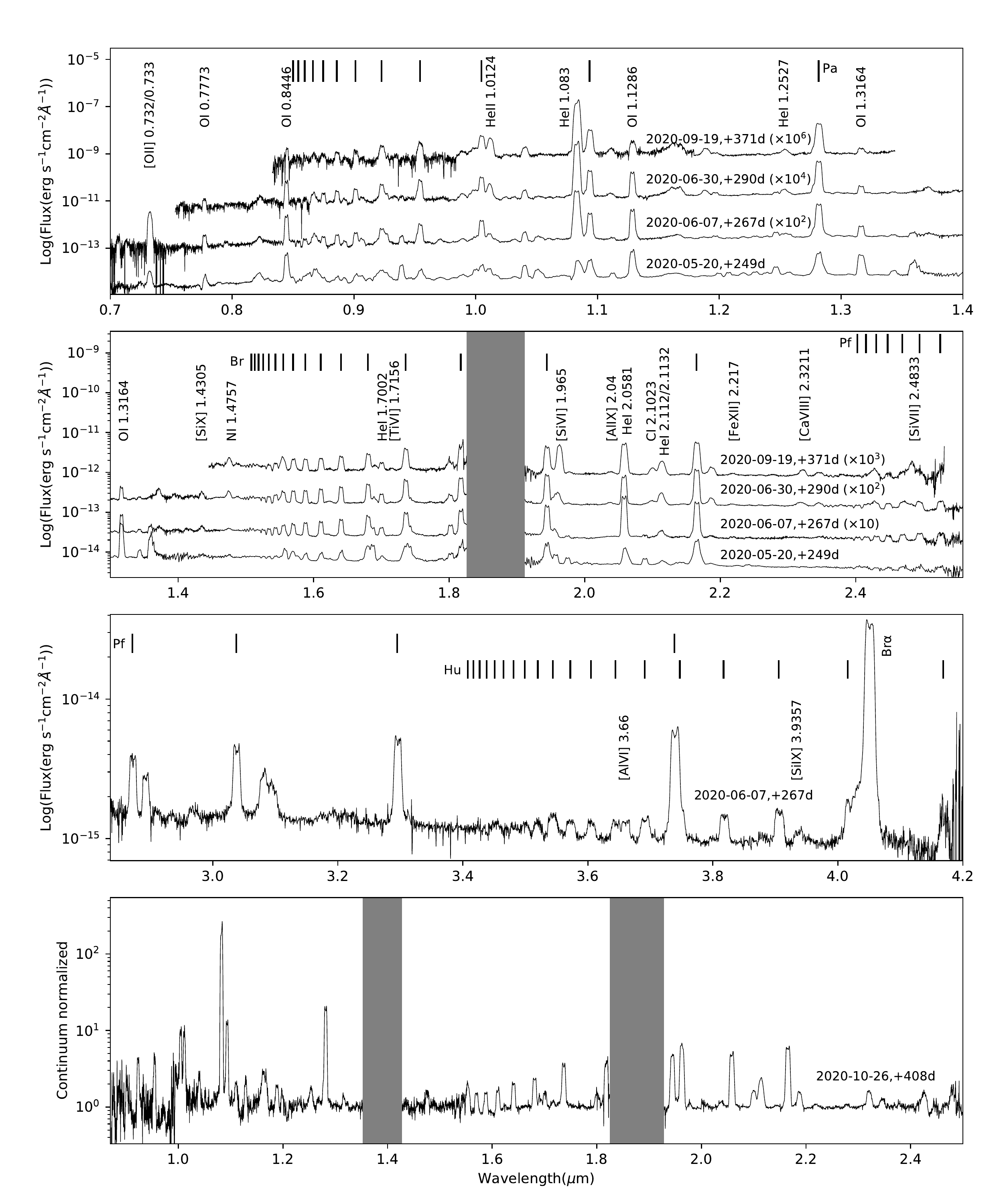}
	\caption[]{The IRTF spectrum (without dereddening) of Nova V2891 Cyg 
	covering 0.67--4.2$\mu$m are shown in top three panels. 
	These were obtained on 2020 May 20, June 07, June 30 and 2020 Sep 19. 
	Lowest panel shows TANSPEC spectrum covering 0.86--2.5 $\mu$m, obtained 
	on 2020 Oct 26.  This spectrum is normalised with respect to the continuum.}
	\label{fig-spec-NIR-IRTF}
\end{figure*}
\begin{figure}
	\centering
	\includegraphics[angle=0,width=0.49\textwidth]{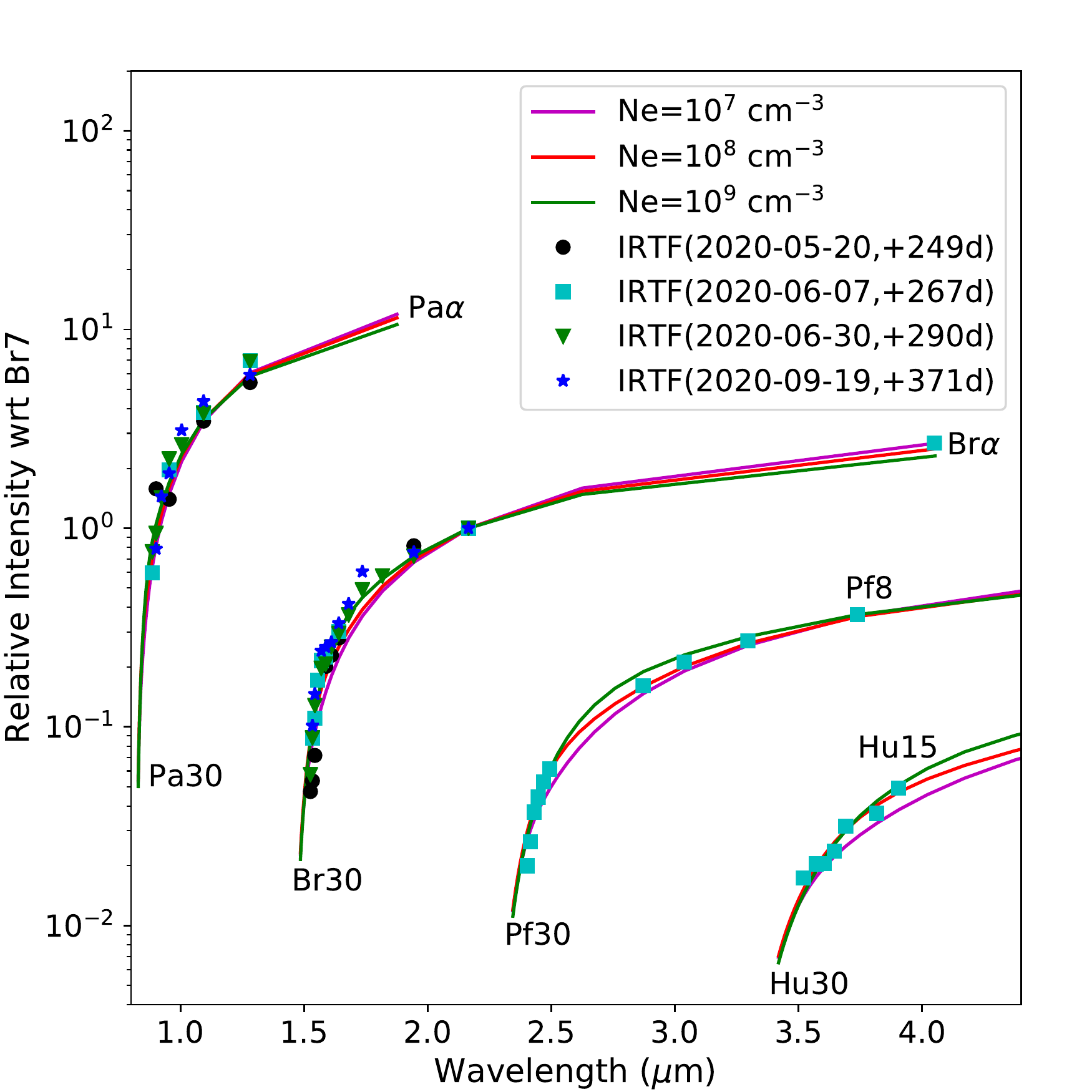}
	\caption[]{Recombination Case B analysis for the  Pa, Br, Pf and Hu 
	lines, with the line intensities normalised  to Br12.  The model Case B
	predictions are also shown for temperature of 10000 K, and electron densities
	of 10$^7$-10$^{9}$ cm$^{-3}$, expected at the epochs under consideration 
	(shown in the inset).}
	\label{fig-CaseBLateEpoch}
\end{figure}

The NIR spectra during the late stages (2020 May - October) are shown in Fig.~\ref{fig-spec-NIR-IRTF}, with the prominent lines identified. The lines of the Pa, Br, Pf and Hu series dominate, while several coronal lines are also detected \citep{Woodward2020b}; these are listed in Table~\ref{table-CoronalLines} and discussed below. A Case B analysis, for these late epochs, now shows good conformity with model predictions (Fig.~\ref{fig-CaseBLateEpoch}) over the gamut of lines from each of the Pa, Br, Pf and Hu series. This indicates that the H lines, at this stage, are optically thin.  The compliance with Case B, with only the ISM reddening of $E(B-V) = 2.20$ being applied, also suggests that the dust event has not influenced the strengths of the lines significantly via additional reddening. This is consistent with the earlier inference that the amount of dust formed is modest. Under these circumstances, the ejecta mass M$_{\rm ej}$ (approximated by the mass in the form of ionised hydrogen) can be estimated from  
\begin{equation}
{M_{ej}} = \frac{4{\pi}{d^{2}}{F(\mbox{line})}{m_{p}}}{{n_{e}}{\epsilon}(\mbox{line})} \:\:,
\end{equation}
where $d$ is the distance, $F(\mbox{line})$ is the observed flux in a H line, $\epsilon$(\mbox{line}) is the emissivity in the line as given in \cite{Storey1995}, n$_{e}$ is 
the electron density, and m$_{p}$ is the proton mass. Table~\ref{table-ejectaMass} gives ejecta mass in M${_\odot}$. Of these, the most reliable estimates are for the lines at longer wavelengths (viz., the  Pf and Hu lines, and Br$\alpha$) because their strengths are least affected by any extra reddening due to the dust (for example, beyond 3.5 $\mu$m, the extinction is only $\sim$ 5$\%$ of any additional A$_{V}$ generated due to the dust in contrast to $\sim$ 25$\%$ at Pa$\beta$). We have assumed an electron density of n$_{e}$ = 10$^{7.5}$ cm$^{-3}$ in 
the mass estimation, for two reasons. First, the smallest critical density of the ions seen during the coronal phase (Table~\ref{table-CoronalLines}, discussed below) suggests this. Second, geometric dilution of the ejecta with time (n$_{e}$ $\propto$ t$^{-2}$) suggests that the lower limit on the electron density of 1.14$\times$10$^{9}$-6.08$\times$10$^{9}$ cm$^{-3}$ at +64d would have decreased to about $\sim$ 10$^{8}$ cm$^{-3}$ during the coronal phase ($\sim$ 300-360d). A homologous flow would create a faster decline in the density. The mass, for any other choice of n$_{e}$, should be scaled accordingly. Based on this analysis, and using the 
Humphreys lines for arriving at a formal estimate, we obtain the mass of the ejecta to be M$_{\rm ej}$  = (8.60$\pm$1.73)$\times$10$^{-5}$ $M_{\odot}$.

\subsubsection{Coronal lines}
\label{subsec_CoronalLines}

By 2020 June, coronal lines had begun to appear in the NIR spectra; these are most clearly seen in the spectrum of 2020 Sep 19, i.e. after 371 days \citep{Woodward2020b}.  Table~\ref{table-CoronalLines} gives the details of the detected NIR lines. It is first necessary to assess whether the coronal lines are truly coronal (in that they are generated by 
collisional ionisation, followed by collisional excitation), or  whether photo-ionisation due to a hot central source is responsible. Which of these two mechanisms is operating has been debated in the literature \citep{Evans2003,Benjamin1990,Greenhouse1990}. In this case, the evidence suggests -- for several independent reasons -- that collisional ionisation is the mechanism responsible for the coronal lines, and photoionisation plays a negligible role. \cite{Greenhouse1990} define nova coronal lines as arising from ground-state fine-structure transitions in species with ionisation potential (IP) $> 100$~eV. Considering that the nova pseudo-photosphere collapses at constant bolometric luminosity, the effective temperature T$_\star$ of the stellar remnant is empirically expected to vary as, $T_\star \simeq T_0 \exp[0.921(t/t_3)]$ \citep{Evans2003,Bath1976,Bath1989}, where $T_0 = 15,280 K$ and $t_3$ is 
the time to fall three magnitudes below the the maximum. However, as novae at maximum show temperature values closer to 8000~K (see discussion in \cite{Evans2005}), we adopt $T_0 = 8000K$ to estimate $T_\star$. However we note that this formula is mostly applicable for a linear (in mag) LC decline from maximum, and our attempt here is to place limits on the temperature of the central ionising source. A similar approach had  been adopted by \cite{Evans2003} for nova V723 Cas, which showed similar erratic LC like nova V2891 Cyg. 
\par 
As discussed above, for $t=371$ days (when the coronal lines were seen) and for $t_3$ in the range 180-230d days (from section~\ref{subsec_Lightcurve}), the temperature of the pseudo-photosphere is estimated to be in the range $\sim$ 35000-54000$K$. For this temperature range, there is a negligible number of photons with energies greater than 100eV (a blackbody at 50000$K$ emits only $\sim9\times 10^{-6}\%$ flux above 100 eV \citep{Allen2000}). This makes it difficult to explain the presence of ions like Si\,{\sc vi} and Al\,{\sc ix}, which are clearly detected. These (and all coronals lines) need IPs $> 150$ eV. However in the supersoft X-ray phase of novae, due to shell burning, the temperature of the hot central source can be $> 10^{5}$K, \citep[see][]{schwarz2011}, however we did not find evidence of such hot source with the X-ray observations.
\par
A Target of Opportunity (ToO) observation on 2020 September 25(+377d), using the XRT instrument on the Neil Gehrels Swift Observatory \citep{Gehrels2004} did not detect X-ray emission in an  exposure time of $\sim$1000 seconds (Target ID 12868\footnote{https://www.swift.psu.edu/toop/summary.php}), with a count rate less than $8.0 \times 10^{-3}$ counts per second. In addition, the UV instrument onboard Swift did not detect any emission in the UV, with the three filters (centered on 190, 220, and 260 nm) all giving limits of $>19.4$ mags. The Swift non-detection argues against the presence of a sufficiently hot central photoionizing source.
	
\par 
In a rather similar case, \cite{Evans2003} explained the early onset of the coronal phase in nova V723 Cas in terms of the presence of collision ionisation caused by the interaction of ejecta parcels at different velocities. In the present case as well, the multiple re-brightenings in the LC are likely caused by the periodic episodes of material ejection -- a premise supported by the 
re-appearance of P-Cygni profiles in the O\,{\sc i} 7773$\AA$ line at different epochs (see Section~\ref{subsec-OxygenLines}). For the gas temperature of $\sim10^4$K, the speed of sound is $\sim13$km~s$^{-1}$. If the relative velocity is in the range of 150-200$kms^{-1}$ the Mach number turns out to be  $\sim$12, indicating a strong shock. In this case, the post-shock temperature $T_{s} (K)$ is related to the single-particle forward shock velocity $v_{s}$ as
\begin{equation}
{v_{s}} = {\left(\frac{16k{T_{s}}}{3{\mu}{m_{H}}}\right)^{1/2} }
\label{equa-vs}
\end{equation}
where $k$ is the Boltzmann constant, and  $\mu{m_{H}}$ is the mean particle mass \citep{Shu1992, Tatischeff2007}. Thus for the shock-heated gas, which is generated by collisions between parcels 
of ejecta having a relative velocity difference in the range $\sim$ 150-200 km$s^{-1}$, the temperatures would be $4.8-9.1 \times 10^{5}$ K. The presence of coronal also lines implies a gas temperature in the same range.

\begin{figure}
	\centering
	\includegraphics[angle=0,width=0.49\textwidth]{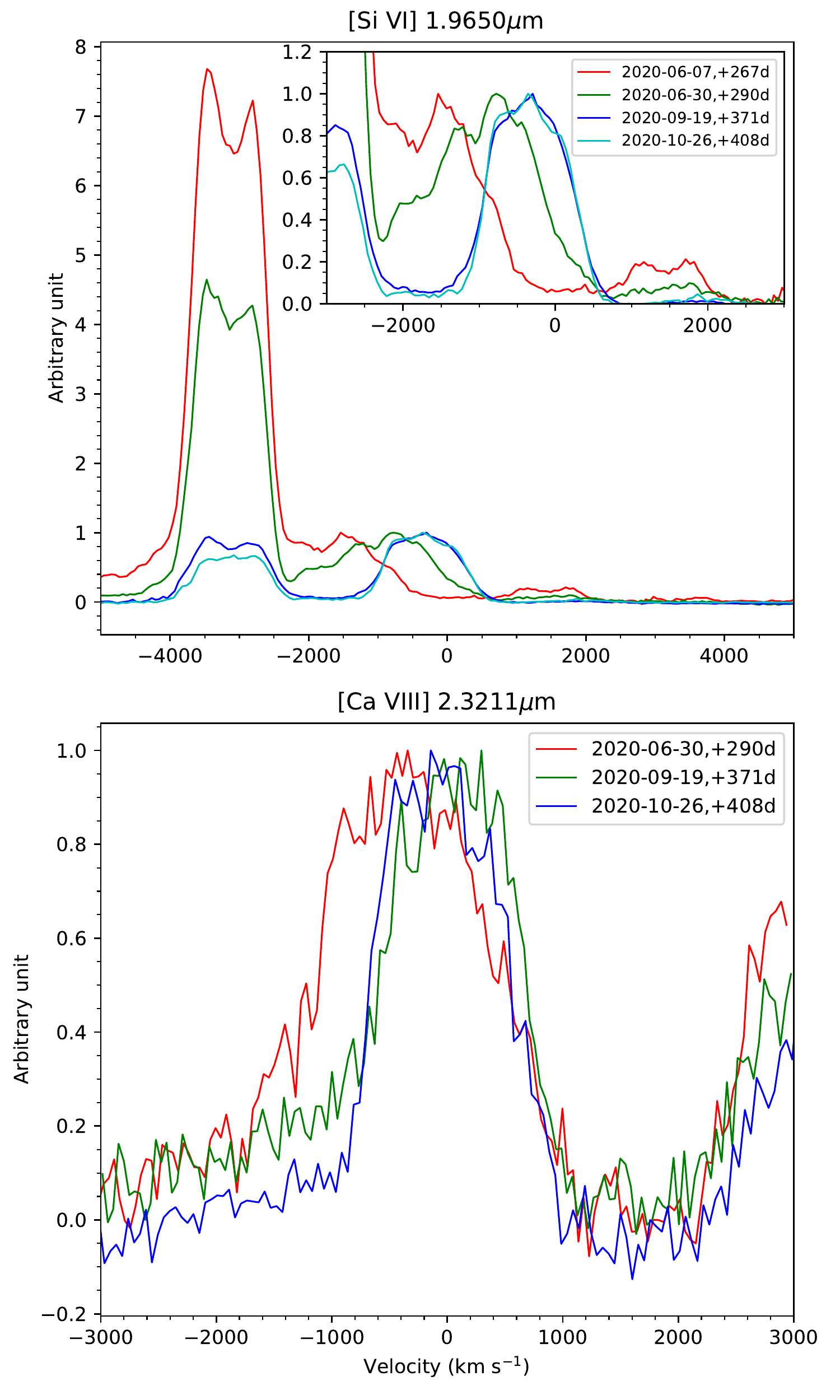}
	\caption[]{The upper and lower panels show the velocity shifts seen in the
	coronal lines as the obscuring dust behind the shocked gas, which is 
	the site of the coronal line emission, dissipates or changes its optical
	depth with time. In contrast to the coronal lines, the H lines remain
	fixed in velocity, as seen for example, from the Br$\delta$ line at 
	1.944 $\mu$m (at $\sim -3000$ km~s$^{-1}$ in the upper panel). 
	Inset in the upper panel shows the shift of the [Si\,{\sc vi}] 1.9650$\mu$m
	line. 2020-10-26 epoch data are from TANSPEC, the remainder are from SpeX. 
	Details are given in section~\ref{subsec_velShift}.}
	\label{fig-Velshift}
\end{figure}


\begin{table*}
	\setlength{\tabcolsep}{3pt} 
	\centering
	\caption{NIR Coronal lines}
	\begin{tabular}{lcccccccc}
		\hline
		Species	&	Wavelength 	&	log(T$_{max}$) $^a$	&	log(n$_{\rm crit}$)$^b$     &	I.P.$^c$	&           & Flux$^d$  &  &Remarks	\\
		&        ($\mu$m)       &                               &                               &       (eV)    & (07 June) & (30 June) & (19 Sept.) &             \\ \hline \hline
		{[S\,{\sc ix}]}  	&	1.2523	&      6.0   &      9.2      &       379.60   &     -            &  -              &       -        &unable to deblend  from He\,{\sc i} 1.2528$\mu$m \\
		{[Si\,{\sc x}]}  	&	1.4305	&      6.2   &      8.8      &       401.40   & 6.91$\pm$0.13   & 6.86$\pm$0.26 & 	   -        &	              \\
		{[Ti\,{\sc vi}]}  	&	1.7156	&   	     &  	         &  	 119.53  &      -           & 0.80$\pm$0.08 & 0.98$\pm$0.04 &	        \\
		{[Si\,{\sc vi}]}	    &	1.9650	&      5.6   &      8.8      &       205.27   &10.97$\pm$0.18   &16.51$\pm$0.19 & 31.76$\pm$0.92&     Deblended from Br 8-4	     \\
		{[Al\,{\sc ix}]}  	&	2.0400	&      6.1   &      8.2      &       330.10   &       -          & 1.94$\pm$0.16 & 1.74$\pm$0.03 &  	                \\
		{[Fe\,{\sc xii}]}	&	2.2170	&      6.2   &      8.6      &       331.00   & 1.72$\pm$0.12   & 1.60$\pm$0.12 & 1.50$\pm$0.21 &	                             \\
		{[Ca\,{\sc viii}]}   &	2.3211	&      5.7   &      7.4$^e$  &       147.24   &     -            & 4.17$\pm$0.07 & 4.18$\pm$0.16 &	                             \\
		{[Si\,{\sc vii}]}	&	2.4833	&      5.8   &      8.2      &       246.49   &     -           & 5.01$\pm$0.06 & 11.20$\pm$0.40&     Deblended from Pf 18-5, 17-5    \\
		{[Al\,{\sc vi}]}  	&	3.6600	&      5.7   &      7.6      &       190.49   & 4.33$\pm$0.06   &    -           &       -        &     Deblended from Hu 19-6  	     \\
		{[Si\,{\sc ix}]}  	&	3.9357	&      6.1   &      7.5      &       351.10   & 3.36$\pm$0.52   &    -        &      -         &	                             \\
		\hline
		\hline
	\end{tabular}
	\begin{list}{}{}
		\item (a) : The equilibrium temperature of the maximum concentration of the ion (~n(X$^{+i}$)/n(X$_{\rm tot}$)~) i
		n a collisionally ionized plasma (Greenhouse et al. 1993)
		\item (b) : Critical densities at $T$ = $T_{\rm max}$  for collisional de-excitation 
		(from Greenhouse et al. 1993, their figure 3a). Somewhat lower values of n$_{\rm crit}$
		are expected in cooler photoionised regions as listed, for example, in \cite{Spinoglio1992} and \cite{Ferguson1997} since n$_{\rm crit}\propto {T}^{1/2}$; equation (1) of  \cite{Greenhouse1993}).
		\item (c): Ionisation Potential
		\item (d): Observed undereddened line fluxes (in units of $\times10^{-14}$erg $s^{-1}$$cm^{-2}$), after deblending 
		with the contaminating lines mentioned in the Remarks column. Where reliable deblending was not possible, or the 
		line was not detected, flux values are not given.
		\item (e) : Estimated from Fig. 2, Greenhouse et al. (1993)
	\end{list}
	\label{table-CoronalLines}
\end{table*}


\par 
The correlation between optical re-brightenings and strong shocks (manifested through $\gamma$-ray emission) has been established in the case of nova V906 Car (ASASSN-18fv). In that nova, the optical and $\gamma$-ray flares were seen to occur simultaneously, implying a common origin in shocks \citep{Aydi2020}. As discussed in Section~\ref{sec_OptSpec}, as well as in this section, the observational evidence for multiple ejections is strong in the nova V2891 Cyg. Thus, we expect physical conditions to be conducive for shock-induced collisional ionisation. In the case of collisional ionisation, the ion temperatures and abundances can be determined in the following manner. The line flux from a region of volume $V$, electron temperature $T$ and distance $d$ is given by \cite{Greenhouse1988, Lang1980, Chandrasekhar1993},

\begin{eqnarray}
{\rm I}_{{\rm x}^{+i}}=\frac{(8.629 \times 10^{-6}){\rm n}_{{\rm
			x}^{+i}}{\rm n}_e {V} {\rm h} \nu \Omega}{{4\pi}{\rm d}^2 {\rm
		T}^{1/2}g_u} {\rm erg.s^{-1}.cm^{-2}}
\label{equa-volume}
\end{eqnarray}

where, ${\rm n}_{\rm x^{+i}}$ is the number density of the x$^{+i}$ ion; h${\nu}$ is the transition energy of the line under consideration, $\Omega$ is the collisional strength, and $g_u$ is the statistical weight of its upper level. For two distinct ion species, or for different ionisation stages of the same ion, this may be recast as:

\begin{eqnarray}
\frac{I_{x}^{+i}}{I_{y}^{+j}}  = \frac{n_{x}^{+i}}{n_{y}^{+j}}\frac{\nu_{x}}{\nu_{y}}\frac{\Omega_{x}}{\Omega_{y}}\frac{g_y}{g_x}
\label{equa-6}
\end{eqnarray}

where $I$, $n$, $\nu$, $\Omega$ and $g$ are the measured fluxes, total numbers of ionised atoms, frequency of the transition, collision strength of the transition and statistical weights of the lower level, respectively. These are defined for species $x$ in its $i^{th}$ ionisation state, and species $y$ in its $j^{th}$ ionisation state.  If $T_e$ is known, the ratio of total relative abundances of the neutral element $n_{x}^{0}/n_{y}^{0}$ can be determined, provided the 
ionisation fraction $n^{i}/n^{0}$ for both the elements are known at $T_e$. The fractional ionisation values of an ion, as a function of temperature, and the collision strengths for the different transitions  have  been taken from various sources in the literature \cite[see][]{Blaha1969,Jain1978,Jordan1969,Greenhouse1988,Zhang1994,Berrington1998,Shull1982}. Using the [Si\,{\sc vi} 1.96$\mu$m and the [Si\,{\sc vii}] 2.48$\mu$m lines, the temperature is estimated to be $4.77[\pm 0.04]\times10^{5}$ K, and $4.54[\pm 0.04]\times10^{5}$ K respectively, using values of the fractional ionisation ratio $n_{Si}^{+5}/n_{Si}^{+6}$ from \cite{Jordan1969} (Table XVII) and \cite{Shull1982} respectively.  For the [Al\,{\sc vi}] 3.06$\mu$m and the 
[Al\,{\sc ix}] 2.04$\mu$m lines, the temperature is estimated to be $4.33[\pm 0.04]\times10^{5}$ K using values of the fractional ionisation for Al from \cite{Jain1978}.  These ion temperatures lie in between the values of $3.2\times10^{5}$ K estimated for the coronal zone in V723 Cas 
\citep{Evans2003}, and $6.3\times10^{5}$ K for Nova Vulpeculae 1984 \citep{Greenhouse1988}.
\par 
The Al/Si abundance n(Al)/n(Si) can be estimated by using Equation~(\ref{equa-6}), where in the left hand side one can use different combinations of the line intensities choosing from the 
[Si\,{\sc vi}] 1.96 $\mu$m, [Si\,{\sc vii}] 2.48 $\mu$m, [Al\,{\sc ix}] 2.04 $\mu$m, and [Al\,{\sc vi}] 3.06 $\mu$m lines. On the right hand side of the equation, the term $n_{\mbox{Al}}^{i}/n_{\mbox{Si}}^{j}$ can be recast as below to allow ($n_{\mbox{Al}}/n_{\mbox{Si}}$) to be determined:
\[  n_{\mbox{Al}}^{i}/n_{\mbox{Si}}^{j} = 
(n_{\mbox{Al}}^{i}/n_{\mbox{Al}})  \times (n_{Al}/n_{Si})\times(n_{Si}/n_{Si}^{i}) \]
A mean value of ($n_{\mbox{Al}}/n_{\mbox{Si}})  = 1.53 \pm 0.15$ is found,
which implies the ratio [$n_{\mbox{Al}}/n_{\mbox{Si}}$](nova)/[$n_{\mbox{Al}}/n_{\mbox{Si}}$]($\odot$)= 18.4,
where we have used the solar abundances of \cite{Lodders2021}. 
This over-abundance with respect to solar is plausible: novae of both the CO and ONe types are known to overproduce Al \citep{Jose1998,Starrfield1997}.
\par
Using Equation~(\ref{equa-volume}), the volume of the coronal gas, derived using the coronal lines of [Si\,{\sc vi}] and [Si\,{\sc vii}] in the spectrum of September 19, is in the range of $9.95\times10^{42}$ -- $1.01\times10^{44}$cm$^{-3}$, and the corresponding mass of the coronal gas is in the range $\sim 8.37\times10^{-7}$ -- $8.42\times10^{-7}$ M$_\odot$. The volume of the coronal gas is almost a factor of 100--1000 smaller than the maximum volume that the ejecta can attain kinematically, (i.e. $ 4/3\pi{R}^{3} \sim 10^{45}$ -- $10^{46}$cm$^{-3}$, with $R$ = $v\times{t}$; for the velocity $v$ in the range 200--500 kms$^{-1}$, and $t\sim 371$d for 2020 Sep 19). This suggests the ejecta are in the form of a thin shell with a very small filling factor. This small volume of the coronal gas (compared with the total volume of the ejecta) is also consistent with the non-detection of the UV/X-ray emission, as well as the modest 
amount of shock-induced dust mass estimated earlier.


\subsection{Velocity shifts of the coronal lines}
\label{subsec_velShift}

A rather curious phenomenon is seen where the coronal lines ``migrate'' from bluer to redder velocities with the passage of time.  This is demonstrated in  Fig.~\ref{fig-Velshift} for the [Si\,{\sc vi}] 1.96$\mu$m and [Ca\,{\sc viii}] 2.32 $\mu$m lines, shown in velocity space. For the [Si\,{\sc vi}] line, the line shift is significantly large, exceeding 1000 kms$^{-1}$, between 2020 June to September. On the other hand, a similar shift is not seen in the H lines (Br$\delta$ 1.944 $\mu$m  at $\sim-3000$ kms$^{-1}$ is shown here to demonstrate this), or in the 
He lines (e.g.  the He\,{\sc i} 2.059 $\mu$m line).  Clearly, there is a spatial stratification of the species.  But the more interesting implication of this behaviour is related to the dust event. As per the model proposed by \cite{Derdzinski2017} (see figure 1 in that paper), a cool, possibly clumpy, dust shell is expected to form behind the forward shock for both the approaching and receding parts of the nova shell. Thus a plausible interpretation of the coronal line shifts is that in the June 7 (+267d) observation, the dust is dense and completely blocks the emission from the receding part of the shell. Hence the line profile is blue-shifted to the maximum. As the dust dissipates with time, more emission from the receding red component reaches the 
observer (as in the profiles of June 30 (+290d) and September 19(+371d)). This is a simplified interpretation, because the dust geometry could be more complex, and dust clumps could drift in and out of the line of sight, causing time-varying obscuration of the red shell (and hence evolving profile shapes as seen between the June 7(+267d), June 30(+290d), Sep 19(+371d) and Oct 26(+408d) profiles shown here). The line shift in the [Si\,{\sc vi}] 1.96 $\mu$m line is also seen in the [Ca\,{\sc viii}] 2.32 $\mu$m line. This behaviour may thus be confirmation that dust formation can be triggered by a radiative shock, as proposed by \cite{Derdzinski2017}, wherein as gas behind the radiative shock cools, it compresses, reaching high densities. The shocked ejecta is expected to collect in a geometrically thin, clumpy shell. If the formation of the dust in V2891 Cyg is indeed shock-induced, then this observation is likely the first time that dust formation via this route is being witnessed in a nova. Interestingly, \cite{Harvey2018} (following \cite{Derdzinski2017} and references therein) had further suggested that dust formed in the forward shock zone can subsequently be destroyed as the shock passes
through this small amount of dust. However, this process retains the seed nuclei responsible for dust formation. Thus, this process of shock-induced dust formation and destruction may be repeated over the successive shock cycles. For the case of nova V339 Del, \cite{Evans2017} and \cite{Gehrz2015} considered grain shattering by electrostatic stress due to charging of grains by UV/X-ray emission; however, as we did not notice any UV-X-ray emission in the case of nova V2891 Cyg so this mechanism is unlikely to be important.

\section{Summary}
\label{sec_summary}

In this study, we have explored the optical and NIR spectroscopic evolution of the nova V2891 Cyg, one of the slowest novae recorded in recent times.  Our optical and NIR spectroscopy campaign lasted around 13 months (2019 Nov -- 2020 Dec) and utilised several facilities worldwide. The nova's light curve is of the ``jitter'' class, having a couple of episodes of re-brightening, which we propose is due to periodic mass ejection, as manifested in the line profile of the O\,{\sc i} 7773$\AA$ line. The temperature and mass of the gas containing the O\,{\sc i} are consistent with values typically seen in a similar class of novae. The absorption components of the H-$\alpha$ are studied, and their evolution is found most likely 
to be due to interaction between gaseous components associated with separate episodes of previous mass ejection. A brief period of dust formation occurred during the evolution, along with the production of coronal line emission in the NIR. The analysis presented here shows 
that the coronal emission, most likely, is due to shock heating rather than photoionisation, and the episode of dust formation  is shock-induced. These phenomena are rare in the evolution of novae. Thus, the present data set and our associated analysis would be of interest to the community to explore the physics of the nova phenomenon.


\section{Acknowledgments}
We are grateful to the anonymous reviewer for several useful suggestions. The research work at the Physical Research Laboratory is funded by the Department of Space, Government of India. VK thanks PRL for his PhD research fellowship.  DPKB is supported by a CSIR Emeritus Scientist grant. CEW acknowledges partial support from NASA grant 80NSSC19K0868. KLP acknowledges the support of the UK Space Agency. We thank the staff of 3.6m Devasthal Optical Telescope (DOT) (which is a National Facility run and managed by Aryabhatta Research Institute of Observational Sciences (ARIES), an autonomous Institute under Department of Science and Technology, Government of India) for service observations with TANSPEC instrument.We acknowledge the use of data from the  Zwicky Transient Facility (ZTF) database. \\

\par 

{\bf DATA AVAILABILITY}: The data may be made available on reasonable request. The corresponding authors may be contacted for that.


\bibliography{BibList2}{}

\begin{thebibliography}{}
\makeatletter
\relax
\def\mn@urlcharsother{\let\do\@makeother \do\$\do\&\do\#\do\^\do\_\do\%\do\~}
\def\mn@doi{\begingroup\mn@urlcharsother \@ifnextchar [ {\mn@doi@}
  {\mn@doi@[]}}
\def\mn@doi@[#1]#2{\def\@tempa{#1}\ifx\@tempa\@empty \href
  {http://dx.doi.org/#2} {doi:#2}\else \href {http://dx.doi.org/#2} {#1}\fi
  \endgroup}
\def\mn@eprint#1#2{\mn@eprint@#1:#2::\@nil}
\def\mn@eprint@arXiv#1{\href {http://arxiv.org/abs/#1} {{\tt arXiv:#1}}}
\def\mn@eprint@dblp#1{\href {http://dblp.uni-trier.de/rec/bibtex/#1.xml}
  {dblp:#1}}
\def\mn@eprint@#1:#2:#3:#4\@nil{\def\@tempa {#1}\def\@tempb {#2}\def\@tempc
  {#3}\ifx \@tempc \@empty \let \@tempc \@tempb \let \@tempb \@tempa \fi \ifx
  \@tempb \@empty \def\@tempb {arXiv}\fi \@ifundefined
  {mn@eprint@\@tempb}{\@tempb:\@tempc}{\expandafter \expandafter \csname
  mn@eprint@\@tempb\endcsname \expandafter{\@tempc}}}

\bibitem[\protect\citeauthoryear{{Arai}, {Kawakita}, {Shinnaka}  \&
  {Tajitsu}}{{Arai} et~al.}{2016}]{Arai2016}
{Arai} A.,  {Kawakita} H.,  {Shinnaka} Y.,   {Tajitsu} A.,  2016, \mn@doi
  [\apj] {10.3847/0004-637X/830/1/30}, \href
  {https://ui.adsabs.harvard.edu/abs/2016ApJ...830...30A} {830, 30}

\bibitem[\protect\citeauthoryear{{Aydi} et~al.,}{{Aydi}
  et~al.}{2020}]{Aydi2020}
{Aydi} E.,  et~al., 2020, \mn@doi [Nature Astronomy]
  {10.1038/s41550-020-1070-y}, \href
  {https://ui.adsabs.harvard.edu/abs/2020NatAs...4..776A} {4, 776}

\bibitem[\protect\citeauthoryear{{Banerjee} \& {Ashok}}{{Banerjee} \&
  {Ashok}}{2012}]{Banerjee2012}
{Banerjee} D.~P.~K.,  {Ashok} N.~M.,  2012, Bulletin of the Astronomical
  Society of India, \href
  {https://ui.adsabs.harvard.edu/abs/2012BASI...40..243B} {40, 243}

\bibitem[\protect\citeauthoryear{{Banerjee}, {Srivastava}, {Ashok}, {Munari},
  {Hambsch}, {Righetti}  \& {Maitan}}{{Banerjee} et~al.}{2018}]{Banerjee2018}
{Banerjee} D.~P.~K.,  {Srivastava} M.~K.,  {Ashok} N.~M.,  {Munari} U.,
  {Hambsch} F.~J.,  {Righetti} G.~L.,   {Maitan} A.,  2018, \mn@doi [\mnras]
  {10.1093/mnras/stx2459}, \href
  {https://ui.adsabs.harvard.edu/abs/2018MNRAS.473.1895B} {473, 1895}

\bibitem[\protect\citeauthoryear{{Bath} \& {Harkness}}{{Bath} \&
  {Harkness}}{1989}]{Bath1989}
{Bath} G.~T.,  {Harkness} R.~P.,  1989, in {Bode} M.,  {Evans} A.,  eds,
  Classical Novae, First Edition. Wiley, Chichester, pp 61--72

\bibitem[\protect\citeauthoryear{{Bath} \& {Shaviv}}{{Bath} \&
  {Shaviv}}{1976}]{Bath1976}
{Bath} G.~T.,  {Shaviv} G.,  1976, \mn@doi [\mnras] {10.1093/mnras/175.2.305},
  \href {https://ui.adsabs.harvard.edu/abs/1976MNRAS.175..305B} {175, 305}

\bibitem[\protect\citeauthoryear{{Benjamin} \& {Dinerstein}}{{Benjamin} \&
  {Dinerstein}}{1990}]{Benjamin1990}
{Benjamin} R.~A.,  {Dinerstein} H.~L.,  1990, \mn@doi [\aj] {10.1086/115619},
  \href {https://ui.adsabs.harvard.edu/abs/1990AJ....100.1588B} {100, 1588}

\bibitem[\protect\citeauthoryear{{Berrington}, {Saraph}  \&
  {Tully}}{{Berrington} et~al.}{1998}]{Berrington1998}
{Berrington} K.~A.,  {Saraph} H.~E.,   {Tully} J.~A.,  1998, \mn@doi [\aaps]
  {10.1051/aas:1998394}, \href
  {https://ui.adsabs.harvard.edu/abs/1998A&AS..129..161B} {129, 161}

\bibitem[\protect\citeauthoryear{{Blaha}}{{Blaha}}{1969}]{Blaha1969}
{Blaha} M.,  1969, \aap, \href
  {https://ui.adsabs.harvard.edu/abs/1969A&A.....1...42B} {1, 42}

\bibitem[\protect\citeauthoryear{{Chambers} \& {Pan-STARRS Team}}{{Chambers} \&
  {Pan-STARRS Team}}{2016}]{Chambers2016}
{Chambers} K.~C.,  {Pan-STARRS Team} 2016, in American Astronomical Society
  Meeting Abstracts \#227. p. 324.07

\bibitem[\protect\citeauthoryear{{Chambers} et~al.,}{{Chambers}
  et~al.}{2016}]{Chambers2016b}
{Chambers} K.~C.,  et~al., 2016, arXiv e-prints, \href
  {https://ui.adsabs.harvard.edu/abs/2016arXiv161205560C} {p. arXiv:1612.05560}

\bibitem[\protect\citeauthoryear{{Chandrasekhar}, {Ashok}  \&
  {Ragland}}{{Chandrasekhar} et~al.}{1993}]{Chandrasekhar1993}
{Chandrasekhar} T.,  {Ashok} N.~M.,   {Ragland} S.,  1993, \mn@doi [Journal of
  Astrophysics and Astronomy] {10.1007/BF02702277}, \href
  {https://ui.adsabs.harvard.edu/abs/1993JApA...14....7C} {14, 7}

\bibitem[\protect\citeauthoryear{{Cohen}}{{Cohen}}{1985}]{Cohen1985}
{Cohen} J.~G.,  1985, \mn@doi [\apj] {10.1086/163135}, \href
  {https://ui.adsabs.harvard.edu/abs/1985ApJ...292...90C} {292, 90}

\bibitem[\protect\citeauthoryear{{Cox}}{{Cox}}{2000}]{Allen2000}
{Cox} A.~N.,  2000, {Allen's astrophysical quantities; 4th ed.}.
AIP, New York, NY

\bibitem[\protect\citeauthoryear{{Cs{\'a}k}, {Kiss}, {Retter}, {Jacob}  \&
  {Kaspi}}{{Cs{\'a}k} et~al.}{2005}]{Csak2005}
{Cs{\'a}k} B.,  {Kiss} L.~L.,  {Retter} A.,  {Jacob} A.,   {Kaspi} S.,  2005,
  \mn@doi [\aap] {10.1051/0004-6361:20035751}, \href
  {https://ui.adsabs.harvard.edu/abs/2005A&A...429..599C} {429, 599}

\bibitem[\protect\citeauthoryear{{Cushing}, {Vacca}  \& {Rayner}}{{Cushing}
  et~al.}{2004}]{Cushing2004}
{Cushing} M.~C.,  {Vacca} W.~D.,   {Rayner} J.~T.,  2004, \mn@doi [\pasp]
  {10.1086/382907}, \href
  {https://ui.adsabs.harvard.edu/abs/2004PASP..116..362C} {116, 362}

\bibitem[\protect\citeauthoryear{{Das}, {Banerjee}, {Ashok}  \&
  {Chesneau}}{{Das} et~al.}{2008}]{das2008}
{Das} R.~K.,  {Banerjee} D.~P.~K.,  {Ashok} N.~M.,   {Chesneau} O.,  2008,
  \mn@doi [\mnras] {10.1111/j.1365-2966.2008.13998.x}, \href
  {https://ui.adsabs.harvard.edu/abs/2008MNRAS.391.1874D} {391, 1874}

\bibitem[\protect\citeauthoryear{{Das}, {Banerjee}, {Nandi}, {Ashok}  \&
  {Mondal}}{{Das} et~al.}{2015}]{Das2015}
{Das} R.,  {Banerjee} D. P.~K.,  {Nandi} A.,  {Ashok} N.~M.,   {Mondal} S.,
  2015, \mn@doi [\mnras] {10.1093/mnras/stu2488}, \href
  {https://ui.adsabs.harvard.edu/abs/2015MNRAS.447..806D} {447, 806}

\bibitem[\protect\citeauthoryear{{De} \& {Palomar Gattini-IR
  Collaboration}}{{De} \& {Palomar Gattini-IR Collaboration}}{2020}]{De2020}
{De} K.,  {Palomar Gattini-IR Collaboration} 2020, The Astronomer's Telegram,
  \href {https://ui.adsabs.harvard.edu/abs/2020ATel13732....1D} {13732, 1}

\bibitem[\protect\citeauthoryear{{De Gennaro Aquino} et~al.,}{{De Gennaro
  Aquino} et~al.}{2015}]{DeGennaroAquino2015}
{De Gennaro Aquino} I.,  et~al., 2015, \mn@doi [\aap]
  {10.1051/0004-6361/201525810}, \href
  {https://ui.adsabs.harvard.edu/abs/2015A&A...581A.134D} {581, A134}

\bibitem[\protect\citeauthoryear{{De} et~al.,}{{De} et~al.}{2019}]{De2019}
{De} K.,  et~al., 2019, The Astronomer's Telegram, \href
  {https://ui.adsabs.harvard.edu/abs/2019ATel13130....1D} {13130, 1}

\bibitem[\protect\citeauthoryear{{Derdzinski}, {Metzger}  \&
  {Lazzati}}{{Derdzinski} et~al.}{2017}]{Derdzinski2017}
{Derdzinski} A.~M.,  {Metzger} B.~D.,   {Lazzati} D.,  2017, \mn@doi [\mnras]
  {10.1093/mnras/stx829}, \href
  {https://ui.adsabs.harvard.edu/abs/2017MNRAS.469.1314D} {469, 1314}

\bibitem[\protect\citeauthoryear{{Evans} et~al.,}{{Evans}
  et~al.}{2003}]{Evans2003}
{Evans} A.,  et~al., 2003, \mn@doi [\aj] {10.1086/377618}, \href
  {https://ui.adsabs.harvard.edu/abs/2003AJ....126.1981E} {126, 1981}

\bibitem[\protect\citeauthoryear{{Evans}, {Tyne}, {Smith}, {Geballe},
  {Rawlings}  \& {Eyres}}{{Evans} et~al.}{2005}]{Evans2005}
{Evans} A.,  {Tyne} V.~H.,  {Smith} O.,  {Geballe} T.~R.,  {Rawlings} J.~M.~C.,
    {Eyres} S.~P.~S.,  2005, \mn@doi [\mnras]
  {10.1111/j.1365-2966.2005.09146.x}, \href
  {https://ui.adsabs.harvard.edu/abs/2005MNRAS.360.1483E} {360, 1483}

\bibitem[\protect\citeauthoryear{{Evans} et~al.,}{{Evans}
  et~al.}{2017}]{Evans2017}
{Evans} A.,  et~al., 2017, \mn@doi [\mnras] {10.1093/mnras/stw3334}, \href
  {https://ui.adsabs.harvard.edu/abs/2017MNRAS.466.4221E} {466, 4221}

\bibitem[\protect\citeauthoryear{{Ferguson}, {Korista}  \&
  {Ferland}}{{Ferguson} et~al.}{1997}]{Ferguson1997}
{Ferguson} J.~W.,  {Korista} K.~T.,   {Ferland} G.~J.,  1997, \mn@doi [\apjs]
  {10.1086/312998}, \href
  {https://ui.adsabs.harvard.edu/abs/1997ApJS..110..287F} {110, 287}

\bibitem[\protect\citeauthoryear{{Gaia Collaboration} et~al.,}{{Gaia
  Collaboration} et~al.}{2016}]{Gaia2016}
{Gaia Collaboration} et~al., 2016, \mn@doi [\aap]
  {10.1051/0004-6361/201629272}, \href
  {https://ui.adsabs.harvard.edu/abs/2016A&A...595A...1G} {595, A1}

\bibitem[\protect\citeauthoryear{{Gehrels} et~al.,}{{Gehrels}
  et~al.}{2004}]{Gehrels2004}
{Gehrels} N.,  et~al., 2004, \mn@doi [\apj] {10.1086/422091}, \href
  {https://ui.adsabs.harvard.edu/abs/2004ApJ...611.1005G} {611, 1005}

\bibitem[\protect\citeauthoryear{{Gehrz} et~al.,}{{Gehrz}
  et~al.}{2015}]{Gehrz2015}
{Gehrz} R.~D.,  et~al., 2015, \mn@doi [\apj] {10.1088/0004-637X/812/2/132},
  \href {https://ui.adsabs.harvard.edu/abs/2015ApJ...812..132G} {812, 132}

\bibitem[\protect\citeauthoryear{Glass}{Glass}{1999}]{Glass1999}
Glass I.~S.,  1999, Handbook of Infrared Astronomy.
Cambridge Observing Handbooks for Research Astronomers, Cambridge University
  Press, \mn@doi{10.1017/CBO9780511564949}

\bibitem[\protect\citeauthoryear{{Greenhouse}, {Grasdalen}, {Hayward}, {Gehrz}
  \& {Jones}}{{Greenhouse} et~al.}{1988}]{Greenhouse1988}
{Greenhouse} M.~A.,  {Grasdalen} G.~L.,  {Hayward} T.~L.,  {Gehrz} R.~D.,
  {Jones} T.~J.,  1988, \mn@doi [\aj] {10.1086/114625}, \href
  {https://ui.adsabs.harvard.edu/abs/1988AJ.....95..172G} {95, 172}

\bibitem[\protect\citeauthoryear{{Greenhouse}, {Grasdalen}, {Woodward},
  {Benson}, {Gehrz}, {Rosenthal}  \& {Skrutskie}}{{Greenhouse}
  et~al.}{1990}]{Greenhouse1990}
{Greenhouse} M.~A.,  {Grasdalen} G.~L.,  {Woodward} C.~E.,  {Benson} J.,
  {Gehrz} R.~D.,  {Rosenthal} E.,   {Skrutskie} M.~F.,  1990, \mn@doi [\apj]
  {10.1086/168537}, \href
  {https://ui.adsabs.harvard.edu/abs/1990ApJ...352..307G} {352, 307}

\bibitem[\protect\citeauthoryear{{Greenhouse}, {Feldman}, {Smith}, {Klapisch},
  {Bhatia}  \& {Bar-Shalom}}{{Greenhouse} et~al.}{1993}]{Greenhouse1993}
{Greenhouse} M.~A.,  {Feldman} U.,  {Smith} H.~A.,  {Klapisch} M.,  {Bhatia}
  A.~K.,   {Bar-Shalom} A.,  1993, \mn@doi [\apjs] {10.1086/191813}, \href
  {https://ui.adsabs.harvard.edu/abs/1993ApJS...88...23G} {88, 23}

\bibitem[\protect\citeauthoryear{{Harvey}, {Redman}, {Darnley}, {Williams},
  {Berdyugin}, {Piirola}, {Fitzgerald}  \& {O'Connor}}{{Harvey}
  et~al.}{2018}]{Harvey2018}
{Harvey} E.~J.,  {Redman} M.~P.,  {Darnley} M.~J.,  {Williams} S.~C.,
  {Berdyugin} A.,  {Piirola} V.~E.,  {Fitzgerald} K.~P.,   {O'Connor} E.~G.~P.,
   2018, \mn@doi [\aap] {10.1051/0004-6361/201731741}, \href
  {https://ui.adsabs.harvard.edu/abs/2018A&A...611A...3H} {611, A3}

\bibitem[\protect\citeauthoryear{{Henden} \& {Munari}}{{Henden} \&
  {Munari}}{2014}]{Henden2014}
{Henden} A.,  {Munari} U.,  2014, Contributions of the Astronomical Observatory
  Skalnate Pleso, \href {https://ui.adsabs.harvard.edu/abs/2014CoSka..43..518H}
  {43, 518}

\bibitem[\protect\citeauthoryear{{Hummer} \& {Storey}}{{Hummer} \&
  {Storey}}{1987}]{Hummer1987}
{Hummer} D.~G.,  {Storey} P.~J.,  1987, \mn@doi [\mnras]
  {10.1093/mnras/224.3.801}, \href
  {https://ui.adsabs.harvard.edu/abs/1987MNRAS.224..801H} {224, 801}

\bibitem[\protect\citeauthoryear{{Jain} \& {Narain}}{{Jain} \&
  {Narain}}{1978}]{Jain1978}
{Jain} N.~K.,  {Narain} U.,  1978, \aaps, \href
  {https://ui.adsabs.harvard.edu/abs/1978A&AS...31....1J} {31, 1}

\bibitem[\protect\citeauthoryear{{Jordan}}{{Jordan}}{1969}]{Jordan1969}
{Jordan} C.,  1969, \mn@doi [\mnras] {10.1093/mnras/142.4.501}, \href
  {https://ui.adsabs.harvard.edu/abs/1969MNRAS.142..501J} {142, 501}

\bibitem[\protect\citeauthoryear{{Jos{\'e}} \& {Hernanz}}{{Jos{\'e}} \&
  {Hernanz}}{1998}]{Jose1998}
{Jos{\'e}} J.,  {Hernanz} M.,  1998, \mn@doi [\apj] {10.1086/305244}, \href
  {https://ui.adsabs.harvard.edu/abs/1998ApJ...494..680J} {494, 680}

\bibitem[\protect\citeauthoryear{{Joshi}, {Banerjee}  \& {Ashok}}{{Joshi}
  et~al.}{2014}]{Joshi2014}
{Joshi} V.,  {Banerjee} D.~P.~K.,   {Ashok} N.~M.,  2014, \mn@doi [\mnras]
  {10.1093/mnras/stu1168}, \href
  {https://ui.adsabs.harvard.edu/abs/2014MNRAS.443..559J} {443, 559}

\bibitem[\protect\citeauthoryear{{Joshi}, {Banerjee}  \& {Srivastava}}{{Joshi}
  et~al.}{2019}]{Joshi2019}
{Joshi} V.,  {Banerjee} D.~P.~K.,   {Srivastava} M.,  2019, The Astronomer's
  Telegram, \href {https://ui.adsabs.harvard.edu/abs/2019ATel13301....1J}
  {13301, 1}

\bibitem[\protect\citeauthoryear{{Landolt}}{{Landolt}}{1992}]{Landolt1992}
{Landolt} A.~U.,  1992, \mn@doi [\aj] {10.1086/116242}, \href
  {https://ui.adsabs.harvard.edu/abs/1992AJ....104..340L} {104, 340}

\bibitem[\protect\citeauthoryear{{Lang}}{{Lang}}{1980}]{Lang1980}
{Lang} K.~R.,  1980, {Astrophysical Formulae. A Compendium for the Physicist
  and Astrophysicist.}.
Springer-Verlag, Berlin (Heidelberg, New York)

\bibitem[\protect\citeauthoryear{{Lee} et~al.,}{{Lee} et~al.}{2019}]{Lee2019}
{Lee} C.-H.,  et~al., 2019, The Astronomer's Telegram, \href
  {https://ui.adsabs.harvard.edu/abs/2019ATel13149....1L} {13149, 1}

\bibitem[\protect\citeauthoryear{{Lodders}}{{Lodders}}{2021}]{Lodders2021}
{Lodders} K.,  2021, \mn@doi [\ssr] {10.1007/s11214-021-00825-8}, \href
  {https://ui.adsabs.harvard.edu/abs/2021SSRv..217...44L} {217, 44}

\bibitem[\protect\citeauthoryear{{Marshall}, {Robin}, {Reyl{\'e}}, {Schultheis}
   \& {Picaud}}{{Marshall} et~al.}{2006}]{Marshall2006}
{Marshall} D.~J.,  {Robin} A.~C.,  {Reyl{\'e}} C.,  {Schultheis} M.,   {Picaud}
  S.,  2006, \mn@doi [\aap] {10.1051/0004-6361:20053842}, \href
  {https://ui.adsabs.harvard.edu/abs/2006A&A...453..635M} {453, 635}

\bibitem[\protect\citeauthoryear{{McLaughlin}}{{McLaughlin}}{1964}]{McLaughlin1964}
{McLaughlin} D.~B.,  1964, Annales d'Astrophysique, \href
  {https://ui.adsabs.harvard.edu/abs/1964AnAp...27..450M} {27, 450}

\bibitem[\protect\citeauthoryear{{Moro} \& {Munari}}{{Moro} \&
  {Munari}}{2000}]{Moro2000}
{Moro} D.,  {Munari} U.,  2000, \mn@doi [\aaps] {10.1051/aas:2000370}, \href
  {https://ui.adsabs.harvard.edu/abs/2000A&AS..147..361M} {147, 361}

\bibitem[\protect\citeauthoryear{{Munari}}{{Munari}}{2014}]{Munari2014}
{Munari} U.,  2014, in {Woudt} P.~A.,  {Ribeiro} V.~A.~R.~M.,  eds,
  Astronomical Society of the Pacific Conference Series Vol. 490, Stellar
  Novae: Past and Future Decades. p.~183

\bibitem[\protect\citeauthoryear{{Munari}}{{Munari}}{2019}]{Munari2019}
{Munari} U.,  2019, The Astronomer's Telegram, \href
  {https://ui.adsabs.harvard.edu/abs/2019ATel13283....1M} {13283, 1}

\bibitem[\protect\citeauthoryear{{Munari} \& {Moretti}}{{Munari} \&
  {Moretti}}{2012}]{Munari2012b}
{Munari} U.,  {Moretti} S.,  2012, \mn@doi [Baltic Astronomy]
  {10.1515/astro-2017-0354}, \href
  {https://ui.adsabs.harvard.edu/abs/2012BaltA..21...22M} {21, 22}

\bibitem[\protect\citeauthoryear{{Munari} et~al.,}{{Munari}
  et~al.}{2012}]{Munari2012}
{Munari} U.,  et~al., 2012, \mn@doi [Baltic Astronomy]
  {10.1515/astro-2017-0353}, \href
  {https://ui.adsabs.harvard.edu/abs/2012BaltA..21...13M} {21, 13}

\bibitem[\protect\citeauthoryear{{Munari}, {Henden}, {Frigo}  \&
  {Dallaporta}}{{Munari} et~al.}{2014}]{Munari2014b}
{Munari} U.,  {Henden} A.,  {Frigo} A.,   {Dallaporta} S.,  2014, Journal of
  Astronomical Data, \href
  {https://ui.adsabs.harvard.edu/abs/2014JAD....20....4M} {20, 4}

\bibitem[\protect\citeauthoryear{{Munari}, {Hambsch}  \& {Frigo}}{{Munari}
  et~al.}{2017}]{Munari2017}
{Munari} U.,  {Hambsch} F.~J.,   {Frigo} A.,  2017, \mn@doi [\mnras]
  {10.1093/mnras/stx1116}, \href
  {https://ui.adsabs.harvard.edu/abs/2017MNRAS.469.4341M} {469, 4341}

\bibitem[\protect\citeauthoryear{{Munari}, {Cotar}, {Andreoli}, {Dallaporta}
  \& {Yalyalieva}}{{Munari} et~al.}{2019}]{Munari2019b}
{Munari} U.,  {Cotar} K.,  {Andreoli} V.,  {Dallaporta} S.,   {Yalyalieva}
  L.~N.,  2019, The Astronomer's Telegram, \href
  {https://ui.adsabs.harvard.edu/abs/2019ATel13340....1M} {13340, 1}

\bibitem[\protect\citeauthoryear{{Ojha} et~al.,}{{Ojha}
  et~al.}{2018}]{Ojha2018}
{Ojha} D.,  et~al., 2018, Bulletin de la Societe Royale des Sciences de Liege,
  \href {https://ui.adsabs.harvard.edu/abs/2018BSRSL..87...58O} {87, 58}

\bibitem[\protect\citeauthoryear{{Osterbrock}}{{Osterbrock}}{1989}]{Osterbrock1989}
{Osterbrock} D.~E.,  1989, {Astrophysics of gaseous nebulae and active galactic
  nuclei}.
University Science Books, Melville, NY, USA

\bibitem[\protect\citeauthoryear{{Pejcha}}{{Pejcha}}{2009}]{Pejcha2009}
{Pejcha} O.,  2009, \mn@doi [\apjl] {10.1088/0004-637X/701/2/L119}, \href
  {https://ui.adsabs.harvard.edu/abs/2009ApJ...701L.119P} {701, L119}

\bibitem[\protect\citeauthoryear{{Rajpurohit}, {Kumar}, {Srivastava}, {Allard},
  {Homeier}, {Dixit}  \& {Patel}}{{Rajpurohit} et~al.}{2020}]{Rajpurohit2020}
{Rajpurohit} A.~S.,  {Kumar} V.,  {Srivastava} M.~K.,  {Allard} F.,  {Homeier}
  D.,  {Dixit} V.,   {Patel} A.,  2020, \mn@doi [\mnras]
  {10.1093/mnras/staa163}, \href
  {https://ui.adsabs.harvard.edu/abs/2020MNRAS.492.5844R} {492, 5844}

\bibitem[\protect\citeauthoryear{{Rayner}, {Toomey}, {Onaka}, {Denault},
  {Stahlberger}, {Vacca}, {Cushing}  \& {Wang}}{{Rayner}
  et~al.}{2003}]{Rayner2003}
{Rayner} J.~T.,  {Toomey} D.~W.,  {Onaka} P.~M.,  {Denault} A.~J.,
  {Stahlberger} W.~E.,  {Vacca} W.~D.,  {Cushing} M.~C.,   {Wang} S.,  2003,
  \mn@doi [\pasp] {10.1086/367745}, \href
  {https://ui.adsabs.harvard.edu/abs/2003PASP..115..362R} {115, 362}

\bibitem[\protect\citeauthoryear{{Schlafly} \& {Finkbeiner}}{{Schlafly} \&
  {Finkbeiner}}{2011}]{Schlafly2011ApJ}
{Schlafly} E.~F.,  {Finkbeiner} D.~P.,  2011, \mn@doi [\apj]
  {10.1088/0004-637X/737/2/103}, \href
  {https://ui.adsabs.harvard.edu/abs/2011ApJ...737..103S} {737, 103}

\bibitem[\protect\citeauthoryear{{Schlegel}, {Finkbeiner}  \&
  {Davis}}{{Schlegel} et~al.}{1998}]{Schlegel1998}
{Schlegel} D.~J.,  {Finkbeiner} D.~P.,   {Davis} M.,  1998, \mn@doi [\apj]
  {10.1086/305772}, \href
  {https://ui.adsabs.harvard.edu/abs/1998ApJ...500..525S} {500, 525}

\bibitem[\protect\citeauthoryear{{Schwarz} et~al.,}{{Schwarz}
  et~al.}{2011}]{schwarz2011}
{Schwarz} G.~J.,  et~al., 2011, \mn@doi [\apjs] {10.1088/0067-0049/197/2/31},
  \href {https://ui.adsabs.harvard.edu/abs/2011ApJS..197...31S} {197, 31}

\bibitem[\protect\citeauthoryear{{Selvelli} \& {Gilmozzi}}{{Selvelli} \&
  {Gilmozzi}}{2013}]{Selvelli2013}
{Selvelli} P.,  {Gilmozzi} R.,  2013, \mn@doi [\aap]
  {10.1051/0004-6361/201220627}, \href
  {https://ui.adsabs.harvard.edu/abs/2013A&A...560A..49S} {560, A49}

\bibitem[\protect\citeauthoryear{{Selvelli} \& {Gilmozzi}}{{Selvelli} \&
  {Gilmozzi}}{2019}]{Selvelli2019}
{Selvelli} P.,  {Gilmozzi} R.,  2019, \mn@doi [\aap]
  {10.1051/0004-6361/201834238}, \href
  {https://ui.adsabs.harvard.edu/abs/2019A&A...622A.186S} {622, A186}

\bibitem[\protect\citeauthoryear{{Shu}}{{Shu}}{1992}]{Shu1992}
{Shu} F.~H.,  1992, {Physics of Astrophysics, Vol. II:Gas Dynamics}.
University Science Books

\bibitem[\protect\citeauthoryear{{Shull} \& {van Steenberg}}{{Shull} \& {van
  Steenberg}}{1982}]{Shull1982}
{Shull} J.~M.,  {van Steenberg} M.,  1982, \mn@doi [\apjs] {10.1086/190769},
  \href {https://ui.adsabs.harvard.edu/abs/1982ApJS...48...95S} {48, 95}

\bibitem[\protect\citeauthoryear{{Sokolovsky} et~al.,}{{Sokolovsky}
  et~al.}{2020}]{Sokolovsky2020}
{Sokolovsky} K.,  et~al., 2020, The Astronomer's Telegram, \href
  {https://ui.adsabs.harvard.edu/abs/2020ATel13653....1S} {13653, 1}

\bibitem[\protect\citeauthoryear{{Sordo} \& {Munari}}{{Sordo} \&
  {Munari}}{2006}]{Sordo2006}
{Sordo} R.,  {Munari} U.,  2006, \mn@doi [\aap] {10.1051/0004-6361:20054619},
  \href {https://ui.adsabs.harvard.edu/abs/2006A&A...452..735S} {452, 735}

\bibitem[\protect\citeauthoryear{{Spinoglio} \& {Malkan}}{{Spinoglio} \&
  {Malkan}}{1992}]{Spinoglio1992}
{Spinoglio} L.,  {Malkan} M.~A.,  1992, \mn@doi [\apj] {10.1086/171943}, \href
  {https://ui.adsabs.harvard.edu/abs/1992ApJ...399..504S} {399, 504}

\bibitem[\protect\citeauthoryear{{Srivastava}, {Ashok}, {Banerjee}  \&
  {Sand}}{{Srivastava} et~al.}{2015}]{Srivastava2015}
{Srivastava} M.~K.,  {Ashok} N.~M.,  {Banerjee} D.~P.~K.,   {Sand} D.,  2015,
  \mn@doi [\mnras] {10.1093/mnras/stv2094}, \href
  {https://ui.adsabs.harvard.edu/abs/2015MNRAS.454.1297S} {454, 1297}

\bibitem[\protect\citeauthoryear{{Srivastava}, {Banerjee}, {Ashok},
  {Venkataraman}, {Sand}  \& {Diamond}}{{Srivastava}
  et~al.}{2016}]{Srivastava2016}
{Srivastava} M.~K.,  {Banerjee} D.~P.~K.,  {Ashok} N.~M.,  {Venkataraman} V.,
  {Sand} D.,   {Diamond} T.,  2016, \mn@doi [\mnras] {10.1093/mnras/stw1807},
  \href {https://ui.adsabs.harvard.edu/abs/2016MNRAS.462.2074S} {462, 2074}

\bibitem[\protect\citeauthoryear{{Srivastava}, {Jangra}, {Dixit}, {Munjal},
  {Arora}  \& {Mavani}}{{Srivastava} et~al.}{2018}]{Srivastava2018}
{Srivastava} M.~K.,  {Jangra} M.,  {Dixit} V.,  {Munjal} B.~S.,  {Arora} H.,
  {Mavani} T.,  2018, in Ground-based and Airborne Instrumentation for
  Astronomy VII. p. 107024I, \mn@doi{10.1117/12.2309306}

\bibitem[\protect\citeauthoryear{{Srivastava}, {Kumar}, {Banerjee}  \&
  {Joshi}}{{Srivastava} et~al.}{2019}]{Srivastava2019}
{Srivastava} M.~K.,  {Kumar} V.,  {Banerjee} D.~P.~K.,   {Joshi} V.,  2019, The
  Astronomer's Telegram, \href
  {https://ui.adsabs.harvard.edu/abs/2019ATel13258....1S} {13258, 1}

\bibitem[\protect\citeauthoryear{{Srivastava}, {Kumar}, {Dixit}, {Patel},
  {Jangra}, {Rajpurohit}  \& {Mathur}}{{Srivastava}
  et~al.}{2021}]{Srivastava2021}
{Srivastava} M.~K.,  {Kumar} V.,  {Dixit} V.,  {Patel} A.,  {Jangra} M.,
  {Rajpurohit} A.~S.,   {Mathur} S.~N.,  2021, \mn@doi [Experimental Astronomy]
  {10.1007/s10686-021-09753-5}, \href
  {https://ui.adsabs.harvard.edu/abs/2021ExA...tmp...45S} {51, 345–382}

\bibitem[\protect\citeauthoryear{{Starrfield}, {Gehrz}  \&
  {Truran}}{{Starrfield} et~al.}{1997}]{Starrfield1997}
{Starrfield} S.,  {Gehrz} R.~D.,   {Truran} J.~W.,  1997, in {Bernatowicz}
  T.~J.,  {Zinner} E.,  eds,  American Institute of Physics Conference Series
  Vol. 402, Astrophysical implications of the laboratory study of presolar
  materials. pp 203--234, \mn@doi{10.1063/1.53312}

\bibitem[\protect\citeauthoryear{{Storey} \& {Hummer}}{{Storey} \&
  {Hummer}}{1995}]{Storey1995}
{Storey} P.~J.,  {Hummer} D.~G.,  1995, \mn@doi [\mnras]
  {10.1093/mnras/272.1.41}, \href
  {https://ui.adsabs.harvard.edu/abs/1995MNRAS.272...41S} {272, 41}

\bibitem[\protect\citeauthoryear{{Strope}, {Schaefer}  \& {Henden}}{{Strope}
  et~al.}{2010}]{Strope2010}
{Strope} R.~J.,  {Schaefer} B.~E.,   {Henden} A.~A.,  2010, \mn@doi [\aj]
  {10.1088/0004-6256/140/1/34}, \href
  {https://ui.adsabs.harvard.edu/abs/2010AJ....140...34S} {140, 34}

\bibitem[\protect\citeauthoryear{{Tanaka}, {Nogami}, {Fujii}, {Ayani}, {Kato},
  {Maehara}, {Kiyota}  \& {Nakajima}}{{Tanaka} et~al.}{2011}]{Tanaka2011}
{Tanaka} J.,  {Nogami} D.,  {Fujii} M.,  {Ayani} K.,  {Kato} T.,  {Maehara} H.,
   {Kiyota} S.,   {Nakajima} K.,  2011, \mn@doi [\pasj]
  {10.1093/pasj/63.4.911}, \href
  {https://ui.adsabs.harvard.edu/abs/2011PASJ...63..911T} {63, 911}

\bibitem[\protect\citeauthoryear{{Tatischeff} \& {Hernanz}}{{Tatischeff} \&
  {Hernanz}}{2007}]{Tatischeff2007}
{Tatischeff} V.,  {Hernanz} M.,  2007, \mn@doi [\apjl] {10.1086/520049}, \href
  {https://ui.adsabs.harvard.edu/abs/2007ApJ...663L.101T} {663, L101}

\bibitem[\protect\citeauthoryear{{Tody}}{{Tody}}{1986}]{Tody1986}
{Tody} D.,  1986, in {Crawford} D.~L.,  ed.,  Society of Photo-Optical
  Instrumentation Engineers (SPIE) Conference Series Vol. 627, Instrumentation
  in astronomy VI. p.~733, \mn@doi{10.1117/12.968154}

\bibitem[\protect\citeauthoryear{{Tonry} et~al.,}{{Tonry}
  et~al.}{2012}]{Tonry2012}
{Tonry} J.~L.,  et~al., 2012, \mn@doi [\apj] {10.1088/0004-637X/750/2/99},
  \href {https://ui.adsabs.harvard.edu/abs/2012ApJ...750...99T} {750, 99}

\bibitem[\protect\citeauthoryear{{UKIDSS Consortium}}{{UKIDSS
  Consortium}}{2012}]{UKIDSS2012}
{UKIDSS Consortium} 2012, VizieR Online Data Catalog, \href
  {https://ui.adsabs.harvard.edu/abs/2012yCat.2316....0U} {p. II/316}

\bibitem[\protect\citeauthoryear{{Vacca}, {Cushing}  \& {Rayner}}{{Vacca}
  et~al.}{2003}]{Vacca2003}
{Vacca} W.~D.,  {Cushing} M.~C.,   {Rayner} J.~T.,  2003, \mn@doi [\pasp]
  {10.1086/346193}, \href
  {https://ui.adsabs.harvard.edu/abs/2003PASP..115..389V} {115, 389}

\bibitem[\protect\citeauthoryear{{Warner}}{{Warner}}{1995}]{Warner1995}
{Warner} B.,  1995, Cambridge Astrophysics Series, \href
  {https://ui.adsabs.harvard.edu/abs/1995CAS....28.....W} {28}

\bibitem[\protect\citeauthoryear{{Williams}}{{Williams}}{1994}]{Williams1994}
{Williams} R.~E.,  1994, \mn@doi [\apj] {10.1086/174062}, \href
  {https://ui.adsabs.harvard.edu/abs/1994ApJ...426..279W} {426, 279}

\bibitem[\protect\citeauthoryear{{Williams} \& {Mason}}{{Williams} \&
  {Mason}}{2010}]{Williams2010}
{Williams} R.,  {Mason} E.,  2010, \mn@doi [\apss] {10.1007/s10509-010-0318-x},
  \href {https://ui.adsabs.harvard.edu/abs/2010Ap&SS.327..207W} {327, 207}

\bibitem[\protect\citeauthoryear{{Williams}, {Mason}, {Della Valle}  \&
  {Ederoclite}}{{Williams} et~al.}{2008}]{Williams2008}
{Williams} R.,  {Mason} E.,  {Della Valle} M.,   {Ederoclite} A.,  2008,
  \mn@doi [\apj] {10.1086/590056}, \href
  {https://ui.adsabs.harvard.edu/abs/2008ApJ...685..451W} {685, 451}

\bibitem[\protect\citeauthoryear{{Woodward}, {Banerjee}  \& {Evans}}{{Woodward}
  et~al.}{2020a}]{Woodward2020}
{Woodward} C.~E.,  {Banerjee} D.~P.~K.,   {Evans} A.,  2020a, The Astronomer's
  Telegram, \href {https://ui.adsabs.harvard.edu/abs/2020ATel13759....1W}
  {13759, 1}

\bibitem[\protect\citeauthoryear{{Woodward}, {Banerjee}  \& {Evans}}{{Woodward}
  et~al.}{2020b}]{Woodward2020b}
{Woodward} C.~E.,  {Banerjee} D.~P.~K.,   {Evans} A.,  2020b, The Astronomer's
  Telegram, \href {https://ui.adsabs.harvard.edu/abs/2020ATel14033....1W}
  {14033, 1}

\bibitem[\protect\citeauthoryear{{Wright} et~al.,}{{Wright}
  et~al.}{2010}]{Wright2010}
{Wright} E.~L.,  et~al., 2010, \mn@doi [\aj] {10.1088/0004-6256/140/6/1868},
  \href {https://ui.adsabs.harvard.edu/abs/2010AJ....140.1868W} {140, 1868}

\bibitem[\protect\citeauthoryear{{Zhang}, {Graziani}  \& {Pradhan}}{{Zhang}
  et~al.}{1994}]{Zhang1994}
{Zhang} H.~L.,  {Graziani} M.,   {Pradhan} A.~K.,  1994, \aap, \href
  {https://ui.adsabs.harvard.edu/abs/1994A&A...283..319Z} {283, 319}

\bibitem[\protect\citeauthoryear{{Zwitter} \& {Munari}}{{Zwitter} \&
  {Munari}}{2000}]{Zwitter2000}
{Zwitter} T.,  {Munari} U.,  2000, {An introduction to analysis of single
  dispersion spectra with IRAF}.
Asiago monografie ; vol. 1, Osservatori Astronomici di Padova e Asiago,
  Dipartimento di Astronomia dell'Universita di Padova

\bibitem[\protect\citeauthoryear{{della Valle} \& {Livio}}{{della Valle} \&
  {Livio}}{1995}]{DellaValle1995}
{della Valle} M.,  {Livio} M.,  1995, \mn@doi [\apj] {10.1086/176342}, \href
  {https://ui.adsabs.harvard.edu/abs/1995ApJ...452..704D} {452, 704}

\bibitem[\protect\citeauthoryear{{van den Bergh} \& {Younger}}{{van den Bergh}
  \& {Younger}}{1987}]{VanDen1987}
{van den Bergh} S.,  {Younger} P.~F.,  1987, \aaps, \href
  {https://ui.adsabs.harvard.edu/abs/1987A&AS...70..125V} {70, 125}

\makeatother
\end{thebibliography}
\bibliographystyle{mnras}

\end{document}